\documentclass[a4paper,12pt]{article}  
\usepackage{graphicx}
\usepackage{amsfonts}
\usepackage{amssymb}
\usepackage{amsmath}
\usepackage[square,comma,numbers,sort&compress]{natbib}
\usepackage{ifthen}
\usepackage{url}

\def\E{{\mathbb E}}
\def\P{{\mathbb P}}
\def\R{{\mathbb R}}

\def\I{{\mathbb I}}

\def\U{{\tilde{U}}}
\def\T{{\mathbb T}}

\def\pa{{\partial\Omega}}

\def\B{{\mathcal B}}

\def\ve{\varepsilon}
\def\r{{\bf r}}
\def\m{c}    

\def\K{W}    

\def\diam{\ell_{\rm diam}}

\def\<{<\hspace*{-0.5mm}}
\def\>{\hspace*{-0.5mm}>}

\def\2{\hspace*{-1.5mm}\raisebox{-0.9mm}{$_2$}}
\def\4{\hspace*{-1.5mm}\raisebox{-0.9mm}{$_4$}}

\newcommand{\flqq}{\raise 0.25ex\hbox{\tiny $\ll$}}
\newcommand{\frqq}{\raise 0.25ex\hbox{\tiny $\gg$}}

\def\r{{\bf r}}

\newcommand{\pagevide}{
\ifthenelse{\boolean{@twoside}}{\thispagestyle{empty}~\newpage}{} }

\title{Surveying Diffusion in Complex Geometries. An Essay. }

\author{Denis~S.~Grebenkov\\ ~ \\
\it Laboratoire de Physique de la Mati\`ere Condens\'ee,\\
\it CNRS -- Ecole Polytechnique, F-91128 Palaiseau, France\\
\it Email: denis.grebenkov@polytechnique.edu}

\begin{document}

\maketitle

\begin{abstract}
The surrounding world surprises us by the beauty and variety of
complex shapes that emerge from nanometric to macroscopic scales.
Natural or manufactured materials (sandstones, sedimentary rocks 
and cement), colloidal solutions (proteins and DNA), biological 
cells, tissues and organs (lungs, kidneys and placenta), they all 
present irregularly shaped ``scenes'' for a fundamental transport
``performance'', that is, diffusion.  Here, the geometrical complexity,
entangled with the stochastic character of diffusive motion, results 
in numerous fascinating and sometimes unexpected effects like 
diffusion screening or localization.  These effects control many
diffusion-mediated processes that play an important role in 
heterogeneous catalysis (reaction rate and overall production), 
biochemical mechanisms (protein search and migration), electrochemistry
(electrode-electrolyte impedance), growth phenomena (solidification,
viscous fingering), oil recovery (structure of sedimentary rocks), 
or building industry (cement hardening).  In spite of a long and 
rich history of academic and industrial research in this field, 
it is striking to see how little we know about diffusion in complex 
geometries, especially the one which occurs in three dimensions.

We present our recent results on restricted diffusion.  We look into
the role of geometrical complexity at different levels, from boundary
microroughness to hierarchical structure and connectivity of the whole
diffusion-confining domain.  We develop a new approach which consists
in combining fast random walk algorithms with spectral tools.  The
main focus is on studying diffusion in model complex geometries (von
Koch boundaries, Kitaoka acinus, etc.), as well as on developing and
testing spectral methods.  We aim at extending this knowledge and at
applying the accomplished arsenal of theoretical and numerical tools
to structures found in nature and industry (X-ray microtomography 3D
scans of a cement paste, 2D slices of human skin and the placenta,
etc.).
\end{abstract}

\newpage

\section*{Preface}

The essay is based on the author's research summary that has been
prepared for defending a Habilitation degree ({\it Habilitation \`a
Diriger des Recherches} in French).  This is a structured overview of
the author's works in different fields in which diffusion, geometry
and complexity are entangled.  As a summary of one scientist's
activity, the essay is unavoidably biased and is not meant to provide
a reader with a complete overview.  Although the bibliography is
extended, it certainly does not include all the references that are
relevant for these research fields.  In spite of this incompleteness,
a reader may hopefully find interesting results and intriguing open
problems for future research.

\vskip 5mm
\tableofcontents

\newpage
\section*{ Introduction }
\addcontentsline{toc}{section}{{Introduction}}

Irregular shapes are ubiquitous in fields as different as biology,
physiology, chemistry, electrochemistry, engineering, and material
sciences.  They play a crucial role in various natural phenomena and
condition multiple industrial processes.  For instance, a human being
needs around hundred square meters of alveolar surface in the lungs
for respiration.  Since this surface has to fit the rib cage, it is
necessarily irregular.  Catalysts are often manufactured to have a
rough boundary in order to increase the overall production because of
a larger reactive surface.  Oil-bearing sedimentary rocks and
sandstones, cement and concretes, biological and textile tissues
exhibit porous, often multiscale structures, in which transport
processes take place.  Either intrinsic or artificially implemented, a
geometrical complexity essentially determines the properties and the
overall functioning of an organ or an industrial device. Even for such
a well-studied transport process as diffusion, an irregular geometry
yields new, often unexpected phenomena.  The central question of this
work is {\it how a geometrical irregularity influences the dynamics of
a complex system and, in turn, what information on geometry can be
extracted from surveying this dynamics}.

We focus on diffusion for many reasons.  First, it provides a reliable
macroscopic framework for describing various microscopic dynamics
according to the central limit theorem \cite{Feller}.  As such,
diffusion turns out to be a fundamental transport mechanism in biology
and chemistry.  Second, diffusion is governed by a simple, though
remarkably rich, second order partial differential equation (PDE)
named diffusion (or heat) equation.  Its interpretation in terms of
Brownian motion provides the whole arsenal of powerful probabilistic
tools as well as a basis for intuitive reasoning.  Third, the spectral
theory of the underlying Laplace operator is probably the most
developed branch of functional analysis.  Although diffusion is
studied for a long time, the presence of a geometrical irregularity
may drastically alter well-established results.  As a matter of fact,
it is striking to see how little we know about diffusion in complex
geometries, especially the one which occurs in three dimensions.  This
is the reason for increasing interest in diffusion in recent decades.

Numerous questions and problems concerning diffusion in complex
geometries can be formally divided in three groups:
\begin{itemize}
\item
{\bf forward problems} which consist in finding transport properties
for a {\it known (defined) geometry} of a diffusion-confining domain.
In mathematical terms, this is equivalent to finding a solution of the
related PDE in this domain.  More specifically, we are interested in
finding a way to understand, both qualitatively and quantitatively,
how various geometrical features of the confining domain change
transport properties.

\item
{\bf inverse problems} which aim at determining an {\it unknown
geometry} of a diffusion-confining domain from measuring the transport
properties of a system.  This is the main purpose of imaging
techniques.  Diffusion-weighted nuclear magnetic resonance (NMR)
imaging is an example.  In this case, the magnetization attenuation
which is caused by restricted diffusion in a porous medium is measured
in order to get its global geometrical characteristics like
surface-to-volume ratio, pore size distribution, etc.  Kac's question
``Can one hear the shape of a drum?'' is the best illustration of
these issues \cite{Kac66}.  Given their practical importance (e.g., in
oil recovery or medical diagnosis), we shall bear in mind inverse
problems in the course of this work.

\item
{\bf optimization problems} which focus on conception and design of an
{\it optimal geometry} to maximize or minimize certain transport
characteristics of a system.  The examples are efficient heat
radiators, gas exchange units (e.g., filters), or catalysts, to name a
few.  Although we are not currently studying these questions,
optimization problems are considered as an important direction for
future research.

\end{itemize}

Throughout this work, we consider forward problems by adapting 
two approaches: Monte Carlo simulations and spectral analysis. 
As for Monte Carlo techniques, they prove to be efficient and 
very flexible tools for generating individual random trajectories 
of diffusing particles in complex geometries, including fractal 
boundaries, porous media, multiscale or branching structures, 
etc. (Section~\ref{sec:numerical}).  As far as spectral analysis 
is concerned, it provides a fundamental explanation and a clearer 
interpretation of the observed effects (Section~\ref{sec:spectral}).  
The goal here is to relate various features of restricted diffusion 
to the Laplace operator eigenbasis and then to investigate the latter 
in depth.  Since the numerical computation of the Laplace operator 
eigenbasis is a difficult task for complex geometries, we mainly
focused on simple domains (e.g., interval, disk, sphere).  For these, 
many characteristics of restricted diffusion can be found either 
analytically, or very accurately via spectral-oriented numerical 
techniques.  As for complex geometries, their systematic study 
is one of the principal research lines in future.

\vskip 2mm

The manuscript is organized as follows.  Section~\ref{sec:background}
briefly summarizes physical concepts and mathematical tools for
describing diffusion in complex media.  We defined two principal
research guidelines in order to present our main results: Monte Carlo
simulations (Section~\ref{sec:numerical}) and spectral analysis
(Section~\ref{sec:spectral}).  All the author's publications are 
available online%
\footnote{
\url{http://pmc.polytechnique.fr/pagesperso/dg/publi/publi_e.htm}}.


\newpage
\section{ Diffusion, geometry and complexity }
\label{sec:background}

\subsection{ Diffusion }
\label{sec:diffusion}

Diffusion is a fundamental transport mechanism, with countless
examples in nature and applications in sciences, from physics to
biology, chemistry, medicine, engineering, and economics.

\subsubsection*{ ``Upscaling'': toward macroscopic (or PDE) description }

The microscopic dynamics of gases and liquids can be described by
kinetic theory, when a collision integral accounts for short-time and
short-range interactions between atoms or molecules
\cite{Liboff,Landau10}.  From this classical point of view, the
trajectory of a particle is a sequence of ``moves'' between successive
``collisions'' (or effective interactions) with other particles or a
reservoir.  In many cases, the mean free path is much shorter than
macroscopic length scales at which the system is being observed.  In
addition, a large number of interactions considerably reduces or fully
eliminates memory effects at macroscopic time scales, so that the
specific motional features of the particle trajectory are averaged
out.  Performing such an ``upscaling'', one aims to introduce a {\it
macroscopic} density of particles and study its properties instead of
looking at the {\it microscopic} dynamics of these particles.  While
the microscopic dynamics may be defiantly complex, its coarser
macroscopic description is often appropriate and very accurate.

\subsubsection*{ Diffusion equation }

The diffusion (or heat) equation provides the simplest, though 
very general, coarser description of the microscopic dynamics:
\begin{equation}
\label{eq:diffusion}
\frac{\partial}{\partial t} c(\r, t) = D \Delta c(\r, t) ,
\end{equation}
where $\Delta = \partial^2/\partial x_1^2 + ... + \partial^2/\partial
x_d^2$ is the Laplace operator in $d$ dimensions, and $D$ the free
diffusion coefficient.  This equation states that the evolution of the
density of particles $c(\r, t)$ in time (the left-hand side term) is
only caused by local displacements of particles in space (the
right-hand side term).  The differential form of this equation
reflects the local character, in space and time, of the microscopic
dynamics.  If large moves were allowed during a short time, the
Laplace operator $\Delta$ would be replaced by an integral operator
accounting for particles coming from distant regions (spatially
nonlocal macroscopic dynamics).  If the microscopic moves were highly
correlated, the time derivative $\partial/\partial t$ would be
replaced by an integral operator accounting for the memory of the
whole preceding evolution (temporarily nonlocal macroscopic dynamics).
Although both situations can be encountered, we shall not consider
such anomalous diffusions (see
\cite{Klafter,Bouchaud90,Shlesinger99,Metzler00,Havlin02,Kimmich02}
and references therein).

\subsubsection*{ Diffusive propagator }

The diffusion equation describes the time evolution of the spatial
density $c(\r, t)$ from a given initial state $c(\r = \r_0, t = 0) =
\rho(\r_0)$.  Among various initial conditions, a point-like source
plays a special role.  In fact, it is natural to ask how the density 
of particles started from a given point $\r_0$ evolves?  The family 
$G_t(\r_0, \r)$ of these densities, parameterized by the
starting point $\r_0$, bears different names: diffusive propagator,
heat kernel, or Green function of diffusion equation.  These densities
satisfy the diffusion equation with respect to $\r$,
\begin{equation}
\label{eq:propagator}
\left(\frac{\partial}{\partial t} - D \Delta \right)G_t(\r_0, \r) = 0 ,
\end{equation}
with the initial condition for all particles to be concentrated in one
point $\r_0$ which is mathematically represented by the Dirac
distribution
\begin{equation}
\label{eq:delta}
G_{t=0}(\r, \r') = \delta(\r - \r') .
\end{equation}
The symmetry property $G_t(\r, \r') = G_t(\r', \r)$ holds.

The propagator is an elementary ``block'' describing the dynamics of
particles at macroscopic level.  As such, the propagator contains all
the available information about diffusion.  In particular, the
solution $c(\r, t)$ of diffusion equation (\ref{eq:diffusion}) with a
given initial density $\rho(\r_0)$ is simply expressed through the
propagator:
\begin{equation*}
c(\r, t) = \int\limits d\r_0 ~ \rho(\r_0) ~ G_t(\r_0, \r) .
\end{equation*}

\subsubsection*{ Extensions }

Apart from the aforementioned global modifications (spatially or
temporarily nonlocal dynamics), the diffusion equation
(\ref{eq:diffusion}) can be extended in many ways:
\begin{enumerate}
\item
In anisotropic media, diffusion in some spatial directions may be
faster or slower than in the others.  This can be easily incorporated
by adding weighting coefficients in front of the partial derivatives
$\partial^2/\partial x_k^2$.  In general, the Laplace operator can be
replaced by a second-order elliptic differential operator,
\begin{equation*}
{\mathcal D} = \sum\limits_{i,j=1}^d a_{i,j}(\r) \frac{\partial^2}{\partial x_i \partial x_j}
\end{equation*}
where the coefficients $a_{i,j}(\r)$ satisfy some requirements to
ensure the ellipticity \cite{Birman}.  Clearly, this extension
includes a spatially inhomogeneous diffusion coefficient.  It is worth
noting that many results for the Laplace operator relying on the
spectral analysis are directly applicable to this extension.  In what
follows, we shall focus on the Laplace operator.

\item
In some cases, the diffusion coefficient depends on time $t$ or/and
density $c(\r, t)$.  The latter dependence may describe saturation
effects in catalytic reactions.  However, this dependence would make
the diffusion equation nonlinear, significantly reducing the arsenal
of analytical tools.  For this reason, we shall not consider such
extensions.

\item
A bulk source or sink term can be added to the diffusion equation.
Its practical implementation is straightforward, especially when using
the spectral description. 

\item
An external scalar field $B(\r, t)$ can be included
\begin{equation}
\label{eq:diffusion2}
\frac{\partial}{\partial t} c(\r, t) = D \Delta c(\r, t) - \kappa B(\r, t) c(\r, t),
\end{equation}
where $\kappa$ is a parameter.  The last term accounts for the local
effects: loss or production of the particles, modification of their
state, etc.  The proportionality to the density $c(\r, t)$ reflects
that these local effects act on each particle individually and
independently from the other particles.  The second and higher-order
powers of the density may represent various chemical reactions, but we
shall not consider these nonlinear equations.

Depending on the application field, $B(\r, t)$ incorporates various
mechanisms.  If the bulk contains absorbing sinks or traps for
diffusing particles, $B(\r, t)$ represents the distribution of their
absorption or trapping rates.  If the particles possess some activity
function which can relax on bulk impurities (e.g., a magnetization of
a polarized spin carried by a diffusing nucleus), $B(\r, t)$ describes
bulk relaxation rate.  If the particles can be chemically transformed
by catalytic germs (or biological cells) distributed in the bulk,
$B(\r, t)$ characterizes the distribution of reaction rates.  In NMR,
$B(\r, t)$ with imaginary parameter $\kappa$ represents the effect of
the applied magnetic field onto the transverse magnetization
(Sect.~\ref{sec:NMR}).  In what follows, our main focus will be onto
Eq.~(\ref{eq:diffusion2}).

\end{enumerate}

\subsubsection*{ ``Downscaling'': toward probabilistic description }

In a coarser description via diffusion equation, all the information
about individual trajectories of the particles is lost.  Nevertheless,
thinking of diffusion in terms of individual particles can still be
instructive to better understand or interpret various characteristics
of diffusion.  As we already mentioned earlier, the real microscopic
dynamics is in general very complicated and intractable analytically.
To overcome this problem, one employs the concept of ``downscaling''
which consists in finding a simpler {\it microscopic} dynamics that
would lead to the chosen {\it macroscopic} description.  A Brownian
motion, mathematically defined as a stochastic process with
independent Gaussian increments \cite{Feller}, is a very natural
candidate for substituting the real microscopic dynamics.  In fact,
the diffusive propagator $G_t(\r_0, \r)$, satisfying
Eqs.~(\ref{eq:propagator},~\ref{eq:delta}), turns out to be the
probability density for Brownian motion to move from $\r_0$ to $\r$
during time $t$.  This is a direct link between probabilistic and PDE
descriptions.  Strictly speaking, this ``substitution'' is neither
unique, nor physically justified.  It is rather a useful model that
offers efficient probabilistic tools for theoretical and numerical
analysis.  In particular, random walks on a lattice (a discretized
version of Brownian motion) are broadly employed to model many
physical processes \cite{Hughes,Weiss94}.

\subsubsection*{ Feynman-Kac formula }

In practice, it is difficult to monitor the individual Brownian
trajectories of the diffusing particles.  It is much easier to measure
or investigate ``an observable'', or a functional, of these
trajectories.  We consider the random variable
\begin{equation}
\label{eq:phi}
\varphi = \int\limits_0^t dt'~ B(X_{t'}, t-t') ,
\end{equation}
where Brownian motion is started with a given initial density
$\rho(\r_0)$.  Intuitively, a given function $B(\r, t)$ can be thought
of as a distribution of ``markers'' to ``color'' the trajectory $X_t$
exploring different points or specific regions in space.  When the
diffusing species passes through these regions, the random variable
$\varphi$ accumulates the corresponding marks.  In other words,
different parts of the trajectory are weighted according to the
function $B(\r, t)$, encoding thus the whole stochastic process.  For
example, if $B(\r, t)$ represents the distribution of absorption or
relaxation rates in the bulk, $\varphi$ is the cumulant absorption
factor penalizing the trajectories that pass through the sinks or
traps.  In NMR, the encoding mechanism is experimentally realized by
applying an inhomogeneous magnetic field $B(\r, t)$, and $\varphi$ is
the total phase accumulated by an individual spin-bearing particle
during its diffusion in this field \cite{Grebenkov07}.

Since $\varphi$ is a random variable, it is convenient to consider the
Fourier or Laplace transform of its probability distribution which can
be respectively interpreted as characteristic function $\E\{
e^{iq\varphi}\}$ and survival probability $\E\{ e^{-\kappa
\varphi}\}$.  The Feynman-Kac formula relates these probabilistic
quantities to the solution of Eq.~(\ref{eq:diffusion2})
\begin{equation}
\label{eq:Feynman-Kac}
\E\{ e^{-\kappa \varphi} \} = \int\limits d\r ~ c(\r, t) 
\end{equation}
(if $\kappa$ is replaced by $-iq$, one obtains the characteristic
function) \cite{Kac49,Kac51,Freidlin,Simon,Bass}.  In what follows, we
shall frequently employ this relation to ``switch'' between
probabilistic and PDE representations.  In particular, the
probabilistic description suggests using Monte Carlo simulations for
solving PDE problems \cite{Sabelfeld,Sabelfeld2,Milshtein}.  We shall
consider various applications of this numerical technique in
Section~\ref{sec:numerical}.

\subsection{ Boundary condition }
\label{sec:boundary}

When the motion of diffusing species is restricted inside a confining
medium, physico-chemical or biological interactions between the
particles and the interface of the medium should be taken into
account.  For instance, paramagnetic impurities dispersed on the
interface cause surface relaxation in NMR experiments; cellular
membranes allow for a semi-permeable transport through the boundary;
chemical reaction may change the diffusive or magnetic properties of
the particle, etc.  A reliable description of these processes at
microscopic level is a challenging problem, demanding for example
accurate molecular dynamics simulations near the interface, or quantum
mechanics calculations.  At the time scale of the macroscopic
transport process, however, the contact with the interface is very
rapid so that the precise description of the interaction is often
irrelevant.  In analogy with diffusion coefficient $D$, effectively
representing the bulk dynamics, the interactions on the boundary can
macroscopically be described by a surface transport coefficient $\K$
\cite{Sapoval94,Sapoval96}.  In microbiology, this is the permeability 
characterizing the rate of transfer across a semi-permeable membrane.
In heterogeneous catalysis, $\K$ is the reactivity of a catalyst, that
is the rate at which diffusing species are chemically transformed into
other species after hitting the catalytic surface.  In NMR, $\K$ is
the surface relaxivity determining the rate at which the nuclei lose
their magnetization in the vicinity of the boundary.  In the two
latter cases, the transformed or relaxed species still remain inside
the confining domain but they do not participate at the transport
process any more (e.g., they do not contribute to formation of the
macroscopic signal, see Sect. \ref{sec:NMR}).

At macroscopic level, interactions with the interface are effectively
incorporated via a boundary condition, which is a mass conservation law:
the flux of particles from the bulk {\it towards} the boundary, $-D\partial
c(\r, t)/\partial n$, is equal to the flux {\it through} the boundary 
(or the flux of transformed or relaxed species) $\K c(\r, t)$, yielding
Robin (also known as Fourier, mixed, relaxing, radiative, or third)
boundary condition
\begin{equation}
\label{eq:Robin}
D \frac{\partial}{\partial n} c(\r, t) + \K c(\r, t) = 0 ,
\end{equation}
where $\partial/\partial n$ is the normal derivative pointing outward
the bulk (if the domain is a sphere, the normal derivative is equal to
the radial derivative).  When the surface transport coefficient $\K$
is zero (no flux through the boundary), one retrieves Neumann boundary
condition: $\partial c(\r, t)/\partial n = 0$.  The opposite limit of
infinite $\K$ (no resistance to the transfer through the boundary)
yields Dirichlet boundary condition: $c(\r, t) = 0$.  The intermediate
Robin boundary condition is a linear combination of these two extreme
cases, which are ``weighted'' by bulk and surface transport
coefficients $D$ and $\K$, respectively.  In other words, the ratio
$D/\K$ represents ``proportions'' of pure reflections (the first term)
and pure absorptions (the second term), mixing Neumann and Dirichlet
boundary conditions.

The role of Robin boundary condition for Laplacian transport was
thoroughly investigated (see \cite{Sapoval94,Sapoval96} and references
therein).  In particular, the ratio $D/\K$, which is homogeneous to a
length, bears different names: ``unscreened perimeter length'' or
``exploration perimeter or length'' \cite{Felici05}, ``Damk\"ohler
first ratio'' in chemistry \cite{Makhnovskii06}, ``surface relaxation
length'' in NMR \cite{Axelrod01}.  In the latter case, $D/\K$ is the
distance a particle should travel near the boundary before surface
relaxation reduces its expected magnetization.  In
Sect.~\ref{sec:spread}, we shall give another interpretation to this
exploration perimeter.

\subsection{ Reflected Brownian motion }
\label{sec:reflected}

In the above macroscopic description, restricting walls of a confining
medium were introduced through boundary conditions for diffusion
equation.  Alternatively, this effect can be incorporated for Brownian
motion in the probabilistic description.  We first consider a much
simpler situation of fully absorbing interface (Dirichlet boundary
condition).  For a Brownian motion $W_t$ started from a point $\r_0$
inside a confining domain $\Omega$, we define the first hitting time
$\T$ when $W_t$ reaches the boundary $\pa$ of $\Omega$:
\begin{equation}
\label{eq:T0}
\T = \inf\{ t > 0 ~:~ W_t \in \pa \} .
\end{equation}
Since the interface is fully absorbing, the process is stopped (or
killed) at this (random) moment.  Figuratively speaking, we simply
``close our eyes'' on what happens with Brownian motion after $\T$.
It is relatively easy to incorporate Dirichlet boundary condition by
restricting the analysis to times $t$ inferior to $\T$.

The situation with a reflecting interface (Neumann boundary condition)
is completely different.  In this case, one needs to keep Brownian
motion inside the confining domain without stopping it.  In other
words, we have to modify the local dynamics of this process to include
reflections on the boundary.  In sharp contrast with ordinary Brownian
motion, the construction of reflected Brownian motion strongly depends
on the geometry of a confining medium, being especially sophisticated 
for irregular boundaries.  The simplest example of this process is the 
absolute value of a one-dimensional Brownian motion $W_t$: $X_t = |W_t|$.  
Each time a diffusing particle crosses the boundary of the positive 
semi-axis (the end point 0), it is reflected toward the bulk (positive 
semi-axis).  In a general situation of a bounded domain $\Omega\subset\R^d$ 
with twice continuously differentiable boundary $\pa$, reflected Brownian 
motion $X_t$ can be defined as a solution of the stochastic differential 
equation, called Skorokhod equation \cite{Freidlin}:
\begin{equation*}
dX_t = dW_t + {\bf n}(X_t) \I_\pa(X_t) d\ell_t ,
\end{equation*}
where $W_t$ is the (ordinary) Brownian motion, ${\bf n}(\r)$ the
normal inward vector at boundary point $\r$, $\I_\pa(\r)$ the
indicator function of the boundary ($\I_\pa(\r) = 1$ if $\r\in\pa$,
and $0$ otherwise), and $\ell_t$ the local boundary time process, 
satisfying certain conditions \cite{Freidlin}.  The most unusual 
feature of this definition is that the single equation defines two 
processes, $X_t$
and $\ell_t$, strongly dependent on each other.  The intuitive meaning
of the Skorokhod equation is simple.  In the bulk, an infinitesimal
variation $dX_t$ of the reflected Brownian motion inside the confining
domain is only governed by the variation $dW_t$ of the (ordinary)
Brownian motion (the second term vanishes due to the indicator
function $\I_\pa$).  When the particle hits the boundary, the second
term does not allow to leave the domain leading to a variation
directed along the inward unit normal ${\bf n}(\r)$ towards the
interior of the domain.  At the same time, each encounter with the
boundary increases the local time $\ell_t$.  Being defined in this
way, reflected Brownian motion can be related to diffusion equation
with Neumann boundary condition.  It is worth noting that reflected
Brownian motion can alternatively be introduced as a continuous limit
of reflected random walks on a regular lattice, which are easier for
intuitive interpretation \cite{Bossy04,Burdzy08}.  We also mention
that reflected Brownian motion can be rigorously defined for domains
with irregular boundaries using the Dirichlet form method
\cite{Chen93,Fukushima}.

In the intermediate case of partially absorbing/reflecting interface
(Robin boundary condition), one can introduce ``partially reflected
Brownian motion'' \cite{Grebenkov06}.  This is reflected Brownian
motion which is conditioned to stop (i.e., to be absorbed) on the
boundary at random moment $\T_h$ when the local time $\ell_t$ exceeds
an independent exponentially distributed random variable $\xi$:
\begin{equation*}
\T_h = \inf\{ t > 0 ~:~ \ell_t \geq \xi \},  \hskip 10mm
\textrm{where} \hskip 3mm \P\{\xi  > \lambda \} = \exp[-\lambda h] .
\end{equation*}
The absorption can happen whenever partially reflected Brownian motion
hits the boundary, these hits being counted by the local time.  A
positive parameter $h$, a kind of absorption rate, is actually the
ratio between the surface and bulk transport coefficients, $h = L
\K/D$, $L$ being a characteristic size of the domain to get rid off
the dimensional units.  In turn, the exponential character of the
random ``threshold'' $\xi$ is related to the fact that the absorption
event is independent from hit to hit (as the exponential decay of
radioactive nuclei).

When $h$ goes to infinity (Dirichlet boundary condition, $\K =
\infty$), one gets $\xi = 0$ with probability $1$, so that the
stopping time $T_\infty$ describes the first moment when the local
time $\ell_t$ exceeds $0$.  This is exactly the moment when partially 
reflected Brownian motion hits the boundary for the first time.  One 
thus retrieves the above definition (\ref{eq:T0}) for a purely absorbing
boundary.  In the opposite limit $h\to 0$, $\xi = \infty$ with
probability $1$, yielding Neumann boundary condition ($\K = 0$) as
expected.  Varying $h$ allows one to explore various situations
between pure absorptions and pure reflections that makes partially
reflected Brownian motion to be a rich and flexible microscopic model
for diffusive transport.

\subsection{ Spectral analysis }
\label{sec:eigen}

To the diffusion (or heat) equation is associated a set of solutions
for a given confining domain.  A particular solution is fixed by
setting the initial condition.  On the other hand, one can remain in a
general frame and study the diffusion equation or, equivalently, the
Laplace operator.  In this so-called spectral approach, one is
searching for the Laplacian eigenfunctions and eigenvalues.  Instead
of looking for one particular solution, one aims to find all the
``relevant'' solutions at once.  It is not therefore surprising that
the computation of the eigenbasis, fully describing the operator, is a
difficult task.  But once it is worked out, any solution of diffusion
equation and any property related to the Laplace operator can be
deduced.

The eigenfunctions $u_m(\r)$ and eigenvalues $\lambda_m$ of the
Laplace operator $\Delta$ in a bounded domain $\Omega$ are defined 
as
\begin{subequations}
\begin{align}
\label{eq:eigen}
\Delta u_m(\r) + \lambda_m  u_m(\r) & =  0 ~~~ (\r \in \Omega), \\
\label{eq:eigen2}
D \frac{\partial}{\partial n} u_m(\r) + \K u_m(\r) & =  0 ~~~ (\r \in \pa),
\end{align}
\end{subequations}
with an integer index $m = 0, 1, 2, ...$.  The eigenvalues $\lambda_m$
are nonnegative, the eigenfunctions $u_m(\r)$ are orthogonal:
\begin{equation}
\label{eq:normalization}
\int\limits_\Omega d\r ~ u_m(\r) ~ u_{m'}^*(\r) = \delta_{m,m'} ,
\end{equation}
where the asterisk denotes complex conjugate, and $\delta_{m,m'}$ is
the Kronecker symbol ($\delta_{m,m'} = 1$ for $m=m'$, and $0$
otherwise).  Note that we explicitly fixed the normalization in
Eq.~(\ref{eq:normalization}).

The diffusive propagator has the following spectral decomposition
\cite{Arfken,Crank,Carslaw}
\begin{equation}
\label{eq:heat}
G_t(\r,\r') = \sum\limits_m u_m^*(\r) ~u_m(\r') ~e^{- D\lambda_m t} .
\end{equation}
One can easily check that this sum satisfies the diffusion equation
(\ref{eq:propagator}), while the initial condition (\ref{eq:delta})
at $t=0$ is guaranteed by the completeness relation
\begin{equation*}
\delta(\r - \r') = \sum\limits_m u_m(\r) ~u_m^*(\r') .
\end{equation*}
At last, the boundary condition for the diffusive propagator follows
immediately from Eq.~(\ref{eq:eigen2}).

It is clear that eigenfunctions and eigenvalues explicitly
determine the diffusive propagator.  The opposite is also true: the
propagator can be formally used to define the eigenfunctions and
eigenvalues.  In fact, the same amount of information about diffusive
motion is ``stored'' differently in the eigenbasis and in the
propagator, the latter mixing this information in a specific way.

\subsection{ Applications in NMR }
\label{sec:NMR}

\begin{figure}
\begin{center}
\includegraphics[width=150mm]{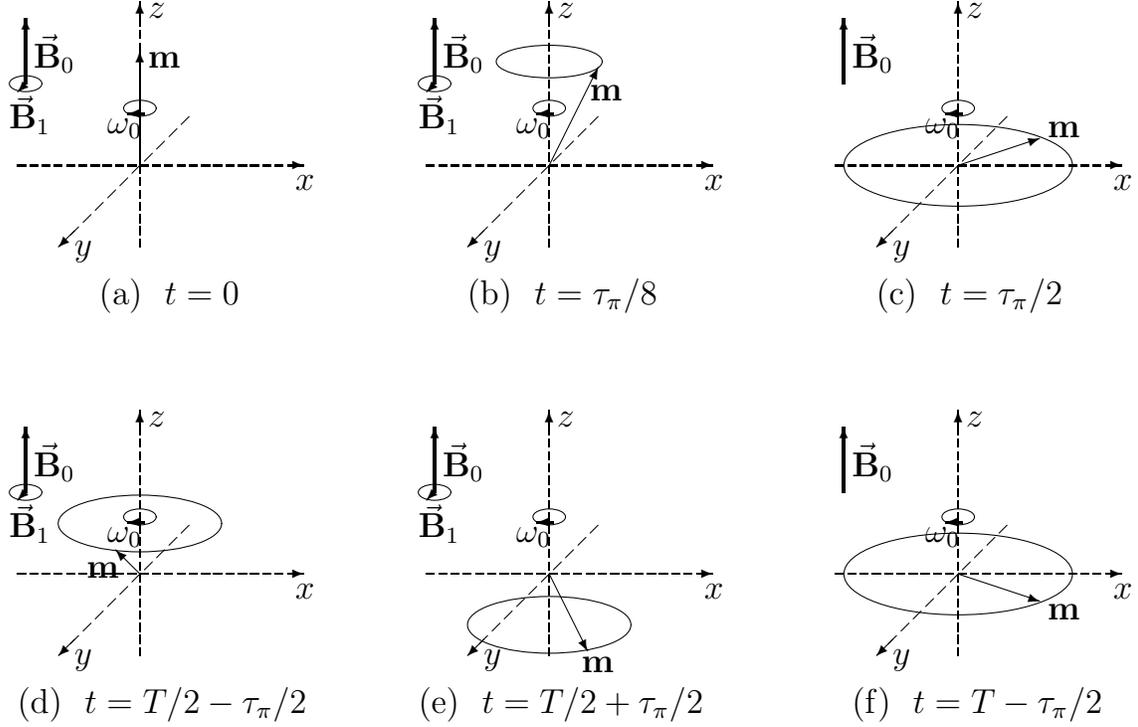}
\end{center}
\caption{
Schematic illustration of echo formation.  In a static field $B_0$
along the $z$ axis, the spins precess around this axis with the Larmor
frequency $\omega_0 = \gamma B_0$, and its magnetization ${\mathfrak
m}$ is directed along $z$.  (a) At time $t=0$, one applies a periodic
magnetic field $B_1$ rotating in the transverse plane $xy$ with
frequency $\omega_0$ (so-called $\pi/2$ or $90^\circ$ radio-frequency
(rf) pulse).  (b) Still precessing, the magnetization is turning,
linearly in time, toward the transverse plane.  (c) At $t =
\tau_\pi/2$ (here $\tau_\pi = \pi/(\gamma B_1)$), the magnetization
lies in the transverse plane, and the periodic magnetic field $B_1$
ceases.  During the following period $\Delta T = T/2-\tau_\pi$, the
magnetization slowly relaxes back to the longitudinal direction (axis
$z$).  (d) At $t = T/2 - \tau_\pi/2$, another magnetic field pulse is
applied during the time $\tau_\pi$ (so-called $\pi$ or $180^\circ$ rf
pulse).  (e) This pulse inverts the longitudinal direction of the
magnetization.  (f) After the $180^\circ$ rf pulse being turned off at
$t = T/2 - \tau_\pi/2$, the slow relaxation during the subsequent time
period $\Delta T$ returns the magnetization into the transverse plane.
Moreover, the magnetizations of various spins are refocused thus
forming a macroscopic signal (an echo) at $t=T-\tau_\pi/2$.  The rf
pulses are very short allowing their durations ($\tau_\pi/2$ and
$\tau_\pi$) to be neglected.  When the magnetic field is spatially
inhomogeneous, the refocusing is not complete because the spins,
precessing with varying Larmor frequencies, acquire different phase
shifts.  The resulting macroscopic signal depends on translational
dynamics of spins, allowing one to survey this dynamics in an
experiment. }
\label{fig:echo}
\end{figure}

Observation of translational dynamics requires a kind of ``marking''
or ``labeling'' of the traveling particles for tracking their
displacements in space.  A magnetic field is a superb experimental
tool for encoding the motion of spin-bearing particles.  For nuclei
with spin $1/2$ (e.g., protons of water), the magnetic field induces
energy splitting into two levels.  At thermal equilibrium, the
population of the nuclei at the lower energy (i.e., the spins parallel
with the magnetic field) is bigger than the population of the nuclei
at the higher energy (i.e., the spins antiparallel with the magnetic
field).  The difference in these populations creates local
magnetization which is parallel with the magnetic field (conventially,
the $z$ axis).  This magnetization can be flipped into the transverse
$xy$ plane by a $90^\circ$ radio-frequency (rf) magnetic field pulse
(Fig. \ref{fig:echo}).  The spins at position $\r$ start to precess
with the Larmor frequency $\gamma B(\r, t)$ which is proportional to
the magnetic field $B(\r, t)$, $\gamma$ being the gyromagnetic ratio
(a fundamental constant of the nucleus).  The use of a spatially
inhomogeneous magnetic field allows one to distinguish points or
regions in space, as the nuclei precess faster or slower in different
regions.  This mechanism is widely applied in experiments to monitor
the translational dynamics and to access the geometry of a confining
medium \cite{Callaghan}.

The $x$ and $y$ projections of the transverse magnetization are
conventionally treated as real and imaginary parts of a complex-valued
``density'' $c(\r, t)$.  The magnetic-field encoding is then
incorporated through Eq.~(\ref{eq:diffusion2}) with an {\it imaginary}
parameter $\kappa = i\gamma$, the last term in this equation being
responsible for describing precession of the transverse magnetization.
This equation, known as Bloch-Torrey equation \cite{Torrey56},
describes the evolution of $c(\r, t)$ according to two independent
mechanisms
\begin{itemize}
\item
diffusive migration of the spin-bearing particles (from $\r$ to
neighboring points), represented by the Laplace operator and
characterized by the free diffusion coefficient $D$;

\item
magnetic field encoding, when the spins at $\r$ acquire the phase
shift $\gamma B(\r, t) \tau$, resulting from their precession in the
transverse plane during a short time $\tau$.
\end{itemize}
The macroscopic signal at time $t$ is formed by the whole ensemble of
the spins diffusing inside the confining domain $\Omega$:
\begin{equation}
\label{eq:signal_integral}
E = \int\limits_\Omega d\r ~ c(\r, t) ~ \tilde{\rho}(\r) ,
\end{equation}
where $\tilde{\rho}(\r)$ is a sampling or pickup function of the
measuring coil or antenna (usually one tries to get $\tilde{\rho}(\r)$
as uniform as possible).  In analogy to Feynman-Kac formula
(\ref{eq:Feynman-Kac}), the signal can alternatively be written as the
characteristic function $\E\{e^{i\gamma \varphi}\}$ of the random 
dephasing $\varphi$ of an individual nucleus diffusing in an applied
magnetic field $B(\r, t)$ (the sampling function $\tilde{\rho}(\r)$
being implicitly incorporated here in the expectation).  Both
formulations are widely used in NMR literature (see \cite{Grebenkov07}
and references therein).

In most practical situations, the encoding term $B(\r, t)$ is a
superposition $B_0 + f(t) ({\bf g}\cdot \r)$ of a static magnetic
field $B_0$ and a linear magnetic field gradient ${\bf g}$, whose
dependence on time is represented through a dimensionless temporal
profile $f(t)$.  The form of the temporal profile can be easily varied
in modern MR scanners.  The simplest choice $f(t) = 1$ corresponds to
a free induction decay (FID) in a constant gradient.  Two identical
gradient pulses of opposite directions can be applied to form a
gradient echo.  If the nuclei were immobile, their dephasing by the
first gradient pulse would be fully compensated by the second gradient
pulse.  When the nuclei diffuse, they experience various magnetic
fields and dephase differently.  As a consequence, the rephasing is
not complete, and the gradient-echo amplitude attenuation
characterizes restricted diffusion (see below).

Measuring the macroscopic signal $E$ as a function of the
experimentally controlled parameters (gradient, diffusion time,
temporal profile, etc.), one aims to retrieve as much information on
the dynamics and the geometry of confinement as possible.  Solving
this inverse problem necessarily requires knowing how a given geometry
influences the macroscopic signal.  This forward problem will be
considered throughout the manuscript.

\subsection{ Complexity and irregular geometry }
\label{sec:irregular}

The diffusion (or heat) equation and the underlying spectral analysis
of the Laplace operator in a bounded domain is a classical problem in
mathematics \cite{Crank,Carslaw}.  What is new in our consideration of
this old problem is {\it an irregular geometry} of a medium or its
boundary.  As a matter of fact, porous materials, concentrated
colloidal suspensions, and physiological organs (such as lungs or
kidneys) are examples of systems developing large specific surfaces.
They all present a rich variety of shapes and exhibit complex
morphologies on a wide range of length scales.  In these systems, the
interfacial confinement strongly influences the diffusive dynamics of 
Brownian particles.

The human lung is a striking example of a complex transport system, in
which the irregular geometry plays a central role.  Its branching
structure guarantees a rapid access of a large quantity of fresh air
towards thirty thousands of pulmonary acini \cite{Weibel}.  Each
acinus is a porous, dichotomously branched gas exchange unit, in which
oxygen molecules diffuse toward the alveolar membranes for further
transfer to the blood.  The complexity of this geometry was shown to
lead to diffusion screening, one of the physical mechanisms for
regulating the physiological efficiency of the lungs \cite{Felici05}.
Many other physiological organs (such as kidneys, intestine, placenta) 
have also hierarchical multiscale structure.  Moreover, biological 
systems often present geometrical complexity even at microscopic 
scales (like the surface of cells \cite{Stone77}).

In material sciences, multiscale porous structures are as well
ubiquitous.  Typical examples are: sedimentary rocks, clays,
sandstones, plasters, cement, etc.  For instance, the geometrical
structure of a cement paste is responsible for mechanical strength and
the overall resistance of buildings, while its alteration in time by
diffusion-mediated reactions is a major reason for aging, cracks,
failures or even collapses \cite{Plassais05}.  A large part of
worldwide reserve of crude oil is ``imprisoned'' in sedimentary rocks,
and their porous morphology is a key factor for oil recovery
\cite{Kleinberg01}.  A microroughness of metallic electrodes
(scratches, abrasions, etc.) may substantially alter their transport
properties and functioning \cite{deLevie65}.  Finally, the tortuous
surface of proteins and DNA molecules, when looked at microscopic
scales, is relevant for diffusion-mediated biochemical reactions.  At
a larger scale, industrial catalysts are made of very irregular shape
to increase the catalytic surface aiming to improve the overall
production rate \cite{Coppens99}.

In summary, irregular shapes are encountered in nature and sciences
more often than one usually expects.  These irregularities span a wide
range of length scales (from nanometers to hundreds of meters) and
exist in a variety of systems, from biology to mineral sciences.  The
diffusive motion of particles inside such a medium is substantially
influenced by its complex geometry.  This is a major research topic of
this manuscript.


\newpage
\section{ Probabilistic insight: Monte Carlo simulations }
\label{sec:numerical}

In physics, speaking about diffusion in complex geometries invokes
numerical analysis by default.  In fact, only for few simple shapes
(such as a disk or a sphere), diffusion equation possesses explicit
solutions.  All other domains require solving diffusion equation on a
computer.  A variety of numerical methods can be roughly divided into
two groups.

In the first group (finite differences, finite elements, boundary
elements, etc.), a domain and/or its boundary is discretized with a
regular or adaptive mesh.  The original continuous problem is then
replaced by a set of linear equations to be solved numerically.  The
solution $c(\r, t)$ is obtained at all mesh nodes at successive time
moments.  Since the accuracy and efficiency of these deterministic
numerical schemes significantly rely on the discretization, mesh
construction for complex geometries turns out to be the key issue and
often a limiting factor, especially in three dimensions.

In the second group (Monte Carlo simulations), a probabilistic
interpretation of PDE is employed. Solving the original continuous
problem is replaced by modeling random trajectories of the underlying
diffusion process.  The solution is then obtained via Feynman-Kac
formulas (Eq.~(\ref{eq:Feynman-Kac}) or similar) by averaging over
random trajectories.  Since there is no discretization neither of the
domain, nor of boundary conditions (the most subtle stage for the
first group), Monte Carlo techniques are in general flexible and easy
to implement.  Two drawbacks should however be mentioned: a slow
convergence to the solution (error typically decreases as
$1/\sqrt{N}$, $N$ being the number of the simulated trajectories) and
the local character of simulations (the solution is obtained at one
spatial point).  For this latter reason, Monte Carlo simulations are
not well appropriate for getting the whole solution $c(\r, t)$.  In
contrast, this drawback is removed when one is interested in some
average of the solution like, for instance, in
Eq.~(\ref{eq:signal_integral}).  As illustrated in this Section, Monte
Carlo techniques become particularly well suited, especially for
studying diffusion in complex geometries.

\subsection*{ Monte Carlo simulations }

In the simplest case, one fixes a small time step $\tau$ and generates
a stochastic trajectory $X_t$ from a chosen starting point $x_0$ by
adding successively random increments $\xi_n$:
\begin{equation*}
X_0 = x_0,   \hskip 10mm   X_{(n+1)\tau} = X_{n\tau} + \xi_n   ~~~ (n \geq 0).
\end{equation*}
For Brownian motion, all the increments are independent identically
distributed random variables with normal (Gaussian) law, with mean
zero and standard deviation $\sigma = \sqrt{2D\tau}$ (an extension to
the multidimensional case is straightforward).  A discretized version
of this process is obtained by taking $\xi_n = \pm \sigma$ with
randomly chosen sign.  These are random walks on a lattice with mesh
$\sigma$.  Restricting walls can be taken into account either by
stopping the process after their hit (Dirichlet boundary condition),
or by reflecting the particle inside the domain (Neumann boundary
condition), or by a combination of both (Robin boundary condition).

Given the simplicity of this procedure, its implementation is easy for
various domains, but its practical applications are in general limited
to simple geometries.  In fact, an accurate modeling of the trajectory
requires to keep a typical step $\sigma$ (or the lattice mesh) smaller
than the smallest geometrical feature of a confining domain.  Since
complex geometries (e.g., fractals) are often multiscale, the
distances that should be explored by diffusing particles inside the
confining domain are by orders of magnitude longer than its smallest
geometrical features, requiring very long and time-consuming
simulations for a single trajectory.

\begin{figure}
\begin{center}
\includegraphics[width=140mm]{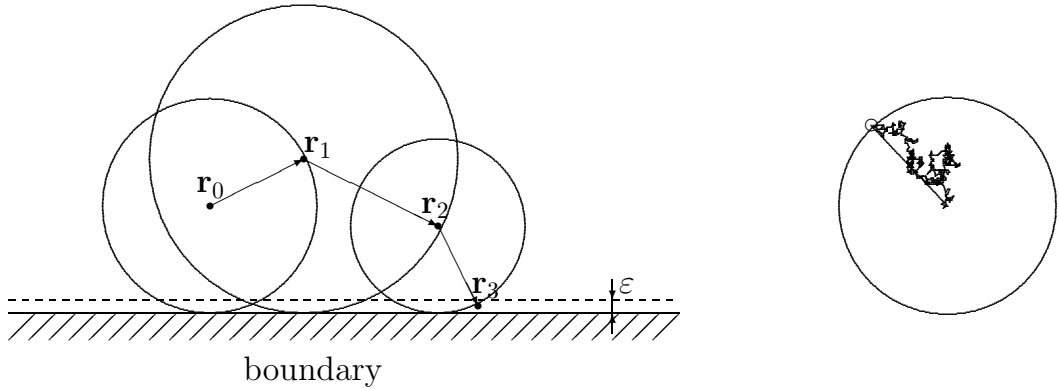}
\end{center}
\caption{
Illustration of a fast random walk algorithm.  From the current
position, one determines the distance to the boundary (here, for a
flat horizontal segment, it is the height of the current point above
the boundary) and draws the circle (or a sphere in 3D).  One then
chooses a uniformly distributed random point on this circle and makes
the jump.  From this new position, one repeats the above steps,
generating a sequence of points $\r_0$, $\r_1$, $\r_2$,..., until the
particle approaches the boundary closer than a chosen threshold $\ve$.
At this point, the simulation stops and the particle is considered as
having hit the boundary.  As illustrated on the right, each jump replaces
a Brownian path inside the circle that speeds up simulations
considerably. }
\label{fig:fast_walks}
\end{figure}

To overcome this difficulty, Muller proposed to replace Brownian
motion by an equivalent ``spherical process'' \cite{Muller56}.  This
method, often called ``fast random walks'', was employed by many
authors (see, e.g., \cite{Torquato89,Zheng89,Ossadnik91}).  The idea
is to explore a confining domain as fast as possible, without
violating probabilistic properties of Brownian motion.  Starting from
a given point, a diffusing particle executes a series of random jumps
in the bulk.  The jump length at each step is taken to be the distance
$\ell$ between the current position and the boundary
(Fig.~\ref{fig:fast_walks}).  Since the sphere (or the disk in 2d) of
radius $\ell$ centered at the current position resides in the interior
of the confining domain, Brownian motion inside this sphere is not
altered by the presence of the restricting walls.  Figuratively
speaking, a diffusing particle can ``see'' only a close neighborhood
of its current position so that it simply does not ``know'' about the
walls far away.  The continuity of Brownian motion implies that the
diffusing particle must intersect somewhere the frontier of this
sphere before hitting the boundary.  The rotational symmetry yields
the uniform distribution of intersection points.  It means that a
lengthy simulation of Brownian trajectory inside the sphere can be
replaced by a single random jump from its center to a uniformly
distributed point on its frontier (Fig.~\ref{fig:fast_walks}).  This
``trick'' drastically speeds up Monte Carlo simulations since at each
step the largest possible exploration is performed.

In practice, the computation is reduced to finding the distance from
any interior point to the boundary.  Depending on the studied
geometry, this problem can be solved in different ways.  In the next
subsection, a geometry-adapted fast random walks (GAFRW) algorithm is
presented.

\subsection{ Geometry-adapted fast random walks (GAFRW) }
\label{sec:GAFRW}

Fractals are often used as a paradigm of complex domains
\cite{Mandelbrot,Feder,Sapoval,Gouyet}.  On one hand, fractals are very
irregular shapes which exhibit geometrical details at various length
scales, ``exploding'' the classical notions of length, surface area or
volume.  On the other hand, self-similar or self-affine hierarchical
structures help to perform accurate theoretical and numerical analysis
on fractals.  In particular, the self-similarity of von Koch curves
and surfaces allowed us for a rapid computation of the distance from
any interior point to these boundaries \cite{Grebenkov05b}.

\begin{figure}
\begin{center}
\includegraphics[width=150mm]{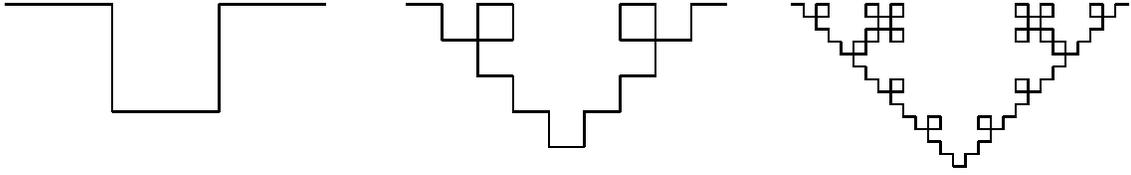}
\end{center}
\caption{
Three generations of the quadratic von Koch curve of fractal dimension
$D_0 = \ln 5/\ln 3\approx 1.465$.  At each iteration, one replaces all
linear segments by the rescaled generator (first generation). }
\label{fig:koch_quad}
\end{figure}

\begin{figure}
\begin{center}
\includegraphics[width=120mm]{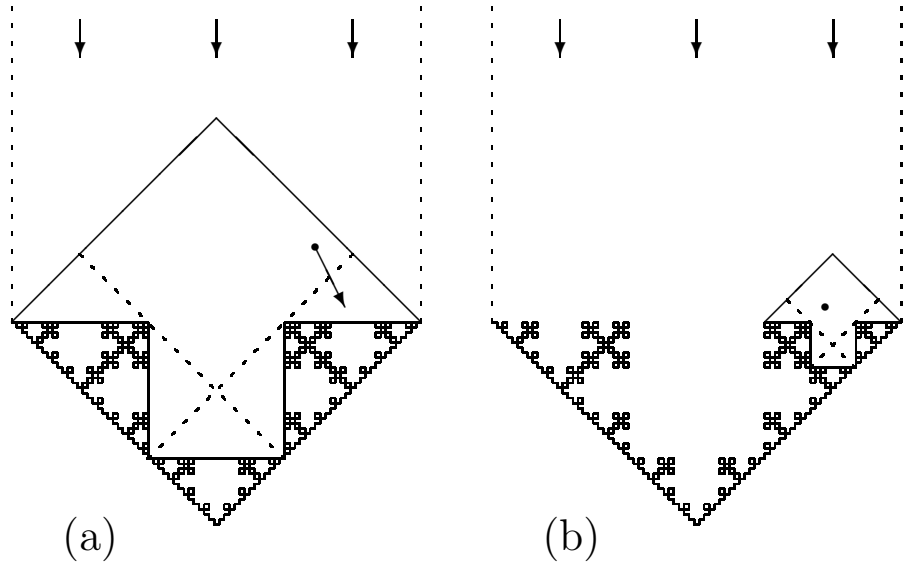}
\end{center}
\caption{
Basic arrow-like cell which is divided into the rotated square and
five small triangles (a).  Once a Brownian particle arrived into the
rotated square, the distance between its current position (full
circle) and the boundary of the arrow-like cell (the generator) can be
computed explicitly.  Random jumps inside the rotated square can
therefore be executed, until the Brownian particle either exits from
the arrow-like cell, or enters into a small triangle.  In the latter
case, the Brownian particle starts to ``see'' the geometrical details
of the next generation.  The rescaled arrow-like cell (b) is then used
to compute the distance to the boundary.  Note that a distant source
is placed on the top of the figure, while two vertical dotted lines
delimit the interior of the confining domain. }
\label{fig:gafrw}
\end{figure}

To illustrate these ideas, we consider the quadratic von Koch curve
(Fig.~\ref{fig:koch_quad}).  When a random walker is far from the
boundary, it does not ``distinguish'' its geometrical details.  One
can thus estimate the distance by considering the coarsest generation
(Fig.~\ref{fig:gafrw}).  Getting closer and closer to the boundary,
the random walker starts to recognize smaller and smaller geometrical
details.  But at the same time, when small details appear in view, the
rest of the boundary becomes ``invisible''.  Consequently, one can
explicitly determine the distance by examining only the local
geometrical environment (see \cite{Grebenkov05b} for details).

When the particle approaches the boundary closer than a chosen
threshold, we say that it hits the boundary.  Depending on the
boundary condition and the problem at hand (see below), the particle
is either absorbed (simulation is stopped), or reflected (simulation
is resumed from a nearby bulk point).

An advantage and eventual drawback of this GAFRW algorithm is the need
for a specific implementation for each studied geometry.  The
algorithm was developed for the above quadratic von Koch curve and the
cubic von Koch surface of fractal dimension $D_0 = \ln 13/\ln 3$, as
well as for triangular von Koch curves of variable fractal dimension
(Sect.~\ref{sec:triangular}).  In the next subsections, we shall
discuss various applications of this algorithm.

\subsection{ Multifractal properties of the harmonic measure }
\label{sec:harmonic}

The initial purpose for developing GAFRW algorithm was for studying
the accessibility of a boundary to diffusing particles, which is of
primary importance in growth and transport phenomena.  Actually, the
transfer or reactive capacity of an interface is crucially limited by
its accessibility for Brownian motion, the effect known as diffusion
screening (similar to electric screening in electrostatics).  A
particle diffusing towards an irregular interface has very little
chance to reach the bottom of a deep ``fjord'' before hitting more
prominent (and easier accessible) points.  The resulting distribution
of diffusive fluxes or arrival probabilities on the boundary is
therefore very uneven, a small fraction of the boundary receiving the
overwhelming majority of the diffusing particles.  The diffusion
screening limits the overall production of species in the
diffusion-limited regime of heterogeneous catalysis but allows one to
design long-working catalysts \cite{Coppens99}.

Mathematically, this accessibility is characterized by a harmonic
measure, which is defined for each (Borel) subset of the boundary as
the probability for Brownian motion to reach the boundary of the
confining domain for the first time at this subset \cite{Garnett}.
This measure governs random growth processes (e.g., diffusion-limited
aggregation, dendric growth, solidification, viscous fingering,
morphogenesis), primary current distribution in electrochemistry,
distribution of particle flows on a membrane, electric charge
distribution on a metallic surface, etc.

\subsubsection{ Multifractal analysis }

\begin{figure}
\begin{center}
\includegraphics[width=60mm]{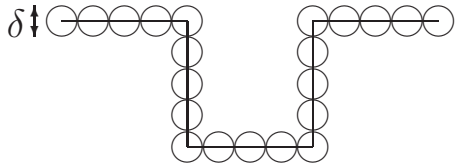}
\end{center}
\caption{
A given boundary (here, the first generation of the quadratic von Koch
curve) is covered by disks of diameter $\delta$. }
\label{fig:cover}
\end{figure}

For a smooth boundary, the harmonic measure is fully characterized by
its density.  When the boundary is irregular (e.g., fractal), there is
no density, though the harmonic measure is still well defined.  In
this case, one uses the multifractal dimensions $D_q$ of the harmonic
measure to characterize its variations over different length scales.

Let a finite generation $g$ of a fractal boundary be covered by
disjoint compact sets of diameter $\delta$ (e.g., disks, spheres,
cubes), as illustrated in Fig.~\ref{fig:cover}.  The harmonic measure
at the scale $\delta$ is represented by a finite set of hitting
probabilities $p_{k,\delta,g}$.  The variation of the harmonic measure
with the scale $\delta$ can be characterized by the behavior of the
moments
\begin{equation}
\label{eq:moments}
\zeta(q,\delta,g) = \sum\limits_k (p_{k,\delta,g})^q .
\end{equation}
When $\delta$ goes to $0$, $\zeta(q,\delta, g)$ scales as
$\delta^{(q-1)D_q}$.  For a smooth $d$-dimensional boundary, all the
$D_q$ are trivially equal to $d$, independently of the real parameter
$q$.  Geometrical irregularities may lead to different scaling of the
harmonic measure for various $q$.  Except for few classes of fractals
\cite{Duplantier99a,Duplantier99b,Duplantier00}, the multifractal
dimensions of the harmonic measure need to be determined numerically.
The GAFRW algorithm was used to simulate Brownian trajectories from a
distant source toward von Koch boundaries and to calculate
approximately the distribution of the hitting probabilities
$p_{k,\delta,g}$ at a given scale $\delta$.  Defining the
scale-dependent multifractal exponents as
\begin{equation*}
D_{q,\delta,g} = \frac{\ln \zeta(q,\delta,g)}{(q-1)\ln \delta},
\end{equation*}
one can obtain the multifractal exponents $D_q$ by interpolation as
$g\to\infty$ and $\delta\to 0$.

The efficiency of the GAFRW algorithm allowed us to simulate up to
$10^{10}$ trajectories, while the generation order could be increased
up to $g=10$ in 2D, and $g=7$ in 3D.  It is worth noting that these
``limiting values'' can be further extended by using parallel computation.

\subsubsection{ Interpolation of the scale-dependent exponents }

First, we investigated the scale-dependent multifractal dimensions
$D_{q,\delta,g}$ and its interpolation as the generation order $g$
goes to infinity and the scale $\delta$ goes to $0$.  This is {\it a
priori} a difficult problem because numerical simulations can only be
realized for small generation orders and limited range of scales.  We
discovered and numerically checked the following interpolation formula
\cite{Grebenkov05b}
\begin{equation*}
D_{q,\delta_g, g} = D_q + c_q \frac{1}{g} + O(e^{-\alpha_q g}) ,
\end{equation*}
where the scale $\delta_g = (1/3)^g$ is chosen to be the smallest
geometrical feature of the $g$-th generation, and $c_q$ and $\alpha_q$
are parameters depending on $q$.  This means that for a moderately
large generation order $g$, $D_{q,\delta_g, g}$ is almost a linear
function of $1/g$, up to exponentially small corrections.  A linear
regression of the values $D_{q,\delta_g,g}$ for several generation
orders allows for a very accurate computation of the limiting values
$D_q$ of the multifractal dimensions.  The accuracy of the whole
computational method was verified by calculating the information
dimension $D_1$ which is equal to $1$ for any planar simply connected
set (a result known as Makarov's theorem) \cite{Makarov85}.  For the
quadratic von Koch curve, we obtained $D_1 = 1.0000\pm 0.0001$, such a
precision being classified as very high for this kind of numerical
computation.

\begin{figure}
\begin{center}
\includegraphics[width=160mm]{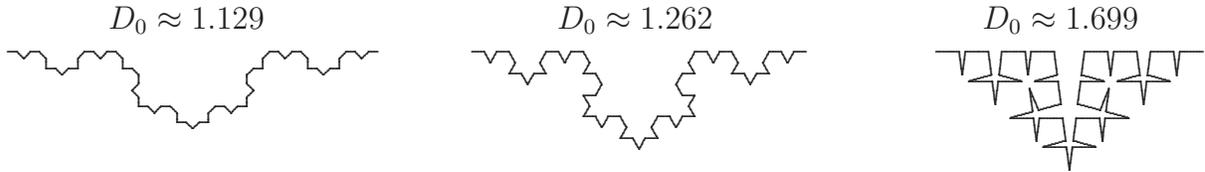}
\end{center}
\caption{
The third generation of three triangular von Koch curves. }
\label{fig:koch_triang_three}
\end{figure}

\subsubsection{ Harmonic measure in 3D }
\label{sec:HM_3d}

The second important result of \cite{Grebenkov05b} is the numerical
computation of the multifractal dimensions for the cubic von Koch
surface of fractal dimension $D_0 = \ln 13/\ln 3$.  For the first
time, the information dimension $D_1$ of a fractal surface was
accurately calculated.  Contrarily to an intuitive (but false)
extension of the Makarov's theorem in three dimensions, we obtained
$D_1 = 2.007\pm 0.002$, providing a concrete numerical counter-example
to such an extension.  It is worth noting that the first mathematical
counter-example, for which $D_1$ was rigorously shown to be strictly
greater than $2$, was given in \cite{Wolff95}, but without providing
the value of $D_1$.  Given that the obtained value $2.007$ is very
close to $2$ for the cubic von Koch surface, one may wonder whether
may exist or not irregular surfaces for which the excess $D_1 - 2$ is
relatively large.  Such surfaces would be particularly promising for
designing efficient exchangers for diffusion-limited transport.

\subsubsection{ What makes a boundary less accessible? }
\label{sec:triangular}

Although the scaling properties of the harmonic measure were largely
studied, the relation between the geometry itself and its multifractal
dimensions is still obscure.  Is a more irregular boundary (greater
fractal dimension $D_0$) more screened?  How does the presence of deep
fjords or a pore network modify the surface accessibility?  More
generally, what makes a boundary less accessible?  This last question,
which is closely related to different optimization problems in
chemical engineering, has been addressed in \cite{Grebenkov05a}.  By
implementing the GAFRW algorithm, we calculated the multifractal
dimensions $D_q$ for two families of self-similar triangular von Koch
curves of variable fractal dimension $D_0$
(Fig.~\ref{fig:koch_triang_three}).  Changing the angle $\alpha$
between two intermediate segments of the generator from $\pi$ to $0$,
the fractal dimension $D_0$ of these curves can be varied continuously
from $1$ to $2$:
\begin{equation*}
D_0 = \frac{\ln 4}{\ln(2 + 2\sin(\alpha/2))} .
\end{equation*}

\begin{figure}
\begin{center}
\includegraphics[width=80mm]{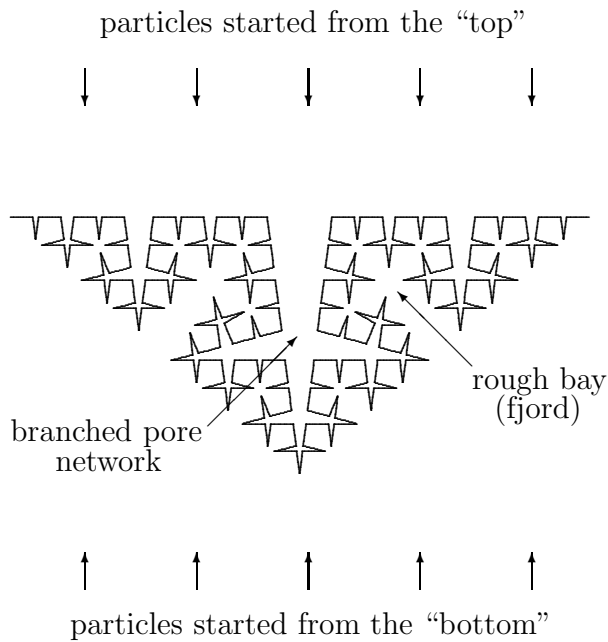}
\end{center}
\caption{
Fourth generation of the self-similar triangular von Koch curve
with Hausdorff dimension $D_0 \simeq 1.699$.  Depending on the
side exposed to diffusing particles, this curve allows one to model
either branched pore networks (source at the ``top'') or fjord-like
rough bays (source at the ``bottom''). }
\label{fig:koch_triang}
\end{figure}

Depending on the side which is exposed to diffusing particles, the
shape of these curves is geometrically very different so that one can
speak about two families.  Starting from a distant source at the top
of Fig.~\ref{fig:koch_triang}, the particles progressively penetrate
into smaller and smaller pores of the material.  Such curves (called
``top-seen'') could mimic the geometry of a branched pore network.  In
contrast, the diffusing particles started at the bottom of
Fig.~\ref{fig:koch_triang} arrive onto a rough surface with a
fjord-like pore structure.  The curves of this family are called
``bottom-seen''.

\begin{figure}
\begin{center}
\includegraphics[width=100mm]{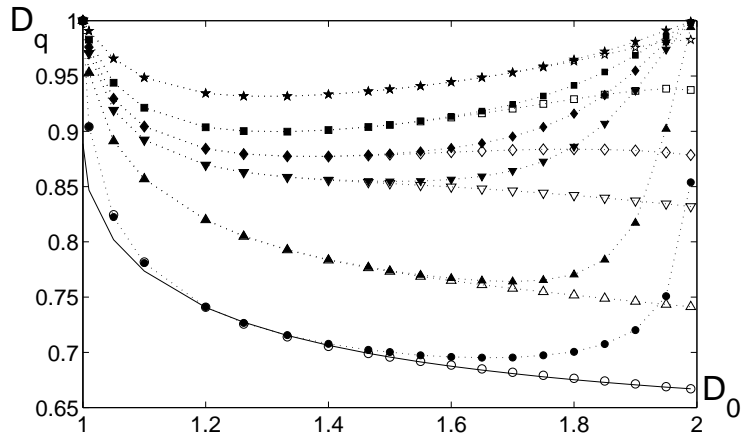}
\end{center}
\caption{
Positive-order multifractal dimensions $D_q$ of the harmonic measure
on the ``top-seen'' (solid symbols) and ``bottom-seen'' (open symbols)
triangular von Koch curves of different fractal dimension $D_0$: $q =
2$ (stars), $q = 3$ (squares), $q = 4$ (diamonds), $q = 5$
(up-pointing triangles), $q = 10$ (down-pointing triangles), and $q =
\infty$ (circles).  Dotted lines are a guide to the eye.  }
\label{fig:koch_triang_Dq}
\end{figure}

The first and quite surprising result is that the multifractal
dimensions $D_q$ (shown on Fig.~\ref{fig:koch_triang_Dq}) turn out to
be almost identical for the ``top-seen'' and ``bottom-seen'' curves of
the same fractal dimension when $D_0\leq 1.3$.  Being apparently
different geometrically, these two types of morphologies are
essentially indistinguishable for diffusing particles, i.e., when they
are ``seen'' by the harmonic measure.  At the same time, the spatial
distributions of hitting probabilities $p_{k,\delta,g}$ over these two
boundaries are very different.  From a certain value of $D_0$, the
multifractal dimensions for the ``top-seen'' and ``bottom-seen''
curves split.  In the former case, these dimensions approach $1$,
while in the latter case, they converge to smaller values.  This
behavior can be qualitatively explained by geometrical arguments (see
\cite{Grebenkov05a} for details).

From this study, we draw attention to two facts.  First, the fractal
dimension $D_0$, as a classical ``measure'' of the geometrical
complexity, is not determinant for the diffusion screening: the
harmonic measure on the ``top-seen'' curves with $D_0$ close to $2$
exhibits almost the same scaling behavior as the harmonic measure on a
smooth boundary since all the $D_q$ are close to $1$.  This result
provides a new insight into practical applications, e.g., in chemical
engineering.  Indeed, it shows that one does not need to explore very
irregular shapes to diminish the multifractal dimensions.  Second, the
multifractal dimensions of the harmonic measure on the ``top-seen''
and ``bottom-seen'' curves are almost identical for moderate values of
$D_0$.  This means that the harmonic measure is not sensitive to
distinguish these boundaries of quite different geometry.  The
hierarchical self-similar structure of these curves is thus more
determinant than their geometrical details.

\subsection{ Spread harmonic measures }
\label{sec:spread}

The harmonic measure characterizes only the accessibility of a
surface: how the particles reach the surface for the first time.  When
the surface is fully absorbing (infinite reaction or transfer rate
$\K$), the motion stops immediately after the first hit.  When $\K$ is
finite, the particle can be reflected from the surface to the bulk to
resume its diffusion.  In this case, the first arrival point and the
final absorption (or reaction) point do not necessarily coincide.  It
is the distribution of the absorption points that is really important
for transport phenomena.

\subsubsection{ Effect of spreading and exploration length }

At microscopic level, a finite transfer rate can be incorporated by
introducing a sticking probability $\sigma$: once a particle reaches
the boundary, it is either transferred (absorbed, relaxed, reacted,
etc.) with probability $\sigma$, or reflected into the bulk with
probability $1-\sigma$ to resume its diffusive motion.  For random
walks on a lattice with mesh $a$, the sticking probability $\sigma$
was related to the exploration length $D/\K$ as $\sigma = a/(a +
D/\K)$ \cite{Grebenkov03}.  The smaller the sticking probability, the
larger the number of reflections that corresponds to ``easy access,
difficult transfer'' situation ($D/\K$ is large).  In this case, the
particle has to explore a certain region of the surface around the
first hitting point before being finally absorbed (or transferred).
While the probability distribution of the arrival points is in general
sharp and uneven due to diffusion screening, the distribution of the
absorption points, at which the particles are transferred across the
boundary, is getting smoother and spread by multiple reflections
\cite{Filoche99}.  Figuratively speaking, reflections ``fight''
against diffusion screening and reduce its effect.  In the ultimate
case of fully reflecting boundary ($\K = 0$ and $\sigma = 0$), each
particle would explore the whole confining domain in an even manner,
without diffusion screening at all.

The exploration length $D/\K$ controls the spreading effect: larger
the $D/\K$, further the particles spread around the first hitting
point until their absorption.  For a flat boundary (straight line in
2D or plane in 3D), we showed that approximately half of particles are
absorbed by the disk of radius $D/\K$ around the first hitting point
\cite{Sapoval05}.  The exploration length $D/\K$ is thus a physical
yardstick for quantifying the geometry.

\subsubsection{ Scaling properties of the spread harmonic measures } 

For a smooth boundary, the ``spread harmonic measure'' is defined for
any (Borel) subset of the boundary as the probability for absorption
of the partially reflected Brownian motion on this subset.  This is a
family of measures which are naturally parameterized by the length
$D/\K$, ranging from the harmonic measure at $D/\K = 0$ to the
Lebesgue measure at $D/\K = \infty$.  Then we studied numerically the
scaling properties of the spread harmonic measures on finite
generations of the quadratic von Koch curve \cite{Grebenkov06c}.

The GAFRW algorithm was well suited for simulating Brownian
trajectories in the bulk.  The new implemented feature was partial
reflections from the boundary.  When the particle arrives on the
boundary, a random number is generated to decide whether the particle
is absorbed or not.  If not, the particle is reflected (jumped at some
small fixed distance $a$ from the boundary) to resume its diffusive
motion.  As the exploration length is typically much larger than $a$,
the sticking probability is very small, leading to a large number of
reflections.  The efficiency and rapidity of the GAFRW algorithm were
therefore crucial for this study.

\begin{figure}
\begin{center}
\includegraphics[width=160mm]{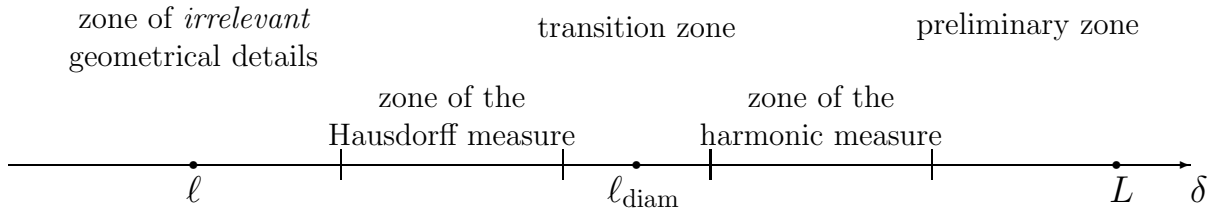}
\end{center}
\caption{
Schematic representation of scale zones of the spread harmonic
measure.  When $\delta$ is of order of the smallest geometrical detail
$\ell$, or smaller (on the left), one is looking at the harmonic
measure of a linear segment.  When $\delta$ is of order of the total
size (diameter) $L$ of the curve, or larger (on the right), the
approximation of the spread harmonic measures is too coarse.  The most
interesting region $\ell \ll \delta \ll L$ includes the transition
between the Hausdorff and harmonic measures.  }
\label{fig:scaling_zones}
\end{figure}

\begin{figure}
\begin{center}
\includegraphics[width=80mm]{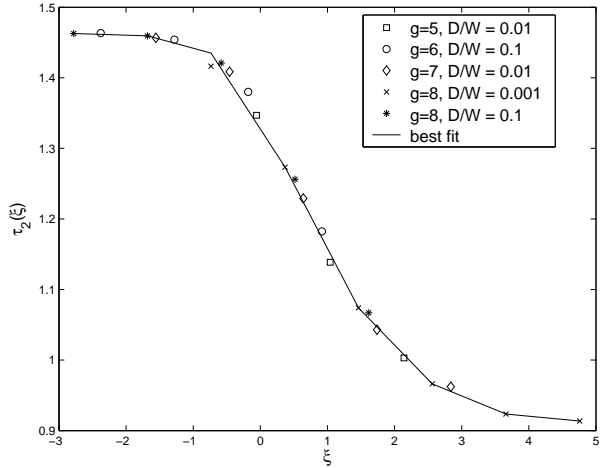}
\end{center}
\caption{
The scale-dependent multifractal dimension (with $q=2$) of the spread
harmonic measures on generations $g = 5, 6, 7, 8$ of the quadratic von
Koch curve with different values of the exploration length $D/\K$.
All these dependencies, considered as functions of the scaling
parameter $\xi = \ln(\delta/\diam)$, fall onto the same master curve
$\tau_2(\xi)$.  For small $\xi$ or $\delta$ (on the left), one
retrieves the multifractal dimension of the Hausdorff measure, which
is equal to the fractal dimension of the curve, $D_0 =
\ln 5/\ln 3 \simeq 1.465$.  For large $\xi$ or $\delta$ (on the
right), $\tau_2(\xi)$ approaches the correlation dimension of the
harmonic measure, $D_2 \simeq 0.8925$ that we found in
\cite{Grebenkov05b}.}
\label{fig:scaling_tau2}
\end{figure}

As in Sect.~\ref{sec:harmonic}, the spread harmonic measures can be
characterized by a set of multifractal exponents showing variations of
the moments in Eq.~(\ref{eq:moments}) with the scale $\delta$.  The
qualitative arguments for determining their scaling properties relied
on comparison between several length scales: the smallest and largest
geometrical lengths $\ell$ and $L$, the scale $\delta$, and the
exploration length $D/\K$.  Knowing for a straight line that the
exploration length characterizes the region around the first hitting
point where a half of particles are absorbed, we conjectured that, for
irregular curves, $D/\K$ represents a perimeter of this region.  For a
fractal curve, the diameter of this region, $\diam$, is related to its
perimeter by a scaling relation, $\diam \sim \ell
[(D/\K)/\ell]^{1/D_0}$, $D_0$ being the fractal dimension and $\ell$
the smallest lengthscale of the boundary.  The cases $\diam \leq \ell$
and $\diam \geq L$ trivially correspond to (almost) harmonic measure
and (almost) Hausdorff measure (the Hausdorff measure is an extension
of the classical Lebesgue measure to fractal objects, see
\cite{Grebenkov06c} for details).  The intermediate case with $\ell
\ll \diam \ll L$ is the most interesting.  Depending on the ratio
$\delta/\diam$, the spread harmonic measures scale differently as
schematically illustrated in Fig.~\ref{fig:scaling_zones}.
Figure~\ref{fig:scaling_tau2} shows a continuous transition from the
multifractal dimensions of the harmonic measure, $D_q$ (studied in
Sect.~\ref{sec:harmonic}), to the multifractal dimension of the
Hausdorff measure, $D_0$.  In addition, the developed concepts brought
us to an alternative derivation of the anomalous constant phase angle
(CPA) frequency behavior in electrochemistry
\cite{deLevie65,Armstrong76,Nyikos85,Halsey92,Sapoval93}.  This new
insight allowed us to explain some disagreements concerning the CPA
exponent (see \cite{Grebenkov06c} for details).

From a mathematical perspective, a continuous transition between the
harmonic and Hausdorff measures is a new interesting phenomenon.
Although the presented study remained qualitative and relied on
numerical analysis of prefractal curves, the observed results
suggested a natural way for extending the spread harmonic measures to
really fractal boundaries.  The fact that a physical parameter $D/\K$
``tunes'' the geometrical scale, at which the spread harmonic measures
are looked at, may have potential applications.  For instance, the
diffusion screening, which is a fundamental obstacle in designing
highly efficient exchangers, is progressively reduced by increasing
the exploration length $D/\K$.  Moreover, the ``functioning'' of a
system depends on the value of $D/\K$.  This means that such
exchangers can be selective with respect to different species.  For
instance, this principle can potentially be used for designing a
filter that would efficiently capture the species with a specific
diffusion coefficient (i.e., specific size).

\newpage
\subsection{ First passage statistics on fractal boundaries }
\label{sec:first_passage}

Reflected Brownian motion is an intermittent process when diffusion
steps in the bulk are altered by encounters with the surface.  In the
previous subsection, the active sites, which could absorb or transfer
the diffusing particles, were supposed to be distributed uniformly
over the boundary (spatially uniform sticking probability).  In this
case, the overall reaction process was shown to be realized in a
region around the first hitting point and of the size of the
exploration length.  When, on the opposite, the active sites for
reaction or transfer are diluted on the boundary, the surface is not
homogeneously reactive any more.  A more detailed study of individual
diffusion steps in the bulk, known as Brownian excursions or bridges,
is then needed \cite{Levitz06}.

\subsubsection{ First passage statistics }

\begin{figure}
\begin{center}
\includegraphics[width=60mm]{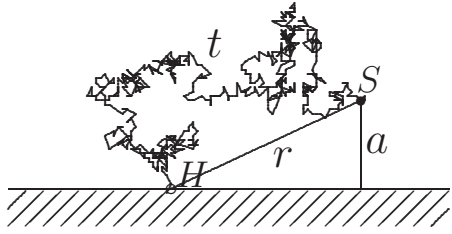}
\end{center}
\caption{
First passage statistics for a flat surface.  Brownian motion started
from a point $S$ above the boundary (at distance $a$) hits this
boundary at some (random) point $H$.  The probability densities
$\psi(t)$ and $\theta(r)$ describe the first hitting time $t$ and the
Euclidean end-to-end displacement $r$, respectively.  }
\label{fig:couchy}
\end{figure}

For a given boundary, we are interested in the two probability
densities (Fig.~\ref{fig:couchy}):
\begin{itemize}
\item
$\psi(t)$ that a particle, started from a close vicinity of the
interface at $t = 0$, returns to this interface, for the first time,
at time between $t$ and $t + dt$;

\item
and $\theta(r)$ that the end-to-end Euclidean distance $r$ between the
starting and hitting points of the corresponding Brownian excursion
(or bridge) lies between $r$ and $r + dr$.

\end{itemize}
For flat surfaces, both densities are known explicitly.  For instance,
one has for a straight line in the plane \cite{Feller}:
\begin{equation*}
\psi(t) = \frac{a}{\sqrt{4\pi D}~ t^{3/2}} \exp\bigl[- a^2/(4Dt)\bigr]  ,
\hskip 15mm  \theta(r) = \frac{a}{\pi r^2} ,
\end{equation*}
where $a$ is the height of the starting point above the flat boundary
(horizontal axis).%
\footnote{
It is important to stress that both densities depend on the parameter
$a$.  In particular, if $a$ goes to $0$, the densities $\psi(t)$ and
$\theta(r)$ approach Dirac distribution $\delta(t)$ and $\delta(r)$,
respectively.  This reflects the well known fact that Brownian motion
started from a smooth boundary crosses this boundary an infinite
number of times during an infinitesimal time interval.  In other
words, Brownian motion immediately returns to the boundary.  At first
thought, this mathematical fact may sound as a puzzling paradox:
Brownian motion simply cannot escape from the boundary.

From a physical of view, one should not forget that Brownian motion is
a mathematical process which leads to the same macroscopic description
as real microscopic dynamics (Sect.~\ref{sec:diffusion}).  In
particular, the parameter $a$ cannot be smaller than an interaction
range between the surface and the diffusing particle (in the order of
at least a few angstroms).  What is important here is that the
parameter $a$, whatever its physically limited value, is much smaller
than the transport scales (variable $r$).  In what follows, $a$ will
be considered as a given small parameter. }
The asymptotic behavior of these densities at long time and large
distance is described by power laws:
\begin{equation}
\label{eq:asympt_psi_theta}
\psi(t) \propto t^{-\alpha}   ~~~ (t\to\infty),  \hskip 15mm
\theta(r) \propto r^{-\beta} ~~~  (r\to\infty),
\end{equation}
with the exponents $\alpha = 3/2$ and $\beta = 2$.  The first moment
(average duration or displacement of an excursion) and the second
moment of the former probability density functions are ill-defined and
diverge mathematically speaking.  But, in most practical situations,
irregular interfaces are encountered and the behavior associated with
flat interfaces is possibly misleading.  It is thus important to know
how an irregular geometry of the interface may affect the asymptotic
behavior (\ref{eq:asympt_psi_theta}).  In particular, what are the
values of the exponents $\alpha$ and $\beta$ for a fractal boundary?
These issues had been addressed in \cite{Levitz06}.

\subsubsection{ Scaling relations }

We first derive two general expressions describing the first-passage
statistics of Brownian excursions (or bridges).  A simple qualitative
argument is the following.  The total number $Q(t)$ of particles which
have diffused out from the source is approximately proportional to the
Minkowski content of the surface.  The size of this content evolves in
time as $t^{1/2}$, yielding $Q(t)\propto [t^{1/2}]^{d-D_0}$ for a
boundary of fractal dimension $D_0$, $d$ being the dimension of the
embedding space \cite{deGennes82}.  On the other hand, the total
number $Q(t)$ is constituted of all the particles survived up to time
$t$, $Q(t) \propto \int_0^t S(t')dt'$, while the survival probability
$S(t)$ is related to the first hitting time density $\psi(t)$ as $S(t)
= 1 - \int_0^t \psi(t')dt'$.  Bringing these together, one gets
$\psi(t) \propto - d^2 Q(t)/dt^2$ and
\begin{equation}
\label{eq:alpha}
\alpha = \frac{D_0 - d + 4}{2} .
\end{equation}
The displacement statistics $\theta(r)$ and the time statistics
$\psi(t)$ are formally related according to
\begin{equation*}
\theta(r) = \int dt ~\psi(t)~ \delta\biggl(\sqrt{<r^2(t)>} - r\biggr) ,
\end{equation*}
where $<r^2(t)>$ is the mean square displacement at time $t$.
Assuming that the mean square displacement of the Brownian motion
evolves as $t$ in the bulk phase, a change of variable $t$ for the
delta distribution in the above equation gives, for a fractal
boundary, $\theta(r) \propto 1/r^{2\alpha - 1}$, and
\begin{equation}
\label{eq:beta}
\beta = 2\alpha - 1   \hskip 10mm \Rightarrow  \hskip 10mm   \beta = D_0 - d + 3 .
\end{equation}
The above qualitative arguments give an idea why and how the exponents
$\alpha$ and $\beta$ are related to the fractal dimension $D_0$,
without pretending for a mathematical rigor.  A more rigorous
derivation for the exponent $\beta$ in 2D is sketched in
\cite{Levitz06}.

\subsubsection{ Numerical verification }

\begin{figure}
\begin{center}
\includegraphics[width=75mm]{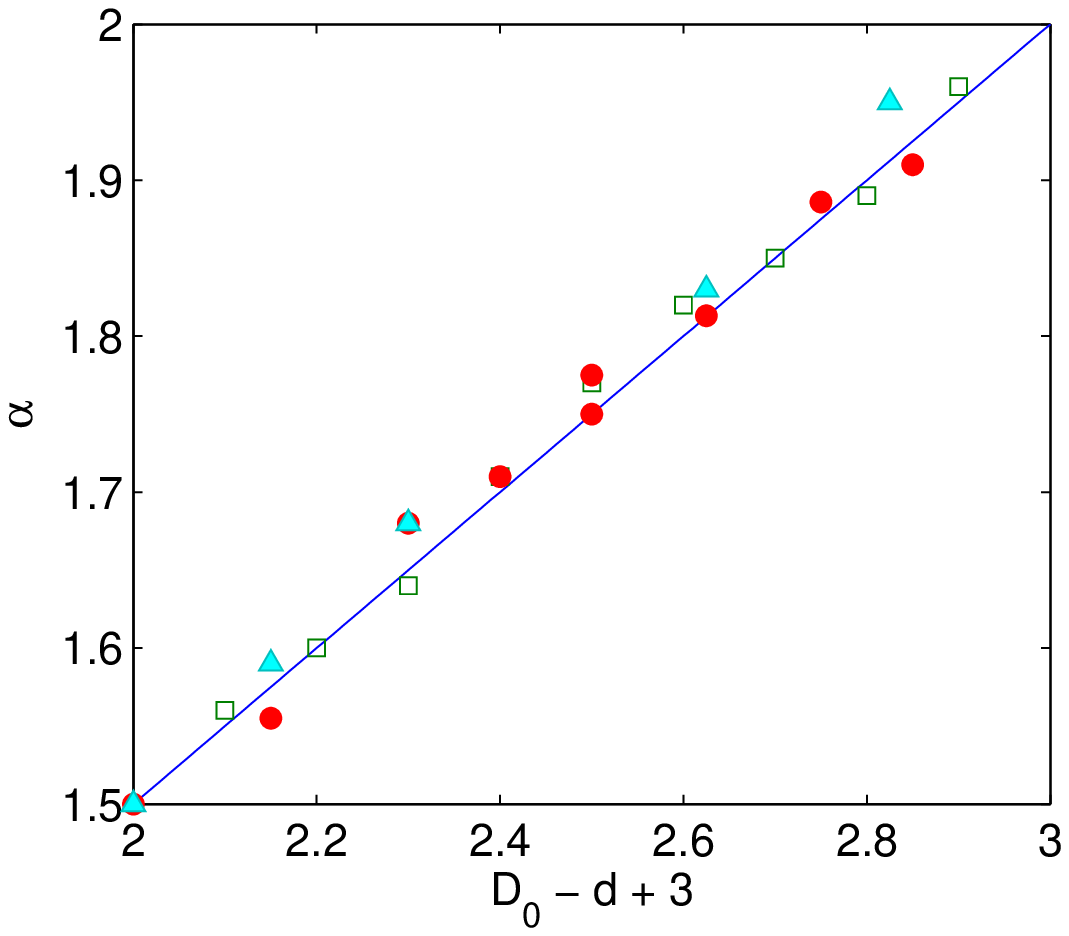}
\includegraphics[width=75mm]{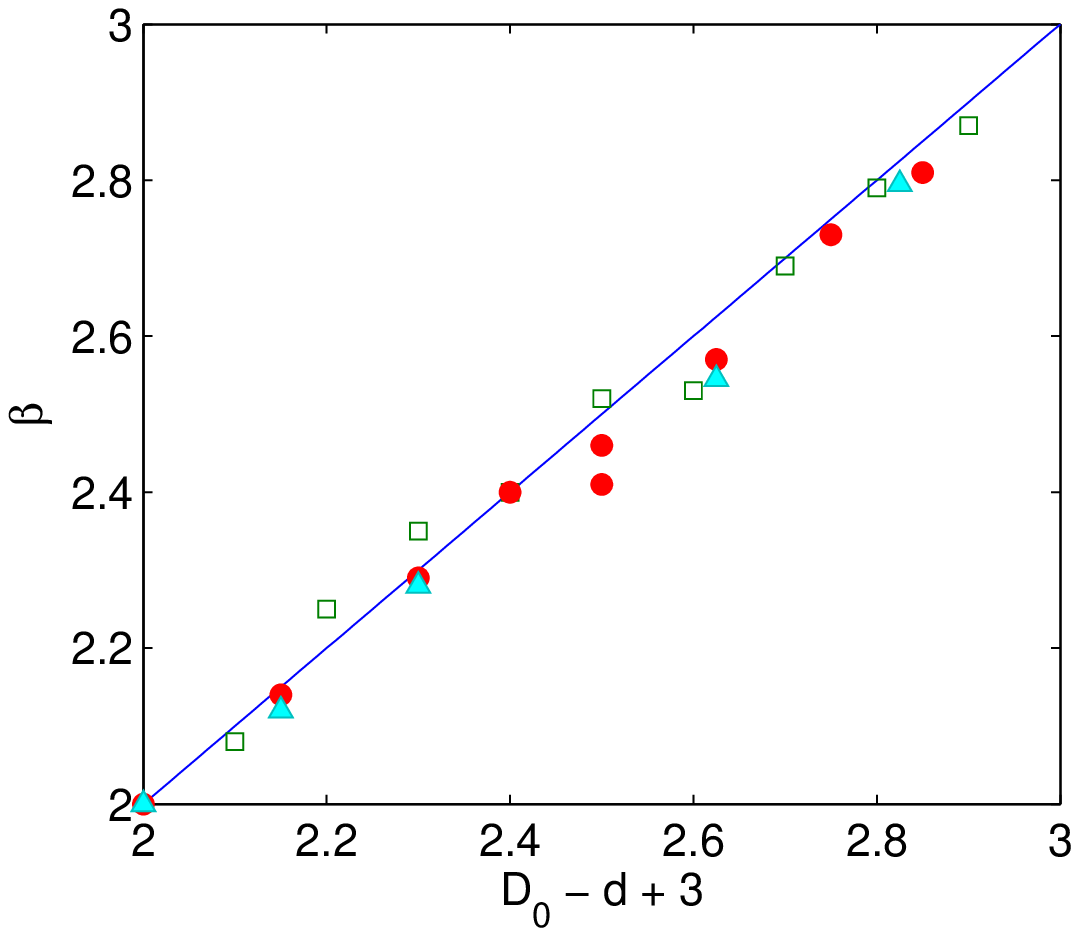}
\end{center}
\caption{
Variation of the exponent $\alpha$ (left) and $\beta$ (right) with the
boundary fractal dimension $D_0$: self-similar curves in two
dimensions (squares); self-similar surfaces in three dimensions
(circles); self-affine surfaces in three dimensions (triangles).  The
solid line follows Eqs.~(\ref{eq:alpha}) and (\ref{eq:beta}),
respectively. }
\label{fig:alpha_beta}
\end{figure}

The scaling relations (\ref{eq:alpha},~\ref{eq:beta}) were checked by
extensive numerical simulations for a class of self-similar and
self-affine interfaces in two and three dimensions.  In particular,
diffusion near various triangular von Koch curves was simulated by the
GAFRW algorithm.  The starting points were chosen uniformly over a
finite generation of the curve, within a small distance $a$.  The
simulation of a Brownian trajectory was terminated when the particle
approached the boundary closer than a chosen threshold.  For each
particle, the duration and the end-to-end Euclidean displacement were
recorded, providing large statistics for $\psi(t)$ and $\theta(r)$
(the number of simulated trajectories was $10^{10}$).  The data were
fitted by power laws in order to determine the exponents $\alpha$ and
$\beta$.  The fact that the fractal dimension of the triangular von
Koch curves can be continuously varied from $1$ to $2$ allowed us for
a careful check of the scaling relations
(\ref{eq:alpha},~\ref{eq:beta}).  A good agreement between theoretical
predictions and numerical results is shown in
Fig.~\ref{fig:alpha_beta}.

A simple dependence of $\alpha$ and $\beta$ on the fractal dimension
is a nontrivial result because diffusion screening and the harmonic
measure properties might influence these exponents.  The simulations
confirmed that the Brownian dynamics in the bulk is not biased by a
geometrical confinement induced by the boundaries.  The exponent
$\beta$ is found to be strictly larger than $2$.  Consequently, the
mean distance for the first-passage encounter is now finite in
opposition with the case of a flat surface.  The fact that Brownian
bridges are sensitive to surface geometrical crossovers at long time
should provide a way to probe colloidal shapes \cite{Levitz05b}.  The
studied first-passage process plays a central role in thermodynamics
of rough colloidal surfaces \cite{Duplantier91}, or in the evaluation
of the mean first exit time from a bounded domain \cite{Benichou05}.
It is also important in nuclear magnetic relaxation in complex fluids
and porous media \cite{Levitz05}.

\subsection{ Passivation processes }
\label{sec:passivation}

In the previous subsections, the surface activity remained constant in
time.  In practice, the boundary itself can be altered in the course
of transport process.  The passivation of the surfaces working under
diffusion-limited conditions is a general phenomenon which appears in
many natural or industrial systems ranging from catalysis
\cite{Froment} to heat transfer \cite{Epstein88}, electrochemistry
\cite{Pricer02}, and physiology \cite{Weibel}.  In such situations, it
is not only diffusion screening which leads to a strongly
inhomogeneous active surface, but part of the surface activity may be
progressively inhibited by phenomena bearing different names,
depending on the application field: passivation, fouling, poisoning or
restricted absorption.  

For instance, the transfer of nutrients from the digestive system to
the blood in humans is mostly realized in the small intestine in which
the major transport mechanisms are passive diffusion and absorption
\cite{DeSesso01}.  The anatomy of the small intestine exhibits a
fractal-like geometry, with finger-like structures at many different
scales of magnification: flexures, plicae, villi, microvilli
\cite{Yan06}.  In this type of geometry, the most exposed parts of the
intestinal membrane are easily accessed by diffusion and thus are the
first to be altered by any inflammatory disorder or any chemical
species that would diffuse in the digestive system.  As a matter of
fact, a wide range of gastrointestinal disorders are associated with
abnormal intestinal permeability \cite{Farhadi03}.

Another important example of passivation is a parallel or serial
``fouling'' in heterogeneous catalysis \cite{Barholomew01}.  During
catalysis, this phenomenon consists in a parasitic reaction that
passivates the catalyst in the regions which are active, and
eventually eliminates the entire activity of these regions.  Diffusion
screening in irregular catalytic grains implies that active regions
represent only a fraction of the total catalytic surface.  The time
evolution of the overall catalytic efficiency will then depend in a
complex way on the accessibility of the more remote regions of the
interface.

Finally, an enhanced efficiency of heat exchangers is often achieved
by building interfaces of very large surface
\cite{Andrade04,Meyer05}.  But the functioning of these interfaces can
be substantially altered by a fouling process, namely the scale
deposition, in which crystalline deposits of low thermal conductivity
locally reduce the heat transfer \cite{Helalizadeh06}.

\subsubsection{ Iterative passivation process }

\begin{figure}
\begin{center}
\includegraphics[width=100mm]{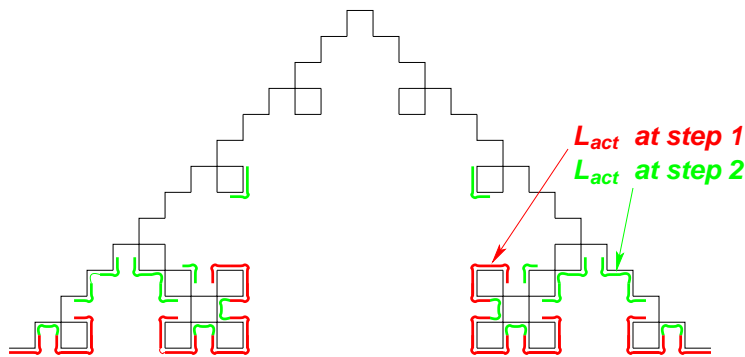}
\end{center}
\caption{
Schematic view of the successive active then passivated regions.  The
diffusing particles are coming from the bottom.  The first active
region is in red and has a length $L_{\rm act}$.  After the first step
of the passivation process, the boundary condition on this region is
set to Neumann, allowing a new region (in green) to become active.
This region will be in turn passivated and so on, until the whole
developed surface is passivated.  In 2D, the active length $L_{\rm
act}$ remains almost constant from one iteration to the next, until
the entire surface is passivated.  }
\label{fig:passiv_2d_pict}
\end{figure}

In this Section, we investigate the process of progressive passivation
of irregular surfaces accessed by diffusion \cite{Filoche08b}.
Passivation, disease, aging or fouling will likely start by damaging
the most accessible part of the active interface, so the question then
naturally arises: what happens after the initially active regions have
been passivated?  After deactivation, most of the diffusing particles
hit a now passivated zone and are reflected, resuming their diffusion
in the bulk to eventually react on an alive but deeper region of the
catalytic surface (Fig.~\ref{fig:passiv_2d_pict}).  Consequently,
regions that were initially poorly active become fully active until
they are in turn passivated.  This passivation process goes on and on
until, finally, the whole catalytic interface is deactivated and the
catalytic process stops.  The question of interest here is: how will
the size of the region, where most of the particles finally react or
are absorbed, evolve as the passivation process gradually deactivates
the initial alive interface.  In other words, how the activity is
gradually transfered from the most accessible to less accessible
regions.

In mathematical terms, the passivation process can be described as
follows: at the beginning of the process, the alive sites are supposed
to be uniformly distributed over the whole irregular surface.  We
suppose that the sticking probability $\sigma$ is equal to $1$ (or $\K
= \infty$) that corresponds to a homogeneous Dirichlet boundary
condition on the concentration of reactant molecules.  On such an
interface, the activity, although existing in principle everywhere, is
distributed in a very uneven manner due to diffusion screening.  One
may then define an ``active zone'' as the smallest part of the
interface carrying a given (large) fraction $p$ of the activity, e.g.,
$80\%$.  The passivation process is then discretized and divided into
the following steps \cite{Sapoval04}:
\begin{itemize}
\item
At first, the entire interface is alive.  The distribution of arrival
probabilities at the interface is calculated by the GAFRW algorithm on
the quadratic and cubic von Koch boundaries as in
Sect.~\ref{sec:harmonic}.

\item
The active region of the interface, which is only a fraction of the
alive interface, is determined.  The activity is proportional here to
the harmonic measure density on the interface.

\item
This first active region is passivated.  In mathematical terms, it
would correspond to a particle concentration obeying Neumann boundary
condition.  Physically, it means that when a particle now hits a site
belonging to this passivated region, it is reflected back and resumes
its bulk diffusion until it reaches the regions that are still alive.
In other words, this passivation process locally transforms a
Dirichlet boundary condition into a Neumann boundary condition.  The
remaining alive interface thus consists in the former alive interface
minus the newly passivated region.

\item
The new distribution of arrival probabilities is now computed using
the new boundary condition (reflecting sites in the former active
region).

\item
A new active region is determined, which is a subset of the remaining
alive interface (the non passivated boundary).  This new active region
will be in turn passivated, and so on. 

\end{itemize}
This passivation process generates at each iteration a new active
region so that the whole interface can be decomposed as a sum of
successive active regions.

\begin{figure}
\begin{center}
\includegraphics[width=100mm]{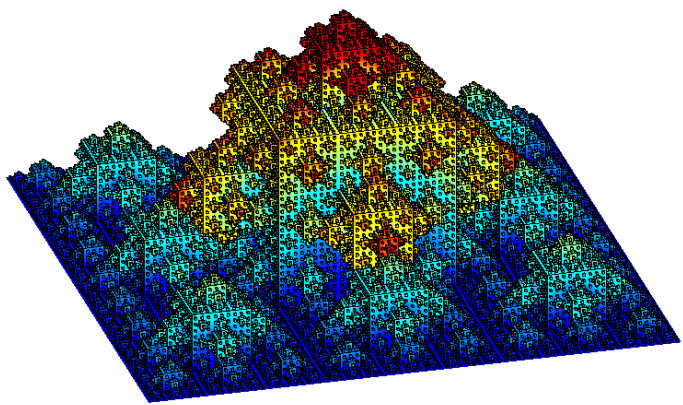}
\end{center}
\caption{
Successive active regions during the passivation process of a $5$th
generation of the cubic von Koch surface.  Diffusing particles are
coming from the bottom and reach first mainly the blue region of the
interface.  Each color in the simulation represents a set of $4$
successive passivated regions (dark blue = regions 1 to 4, light blue
= 5 to 8,...).  One can see that the size of the active region
decreases during the passivation process.  At the end, only the dark
red regions on the tip are active.  }
\label{fig:passiv_3d_5th}
\end{figure}

\subsubsection{ Modification of the GAFRW algorithm }

Although the GAFRW algorithm has proved to be very efficient, its
direct application to a partially passivated boundary is still not
sufficient to solve the problem here.  The problem is the following:
after a few iterations of the passivation process, the remaining alive
regions are highly screened.  So a reactant particle launched from a
distant source has to follow a very long stochastic trajectory, with a
large number of reflections on the passivated sites of the boundary,
before reaching any potentially active region.  The extremely large
computational time required for further passivation iterations makes
then difficult or even impossible to study the whole passivation
process.  This difficulty was overcome by using the distribution of
the activity at step $n$ as the initial source distribution for the
next step $(n+1)$.  As a matter of fact, the distribution of the
activity at step $(n+1)$ is composed of two types of particles: (i)
particles arriving directly on the non passivated part of the boundary
(corresponding to its contribution to the initial harmonic measure)
and (ii) particles arriving after being reflected by the already
passivated part of the boundary up to step $n$.  The distant source of
diffusing particles is thus replaced by a fictitious source on the
boundary itself.  Since the last passivated regions are close to the
remaining alive zones, using them as sources considerably enhances the
efficiency of the computation (see \cite{Filoche08b} for details).

This modification of the GAFRW algorithm has permitted us to study the
passivation of the quadratic von Koch curve (2D) up to the $7$th
generation and of the cubic von Koch surface (3D) up to the $5$th
generation.  One can take note visually of the complexity of the 3D
surface and the evolution of the active zone in
Fig.~\ref{fig:passiv_3d_5th}.

\begin{figure}
\begin{center}
\includegraphics[width=75mm]{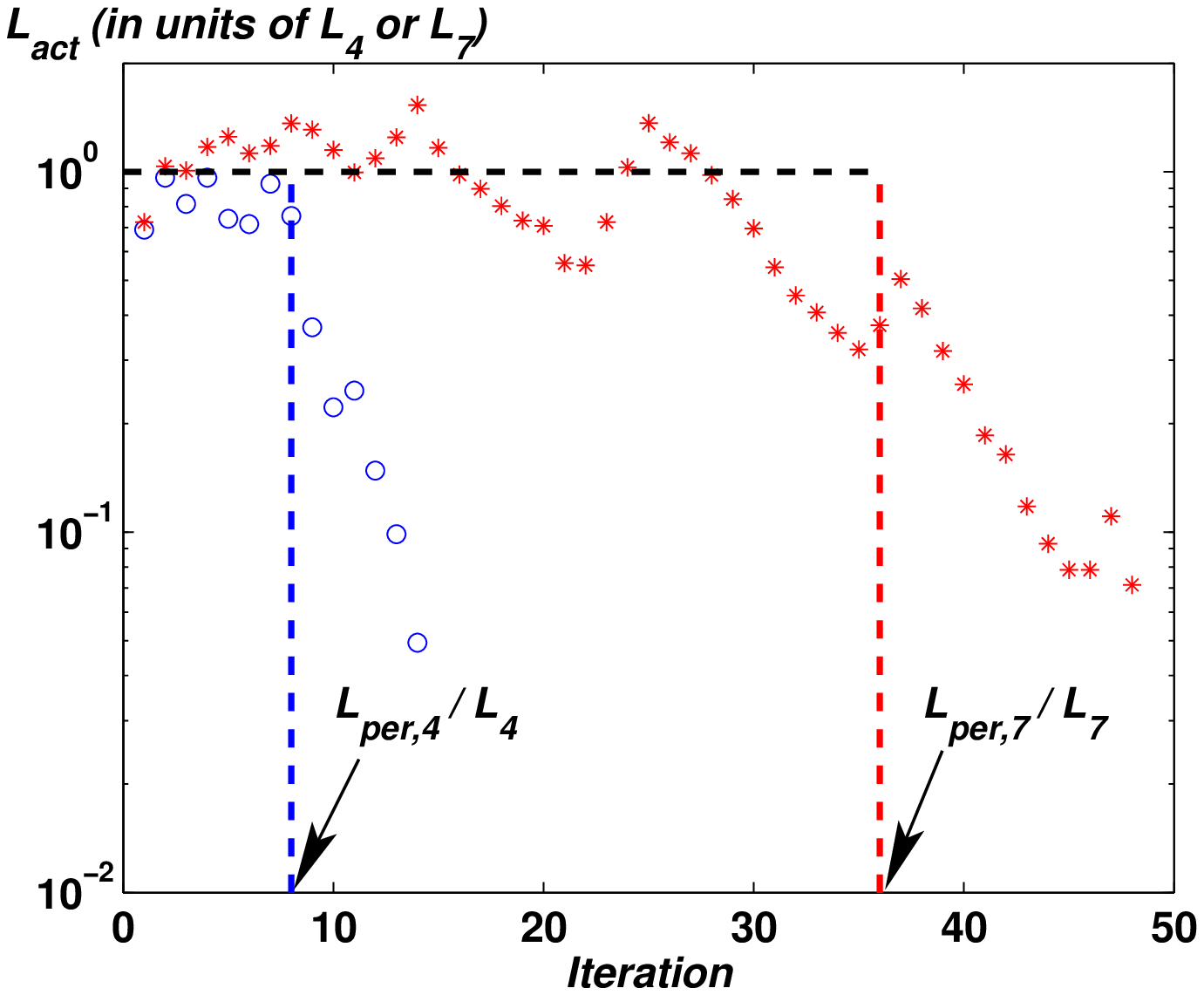}
\includegraphics[width=75mm]{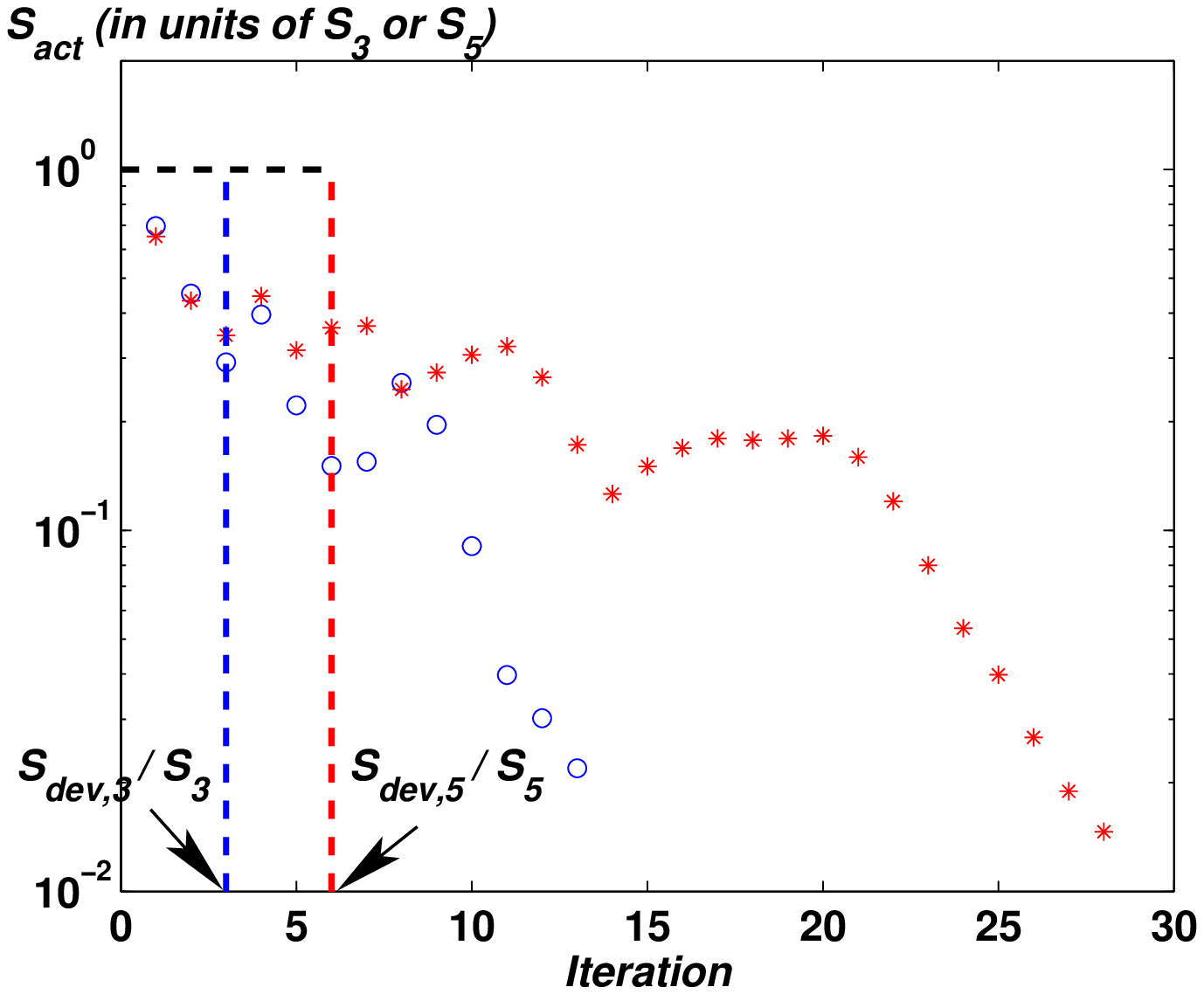}
\end{center}
\caption{
Comparison between passivation processes in 2D (left) and 3D (right).
{\bf Left}.  The size $L_{\rm act}$ of the active region is plotted in
units of the system width ($L_4$ and $L_7$ for generations $4$ and $7$
of the quadratic von Koch curve) at each step of the passivation
process.  Blue circles and red stars represent generations $4$ and $7$
in 2D (resp., $3$ and $5$ in 3D).  One can see that the size $L_{\rm
act}$ remains almost constant throughout the process, and is of the
order of the structure width, until the number of iterations reaches
the ratio between the perimeter of the interface and its width,
$L_{\rm per}/L$.  These ratios are represented by the vertical dashed
lines.  After this threshold, the passivation process rapidly
terminates.  {\bf Right}.  The surface area $S_{\rm act}$ of the
active region is plotted in units of the projected surface ($S_3$ and
$S_5$ for generations $3$ and $5$ of the cubic von Koch surface) at
each step of the passivation process.  In contrast to the left plot,
the active region at each step regularly decreases.  It takes more
than 25 steps to passivate the total surface at $5$th generation,
whereas the total developed surface represents only $6$ times the
projected surface.  The vertical lines correspond to the ratio of the
total developed surface ($S_{{\rm dev},3}$ and $S_{{\rm dev},5}$) on
the projected surface ($S_3$ and $S_5$).  Note the progressive slow
decrease which is very different from the 2D case. }
\label{fig:passiv_act}
\end{figure}

\subsubsection{ Comparison between 2D and 3D cases }

In 2D, the active region has approximately a constant length $L_{\rm
act}$ at each step of the process, as schematically indicated in
Fig.~\ref{fig:passiv_2d_pict}.  As one can see in
Fig.~\ref{fig:passiv_act} (left), this result is confirmed by
numerical simulations for both the $4$th and the $7$th generations.
The observed oscillations can be attributed to the discrete scaling of
the deterministic interface.  Moreover, it is known from previous
studies that the size of the active region is of the order of the
width of the interface \cite{Sapoval04,Makarov85}.  This implies that
the number of passivation steps before the whole interface is
deactivated should be of the order of the ratio between the total
developed perimeter and the size of the interface.  Quantitatively,
the perimeter of the $4$th generation (resp. $7$th generation) of the
quadratic Koch curve is $(5/3)^4 \approx 7.7$ times (resp. $(5/3)^7
\approx 35.7$ times) larger than the size of the system.  Hence, one
can see in Fig.~\ref{fig:passiv_act} that the length of the active
region becomes smaller than $50\%$ of the width of the cell
respectively after $8$ and $37$ passivation iterations.  After that,
in both cases the size of the active region falls rapidly.

In 3D, the striking result is that, unlike in 2D, the surface area of
the active region $S_{\rm act}$ is not constant during the passivation
process, but gradually decreases (as shown in
Fig.~\ref{fig:passiv_act}(right) for $p = 80\%$).  Even more, for the
$5$th generation, the developed surface in our 3D simulation contains
$(13/9)^5 \approx 6.3$ times the projected surface, but $25$
passivation iterations are necessary to completely passivate the
surface.  We can observe here a net discrepancy between the 2D and the
3D cases: due to the properties of Brownian motion, the passivation
process is much steadier in 2D than in 3D, and it stops much more
abruptly.

This observation is significant for the case of catalysis, where the
need of a large surface within a finite volume implies the use of very
irregular surfaces.  As we have shown, the deep parts of the surface
become more and more difficult to reach as the passivation gradually
progresses, and the yield of reaction may decrease, despite the fact
that active regions still exist.  In other words, even when catalyst
grains seem to be exhausted, they may still contain a large amount of
alive surface.  In this case, a large amount of catalyst is wasted
simply because the alive surface is not easily accessed by 3D
diffusion anymore.  From this point of view, an engineered (2+1)D
geometry (with translational invariance) would be preferable.

\subsection{ Diffusion-weighted imaging of the lungs }
\label{sec:Kitaoka}

The above studies were focused on various properties of the diffusive
transport from a distant source towards an irregular boundary.  In
other words, we mainly considered what is the role of a geometrical
irregularity of the interface.  For this purpose, finite generations
of self-similar von Koch boundaries were particularly suited.  In many
situations, however, the internal morphological, topological or
geometrical structure of the confining domain appears to be even more
significant.  The examples are porous matrices of sedimentary rocks
and sandstones, interstitial space of human skin, pulmonary acini, to
name a few.  Although the irregularity of the interface is still
relevant, the structure of the whole diffusion-confining domain (e.g.,
pore size distribution, pore connectivity, etc.) may dominate the
overall transport characteristics.  In spite of the active research in
this field, many questions still remain unanswered.  This is probably
because of a lack of reliable geometrical models that would be at the
same time simple enough for performing numerical simulations and
representative of porous media found in nature and industry.  As an
example, we refer to a recent progress in a quantitative understanding
of the diffusive transport in the lungs
\cite{Felici03,Felici04,Felici05,Grebenkov05c}, which became possible
thanks to the Kitaoka algorithm for generating a model geometry of the
human pulmonary acinus \cite{Kitaoka00}.

\subsubsection{ The human pulmonary acinus and its modeling }

\begin{figure}
\begin{center}
\includegraphics[width=30mm]{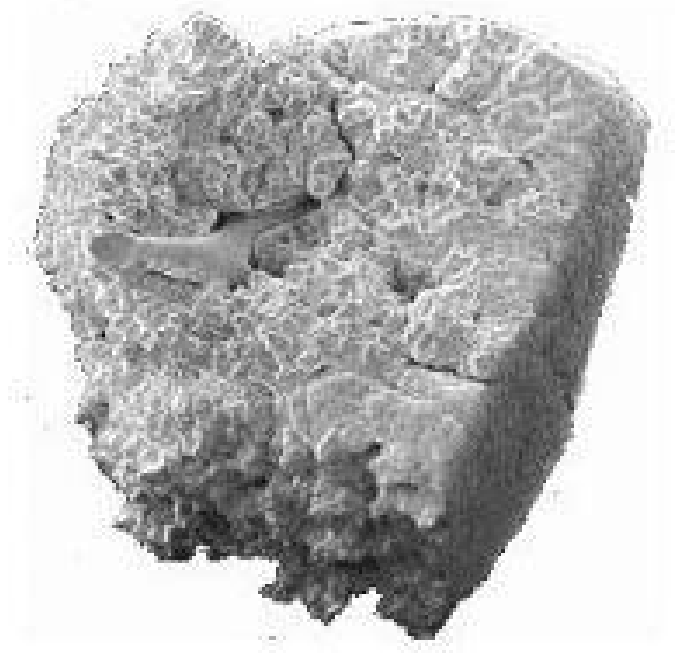}
\includegraphics[width=30mm]{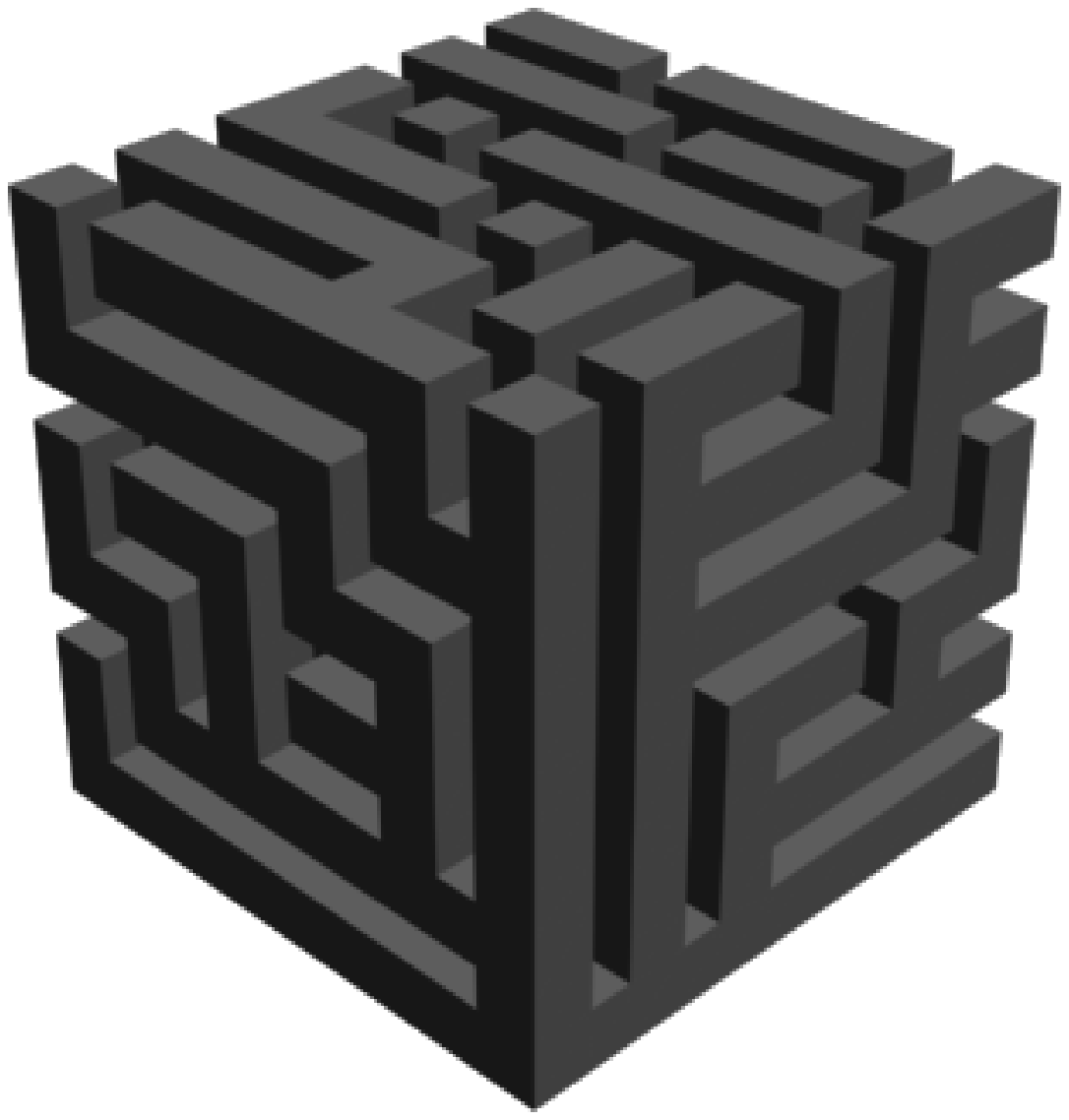}
\includegraphics[width=30mm]{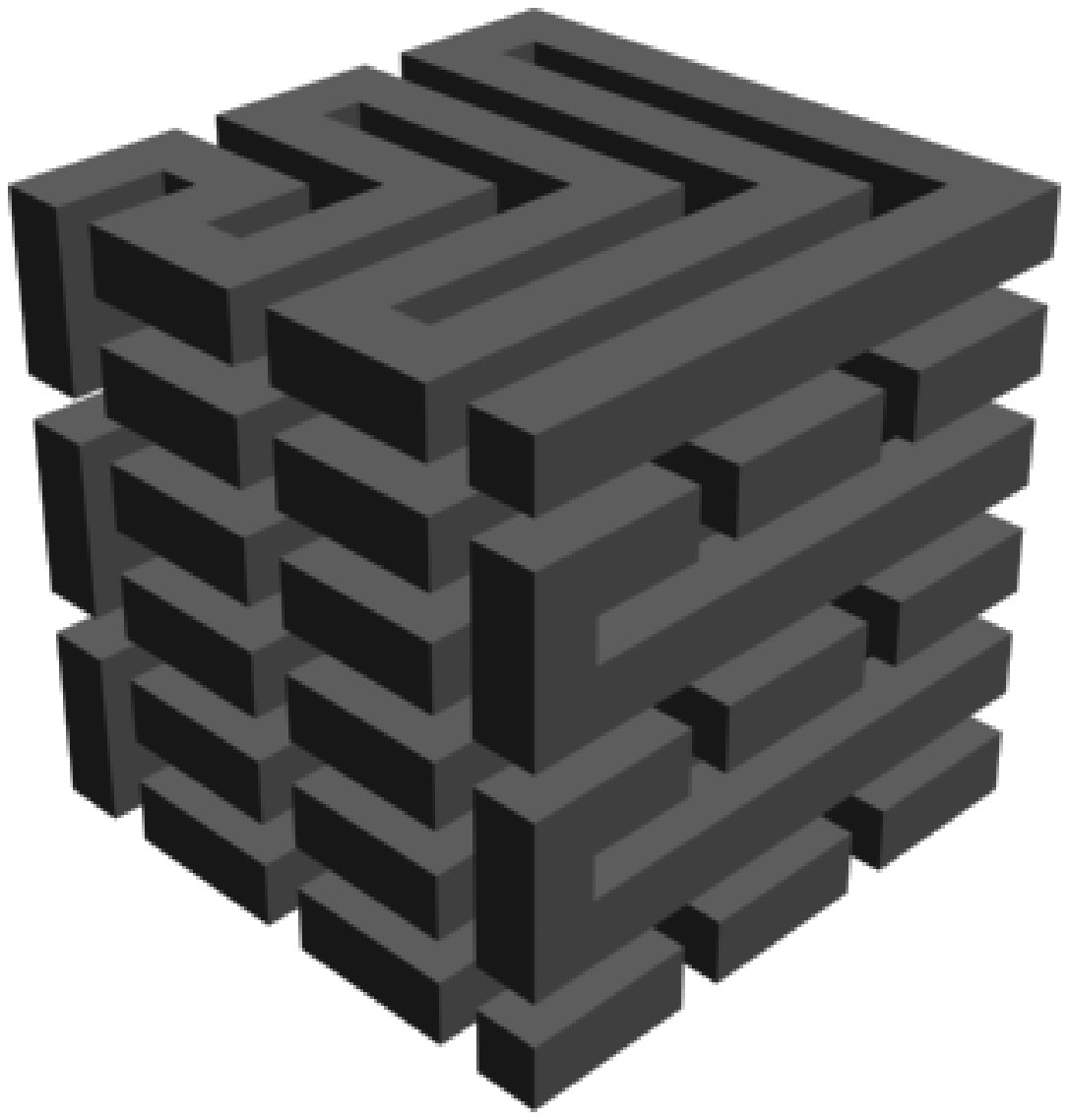}
\includegraphics[width=30mm]{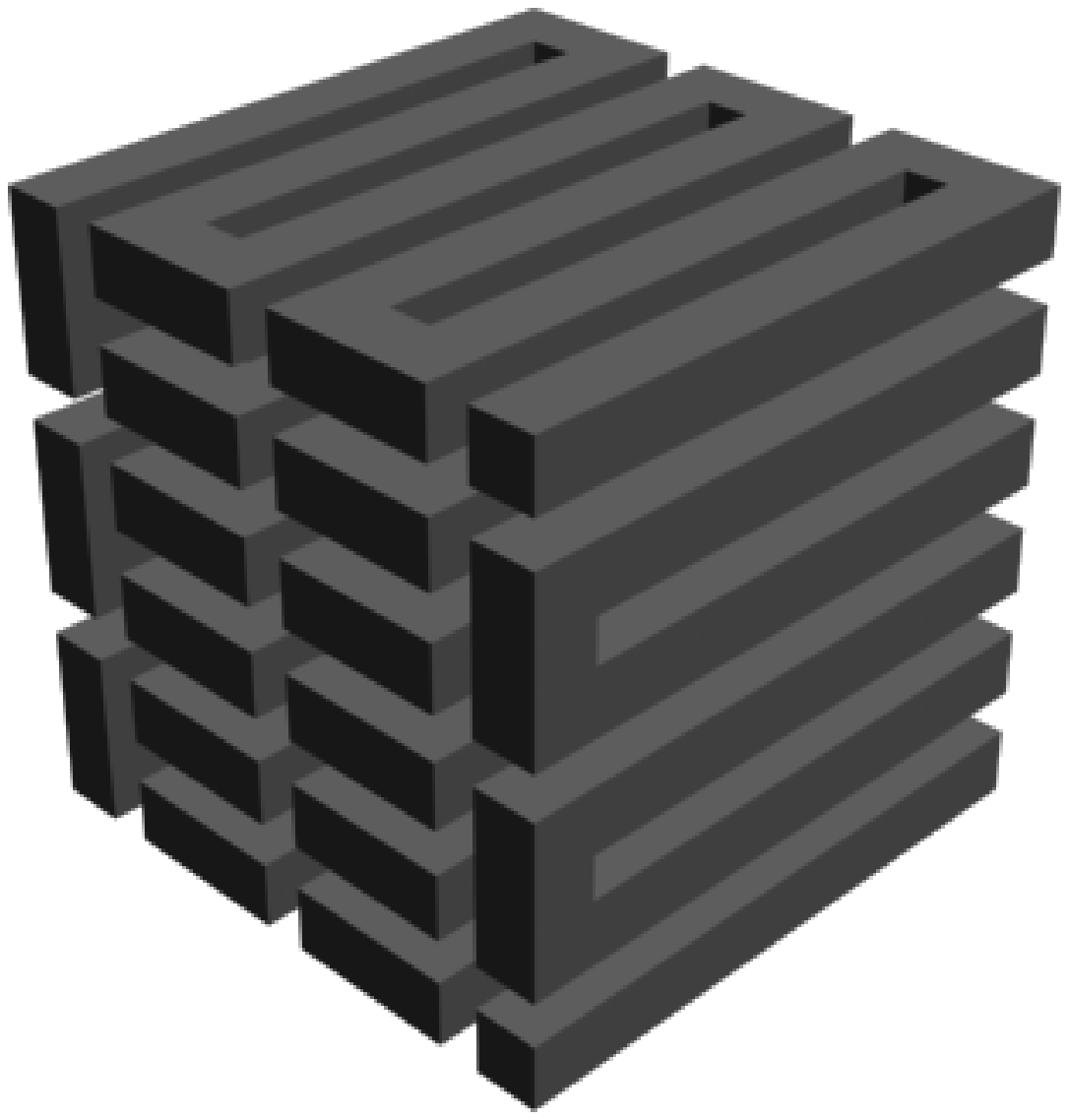}
\includegraphics[width=30mm]{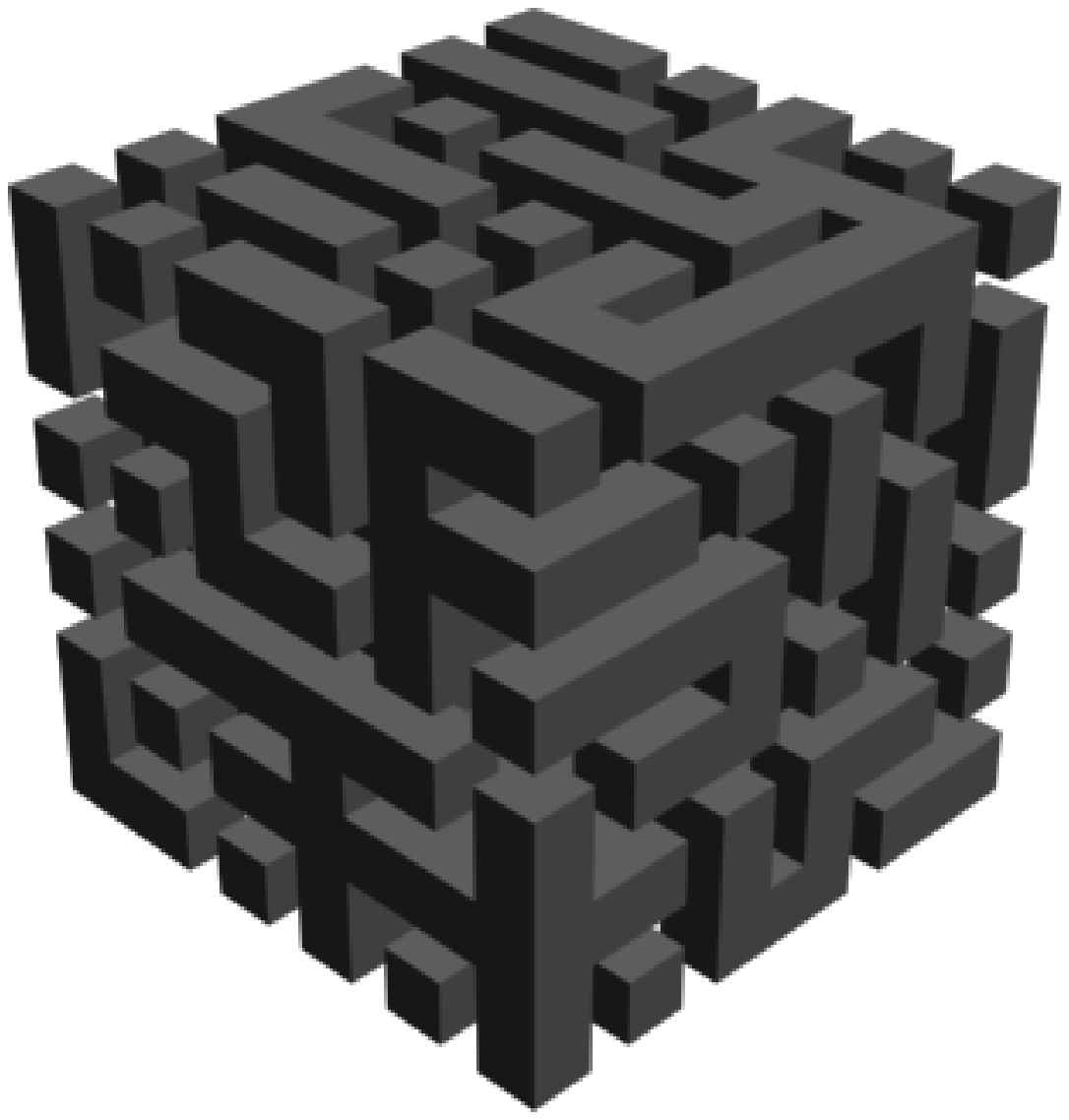}
\setlength{\unitlength}{1mm}
\begin{picture}(0,0)(156,5)  
\put(0,6){(a)}
\put(30,6){(b)}
\put(61,6){(c)}
\put(93,6){(d)}
\put(124,6){(e)}
\put(7,2){\vector(0,1){19}}
\put(44,2){\vector(0,1){2}}
\put(75,2){\vector(0,1){2}}
\put(107,2){\vector(0,1){2}}
\put(138,2){\vector(0,1){2}}
\end{picture}
\end{center}
\caption{
Five different geometrical structures with the same surface-to-volume
ratio: (a) cast of a human acinus, (b) branched Kitaoka labyrinth,
(c,d) two long channels ``packed'' in the cube, and (e) a disordered
porous medium created by random ``digging'' in the cube.  The first
four domains satisfy the connectivity condition (accessibility from
the ``entry'', indicated by an arrow), while the last one does not.
The ``solid'' channels represent the volume of the confining medium
where a gas diffuses.  }
\label{fig:Kitaoka}
\end{figure}

A human pulmonary acinus, the gas exchange unit in the lungs, has a
dichotomic branching structure with tortuous channels densely filling
a given volume (Fig.~\ref{fig:Kitaoka}a).  A ``Kitaoka acinus'' is
designed as a three-dimensional labyrinth of channels with square
profile (Fig.~\ref{fig:Kitaoka}b).  Being well appropriate for
numerical simulations, this simplified geometry captures the essential
features of the pulmonary acinus: a dichotomic tree having
non-symmetric branches of random lengths and filling a given volume.
Moreover, the total surface area and the average length of the
branches correspond to those of real acini.

We shall not discuss former works concerning oxygen diffusion in the
lungs which were clarified by simulations in the Kitaoka acinus
\cite{Felici03,Felici04,Felici05,Grebenkov05c}.  In turn, we present
Monte Carlo simulations that we developed to investigate restricted
diffusion of hyperpolarized gases (like helium-3) and the consequent
signal attenuation in a diffusion-weighted NMR experiment
\cite{Grebenkov07c}.  The aim of this study was to facilitate the
development of a reliable MRI diagnosis of early stage emphysema which
results in partial destruction of the alveolar tissue and enlargement
of the distal airspaces.  The geometry of emphysematous acini was
modeled by removing randomly a fraction of the internal walls from
previously generated Kitaoka labyrinth.  We demonstrated that
diffusion-weighted NMR could be sensitive to destruction of the
branched structure.  In fact, partial removal of the interalveolar
tissue creates ``loops'' in the tree-like acinar architecture that
enhance diffusive motion and the consequent signal attenuation.

\subsubsection{ Monte Carlo simulations accounting for magnetic field gradients }

Given the specific shape of the Kitaoka acinus (long branched channels
of the same square profile), there was no need for using fast random
walks.  An accurate accounting for a linear magnetic field gradient
would further complicate an implementation of such an algorithm.  
For this reason, we limited ourselves to
implementation of simple off-lattice random walks with normal
reflections on the boundary of the Kitaoka acinus
\cite{Grebenkov07c,Grebenkov05d}.  The given temporal profile $f(t)$
of the applied magnetic field gradient was discretized on the
simulation time interval $[0,t]$ with a time step $\tau = t/n$.  The
starting point $\r_0$ is chosen randomly with a uniform distribution
inside the Kitaoka acinus.  For each step $k$, one generates
independent Gaussian displacements $d\r^i$ in the three space
directions ($i = x,y, z$) with mean zero and dispersion
$\sqrt{2D\tau}$, in order to pass from the current position $\r_k$ to
a new position $\r_{k+1}$.  If the linear segment between the current
position and the new position intersects the boundary, a mirror
reflection is applied.  At each step $k$, the term $\tau
f(k\tau)\r^i_k$ is added to the phase counter $\varphi_i$ for each
space direction $i = x, y, z$.  The total phase accumulated during the
whole trajectory for unit gradient is then approximated, for each $i$,
by the sum
\begin{equation*}
\varphi_i = \tau \sum\limits_{k=0}^n f(k\tau) \r_k^i .
\end{equation*}
If the gradient is applied along a given direction ${\bf e} = (e_x,
e_y, e_z)$, the total phase accumulated in this direction for unit
gradient is $\varphi_{\rm \bf e} = e_x \varphi_x + e_y \varphi_y + e_z
\varphi_z$.  Although the gradient direction is fixed in the NMR
scanner, the real acini in the lungs are oriented randomly.  This
effect is taken into account by looking at the signal that is averaged
over all spatial directions.  In this case, the averaged phase is
$\varphi_{\rm av} = \sqrt{\varphi_x^2 + \varphi_y^2 + \varphi_z^2}$.
Repeating the Monte Carlo simulation $N$ times, one records these
phases in order to approximate their probability distribution.  Once
the simulations are terminated, one can find the signal $E$ or the
directionally averaged signal $S_{\rm av}(g)$ as a function of the
gradient amplitude $g$ as
\begin{equation*}
E \simeq \frac{1}{N}\sum\limits_{j=1}^N \exp[i\gamma g\varphi_{\rm \bf e}^j] ,   \hskip 10mm
S_{\rm av}(g) \simeq \frac{1}{N}\sum\limits_{j=1}^N \frac{\sin(\gamma g \phi_{\rm av}^j)}{\gamma g \varphi_{\rm av}^j} .
\end{equation*}
Since the statistical error of Monte Carlo technique is of order of
$1/\sqrt{N}$, the choice of $N = 10^6$ leads to reasonably accurate
results.

\begin{figure}
\begin{center}
\includegraphics[width=100mm]{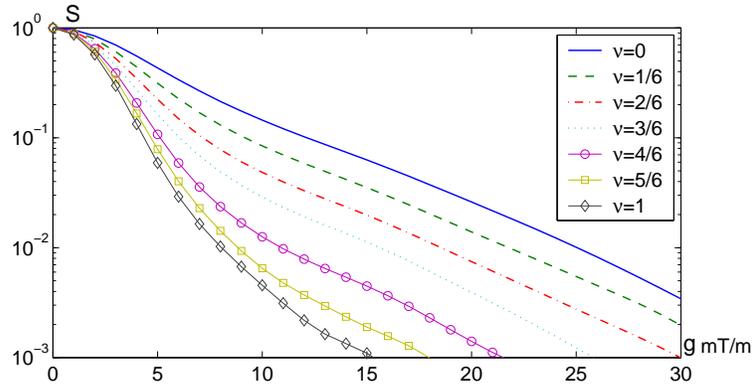}
\end{center}
\caption{
Directionally averaged signal $S_{\rm av}(g)$ as a function of the
gradient intensity $g$ for different destruction factors $\nu$
starting from $0$ (healthy acinus) and ranging up to $1$ to model
progressively damaged emphysematous acini.  The signal is more
attenuated in more damaged structures because diffusion is faster with
a lower restriction.  The ratio between the signals from a healthy
acinus ($\nu = 0$) and that from an early emphysematous acinus ($\nu =
1/6$) is higher for larger gradient intensities. }
\label{fig:Kitaoka_nu}
\end{figure}

\subsubsection{ Diffusion in healthy and emphysematous acini }

Figure~\ref{fig:Kitaoka_nu} illustrates the dependence of the
directionally averaged signal $S_{\rm av}(g)$ on the gradient
intensity $g$ for healthy and emphysematous acini.  For weak gradient
intensities, one recovers the classical (Gaussian) $g^2$ behavior of
the logarithm of the signal, with a single apparent diffusion
coefficient (ADC).  Since partial destruction of the alveolar tissue
by emphysema creates loops in the branched structure, diffusion
becomes faster, and the signal is then more attenuated (larger ADC).
The high sensitivity of diffusion-weighted NMR measurements to this
effect can be potentially employed to diagnose emphysema at early
stages.

The geometrical confinement and branching structure of the acinus lead
to deviations from the $g^2$ behavior at higher gradients.  In
particular, one can see on Fig.~\ref{fig:Kitaoka_nu} a transition to a
stretched-exponential behavior known as localization regime
\cite{Stoller91,Hurlimann95}.  This observation indicates that the
notion and use of ADC should be substantially revised, especially
because experimental measurements at higher gradients appear to be
more sensitive to the acinar structure.  This also explains several
confusions and possible ambiguities in determination of ADC in medical
literature.

The numerical simulations were also performed on a tenfold scale
Kitaoka acinus with a broad range of diffusion coefficients.  The
obtained results were confronted to experimental measurements in a
tenfold phantom made from the standard epoxy resin by
stereolithography \cite{Habib07,Habib08}.

\begin{figure}
\begin{center}
\includegraphics[width=100mm]{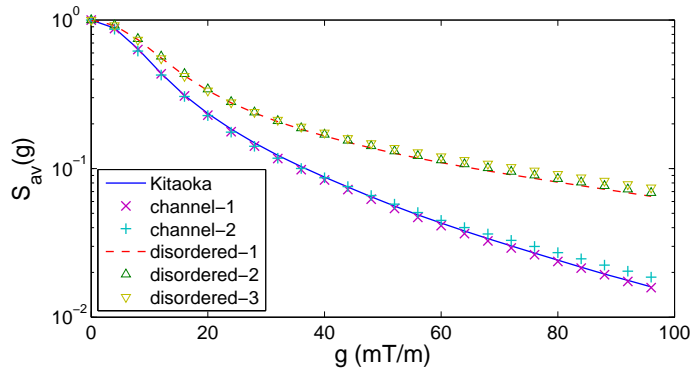}
\end{center}
\caption{
Directionally averaged signal $S_{\rm av}(g)$ as a function of the
gradient intensity $g$ for different porous structures on
Fig.~\ref{fig:Kitaoka} (the domains called ``disordered-2'' and
``disordered-3'' are not shown).  }
\label{fig:Kitaoka_three}
\end{figure}

\subsubsection{ What is the role of a complex internal architecture? }

In a separate work \cite{Grebenkov07g}, we addressed a more general
question: What is the role of a complex internal architecture (e.g.,
branching of the pulmonary acinus or pore network in rocks) for NMR
measurements?  To answer this question, we have performed Monte Carlo
simulations of restricted diffusion in three groups of
three-dimensional structures with the same surface-to-volume ratio.  A
basic domain is a cube of size $L$ divided into $6\times 6\times 6 =
216$ small cubic cells.  The first group is a set of random dichotomic
labyrinths generated inside the cube by the Kitaoka algorithm
(Fig.~\ref{fig:Kitaoka}b).  The second group consists of two
realizations of a long channel filling the same cube
(Fig.~\ref{fig:Kitaoka}c,d).  In the third group, disordered porous
media are generated by connecting a number of randomly chosen adjacent
cells (Fig.~\ref{fig:Kitaoka}e).  The directionally averaged signals
from these structures are computed by the above Monte Carlo
simulations.  Figure~\ref{fig:Kitaoka_three} shows that the signal
attenuation $S_{\rm av}(g)$ for different branched structures (the
first and second groups) are almost identical, while the signal for
disordered media (the third group) is significantly higher.  This is
related to the fact that a disordered medium consists of a number of
small disconnected patterns where the signal is less attenuated.  The
important conclusion is that the internal structure of porous media
significantly influences the restricted diffusion and NMR measurements
and should thus be taken into account for practical applications.

\subsection*{ Summary }
\addcontentsline{toc}{subsection}{{Summary}}

In this Section, we showed that Monte Carlo simulations were flexible
and efficient tools for studying restricted diffusion in complex
geometries, e.g., von Koch fractals.  A geometry-adapted fast random
walk algorithm allowed us to solve different problems, including
multifractal properties of the harmonic measure in 2D and 3D
(Sect.~\ref{sec:harmonic}), scaling properties of the spread harmonic
measures and the role of the exploration length
(Sect.~\ref{sec:spread}), first passage statistics
(Sect.~\ref{sec:first_passage}), passivation processes
(Sect.~\ref{sec:passivation}), etc.  All these problems concern
restricted diffusion with different behaviors on the boundary.  The
GAFRW algorithm simulated Brownian trajectories in the bulk, while the
boundary behavior was specifically implemented for each problem.


\newpage
\section{ Spectral insight: Laplacian eigenfunctions }
\label{sec:spectral}

This section is devoted to the spectral description of restricted
diffusion when the Laplace operator eigenfunctions allow one to
structure the whole information about diffusion in a particularly
useful form.  Most diffusion characteristics (e.g., propagator, total
flux, residence time, etc.) can be explicitly written in terms of the
eigenfunctions.  Specific features of these characteristics originate
somehow from eigenfunctions.  {\it We aim to reveal their relation and
understand various features of restricted diffusion in complex
geometries via the properties of the underlying Laplace operator
eigenfunctions.}  While this spectral description appears to be
natural, it is not always considered as the most efficient way.  In
fact, except few simple domains, for which the Laplace operator
eigenfunctions are known explicitly, a numerical calculation is
required.  This is a difficult time-consuming computational problem,
especially for porous, multiscale, or irregular geometries.  But, as
far as the eigenfunctions are found for a given domain, the whole
range of diffusion characteristics becomes directly accessible at
once.

We used the spectral description as a unifying mathematical language
for presenting the main achievements on restricted diffusion in NMR
\cite{Grebenkov07,Grebenkov08b,Grebenkov08c}.  The diversity and
complexity of diffusive NMR phenomena, observed in experiments, were
shown to result from the specific properties of the reflected Brownian
motion and the underlying Laplace operator eigenfunctions.  Many
classical results were retrieved, extended and critically discussed.
In what follows, we present our main results basing on this spectral
point of view.

\subsection{ Matrix formalism }
\label{sec:matrix}

Several matrix formalisms were developed for numerical analysis
of the macroscopic signal formed by the nuclei diffusing in magnetic
fields
\cite{Caprihan96,Callaghan97,Barzykin98,Barzykin99,Axelrod01}.  
We have reformulated and extended these formalisms to describe
restricted diffusion in any geometrical confinement and arbitrary
magnetic field \cite{Grebenkov07,Grebenkov07f}.  The ``derivation'' of
the compact matrix form for the signal (shown in the next subsection)
simply relies on a representation of the Bloch-Torrey equation
(\ref{eq:diffusion2}) in the Laplace operator eigenbasis.  The use of
this natural basis allows one to get a {\it structured representation
of diffusion} via two governing matrices $\Lambda$ and $\B$.  A
truncation of these matrices for further numerical analysis is well
controlled, leading to negligible computational errors.  The use of
the Laplace operator eigenfunctions is crucial here.

\subsubsection{ Perturbative derivation }

When the applied magnetic field is independent of time, $B(\r,t) =
\beta B(\r)$, Eq.~(\ref{eq:diffusion2}) reads as
\begin{equation}
\label{eq:Bloch-Torrey2}
\left(\frac{\partial}{\partial t} - D \Delta + i\gamma \beta B(\r)\right) \m(\r,t) = 0 ,
\end{equation}
where $B(\r)$ is the normalized (dimensionless) spatial profile of an
inhomogeneous magnetic field, and $\beta$ its intensity.  We are
looking for a solution of this equation in the basis of the Laplace
operator eigenfunctions $u_m(\r)$ (of eigenvalues $\lambda_m$)
satisfying Eqs.~(\ref{eq:eigen}, \ref{eq:eigen2}):
\begin{equation} 
\label{eq:m_decomp}
\m(\r,t) = \sum\limits_{m'} \tilde{c}_{m'}(t)~ u_{m'}(\r) ,
\end{equation}
with unknown time-dependent coefficients $\tilde{c}_{m'}(t)$.
By construction, $\m(\r,t)$ satisfies an appropriate boundary 
condition that was imposed for the eigenfunctions. 
Substitution of this expansion in Eq.~(\ref{eq:Bloch-Torrey2}),
multiplication by $u_{m}^*(\r)$, and integration over the confining
domain $\Omega$ yield a set of ordinary differential equations on the
coefficients $\tilde{c}_m(t)$
\begin{equation}
\label{eq:set_equation_cm}
\frac{\partial \tilde{c}_m(t)}{\partial t} + \sum\limits_{m'} 
\biggl(D\Lambda_{m,m'} + i \gamma \beta \B_{m,m'}\biggr) \tilde{c}_{m'}(t) = 0 ,
\end{equation}
where the infinite-dimensional matrices $\Lambda$ and $\B$ represent
the Laplace operator and the ``perturbing'' magnetic field $B(\r)$ in
the eigenbasis of the Laplace operator:
\begin{align}
\label{eq:B}
\B_{m,m'} &= \int\limits_\Omega d\r ~ u_m^*(\r)~ B(\r)~ u_{m'}(\r) ,\\
\Lambda_{m,m'} &= \delta_{m,m'} \lambda_m . 
\end{align}
Thinking of $\tilde{c}_m(t)$ as components of an infinite-dimensional
vector $C(t)$ leads to a matrix first-order differential equation
\begin{equation}
\left(\frac{d}{dt} + D\Lambda + i\gamma \beta \B\right) C(t) = 0.
\end{equation}
As for a scalar equation, its solution is the (matrix) exponential:
\begin{equation}
\label{eq:Ct}
\sqrt{V} C(t) = U e^{-(D\Lambda + i\gamma \beta \B)t} .
\end{equation}
The supplementary factor $\sqrt{V}$ is put explicitly to get rid of
the dimensional unit, meter$^{-d/2}$, of the vector $C(t)$ ($V$ being
the volume of the domain).  Here the matrix exponential $e^{-(D\Lambda
+ i\gamma \beta\B)t}$ acts on the left on the vector $U$ representing
the initial density $\rho(\r)$ in the basis $\{u_m(\r)\}$:
\begin{equation}
\label{eq:U}
U_m = V^{1/2}\int\limits_\Omega d\r ~ u_m^*(\r)~ \rho(\r) .
\end{equation}
The macroscopic signal is then obtained according to
Eq.~(\ref{eq:signal_integral}) by integrating the transverse
magnetization $\m(\r,t)$ over the whole confining domain $\Omega$ with
a sampling or pickup function $\tilde{\rho}(\r)$ of the measuring coil
or antenna:
\begin{equation}
E = \int\limits_\Omega d\r ~ \m(\r,t) ~ \tilde{\rho}(\r) = 
\sum\limits_m \tilde{c}_m(t)~ \underbrace{V^{-1/2}\int\limits_\Omega d\r ~ u_m(\r) ~ \tilde{\rho}(\r)}_{\tilde{U}_m} .
\end{equation}
The last sum can be interpreted as a scalar product between the vector
$C(t)$ and the vector $\tilde{U}$ representing the pickup function
$\tilde{\rho}(\r)$ in the basis of the eigenfunctions $\{u_m(\r)\}$:
\begin{equation}
\label{eq:Ut}
\tilde{U}_m = V^{-1/2}\int\limits_\Omega d\r ~ u_m(\r)~ \tilde{\rho}(\r)  .
\end{equation}
The macroscopic signal at time $t$ can thus be written in a compact
matrix form of a scalar product:
\begin{equation}
\label{eq:exp-surv}
E = \bigl(U ~ e^{-(D\Lambda + i\gamma \beta \B)t}~ \tilde{U}\bigr) ,
\end{equation} 
matrices being acting on the left.  From a quantum-mechanical point of
view, the matrix $e^{-(D\Lambda + i\gamma \beta\B)t}$ can be thought
of as an evolution operator acting on the initial state $\rho(\r)$
(represented by the vector $U$), while the resulting density
$\m(\r,t)$ is weighted by the pickup or sampling function
$\tilde{\rho}(\r)$ (represented by vector $\U$).  A rapid increase of
the eigenvalues $\lambda_m$ with $m$ allows one to truncate the
matrices $\Lambda$ and $\B$ to moderate sizes in order to perform the
numerical computation.

Although the compact matrix form (\ref{eq:exp-surv}) is derived for
time-independent field $B(\r)$, the approach is easily applicable to
various NMR sequences.  For instance, Carr-Purcell-Meiboom-Gill (CPMG)
sequence consists in repeating the inverting $180^\circ$
radio-frequency pulse $n$ times to acquire successive echoes
\cite{Carr54}.  In the matrix formalism, the amplitude of the $k$th
echo ($k = 1,...,n$) is simply calculated as
\begin{equation*}
E_k = \biggl(U \biggl[e^{-(D\Lambda + i\gamma \beta \B)t/(2n)} ~ 
e^{-(D\Lambda - i\gamma \beta \B)t/(2n)}\biggr]^k \tilde{U}\biggr) .
\end{equation*} 
We developed a spectral analysis of the multiple echo attenuation for
CPMG sequences \cite{Grebenkov06b}.

In general, one can approximate any given temporal profile $f(t)$ by a
piecewise constant function: $f(t) = f_k$ on $[t_k,t_{k+1}]$ ($k =
0,...,K$).  For each interval $[t_k,t_{k+1}]$, the Bloch-Torrey
equation can be solved using the matrix exponential.  In order to
merge the solutions, the ending magnetization of the interval
$[t_{k-1},t_k]$ is taken as the initial condition for solving the
problem on the next interval $[t_k,t_{k+1}]$.  The use of the matrix
representation is particularly efficient to handle such successive
computations:
\begin{equation}
\label{eq:E_general}
E = \biggl(U  \biggl[\prod\limits_{k=0}^K e^{-(D\Lambda + i\gamma \beta f_k\B)(t_{k+1} - t_k)} \biggr] \tilde{U}\biggr) ,
\end{equation} 
where $t_0 = 0$ and $t_{K+1} = t$.  It is worth stressing that this
result is exact, no approximation was involved.  Moreover, since any
function can be approximated by a piecewise-constant function, the
above relation allows one to {\it approximately} compute the signal
for {\it arbitrary} temporal profile $f(t)$.

Conceptually, Eq.~(\ref{eq:exp-surv}) is nothing more than a
representation of the Bloch-Torrey equation in the Laplace operator
eigenbasis.  The matrices $\B$ and $\Lambda$, depending merely on the
eigenbasis and the spatial profile $B(\r)$, have to be computed only
once for a given geometry.  Once this preliminary step is achieved,
the computation of the signal for a given set of physical parameters
is straightforward, accurate and very rapid.  This is the crucial
advantage with respect to conventional numerical techniques for
solving the Bloch-Torrey equation (Monte Carlo simulations, finite
difference or finite element methods, etc.).

As we pointed out in Sect.~\ref{sec:eigen}, the computational
efficiency of the matrix formalism should not be surprising because
finding the eigenbasis is equivalent to solving all the diffusive
problems at once for a given geometry.  Naturally, finding the
eigenfunctions is a difficult task, especially for complex geometries.
But once the eigenbasis is found, the remaining computations are easy.  
We are going to discuss the
recent progress that we achieved by using the matrix formalism for
simple geometries, for which the Laplace operator eigenbasis is known
explicitly: a slab (an interval), a cylinder (a disk), and a sphere,
as well as circular and spherical layers.

\subsubsection{ Moments }
\label{sec:moments}

The compact matrix form (\ref{eq:exp-surv}) of the signal is
particularly suitable for numerical purposes because the computation
of matrix exponentials is rapid and very accurate.  At the same time,
a theoretical analysis with this form faces considerable difficulties
because the governing matrices $\B$ and $\Lambda$ do not commute.  In
this subsection, we follow Ref. \cite{Grebenkov07} and briefly discuss
another strategy, when the signal is represented as a power series
with the moments of the normalized total phase $\phi = \varphi/t$:
\begin{equation}
\label{eq:E_char}
E = \E\{ e^{iq\phi} \} = \sum\limits_{n=0}^\infty \frac{(iq)^n}{n!} \E\{\phi^n\} ,
\end{equation}
where $q = \gamma \beta t$ is a dimensionless parameter characterizing
the intensity of the applied field.  Each moment can be explicitly
written by using the definition (\ref{eq:phi}) of $\phi$:
\begin{equation}
\label{eq:time_integral}
\E\{ \phi^n \} = \frac{n!}{t^n} \int\limits_0^t dt_1~ f(t_1) \int\limits_{t_1}^t dt_2~ f(t_2) ... 
\int\limits_{t_{n-1}}^t dt_n~ f(t_n)~ \E\{ B(X_{t_1}) ... B(X_{t_n})\} .
\end{equation}
where the time moments $t_1$, ... , $t_n$ were put in ascending order.
The multiple correlation function $\E\{ B(X_{t_1}) ... B(X_{t_n})\}$
can be calculated using the Markov property of Brownian motion:
\begin{equation*}
\begin{split}
\E\{ B(X_{t_1}) ... B(X_{t_n})\} = & \int\limits_\Omega d\r_0 ~ \rho(\r_0) 
\int\limits_\Omega d\r_1 ~ G_{t_1}(\r_0, \r_1) ~ B(\r_1) ~ ... \\
& \int\limits_\Omega d\r_n ~ G_{t_n-t_{n-1}}(\r_{n-1}, \r_n) ~ B(\r_n)
\int\limits_\Omega d\r_{n+1} ~ G_{t-t_n}(\r_n, \r_{n+1}) ~ \tilde{\rho}(\r_{n+1}) . \\
\end{split}
\end{equation*}
This lengthy expression has a simple probabilistic interpretation.
Starting at time $0$ from a randomly chosen point $\r_0$ (with a given
initial density $\rho(\r_0)$), a particle diffuses up to time $t_1$ to
point $\r_1$ where it ``experiences'' the magnetic field $B(\r_1)$.
The diffusive propagator $G_{t_1}(\r_0,\r_1)$ describes this process.
From $\r_1$, the particle diffuses to a new point $\r_2$, and so on.
At last, the particle arrives at point $\r_n$ at time $t_n$ and
experiences the magnetic field $B(\r_n)$.  During the the remaining
time $t-t_n$, the particle can diffuse to another $\r_{n+1}$ where it
is ``sampled'' with a given function $\tilde{\rho}(\r_{n+1})$.  Using
the spectral decomposition of the diffusive propagator $G_t(\r,\r')$,
one can deduce a compact matrix form for the multiple correlation
function \cite{Grebenkov07}
\begin{equation}
\label{eq:n-correlation}
\E\{ B(X_{t_1}) ... B(X_{t_n})\} = \bigl(U  e^{-D\Lambda t_1} \B e^{-D\Lambda (t_2-t_1)} \B ~ ... ~ 
\B e^{-D\Lambda(t_n-t_{n-1})} \B e^{-D\Lambda(t-t_n)} \tilde{U}\bigr) ,
\end{equation}
with the same matrices $\B$ and $\Lambda$ and vectors $U$ and
$\tilde{U}$ as earlier.  This is simply a representation of the above
integral form in the Laplace operator eigenbasis.  

In many practical cases, the boundary is purely reflecting (Neumann
boundary condition), while the initial density $\rho(\r)$ and sampling
function $\tilde{\rho}(\r)$ are uniform.  It is easy to check that the
components of the vectors $U$ and $\tilde{U}$ are then zero, except
for the ground mode: $U_m = \tilde{U}_m = \delta_{m,0}$.  Given that
the matrices $e^{-D\Lambda t_1}$ and $e^{-D\Lambda(t-t_n)}$ are
diagonal, the multiple correlation function reads as
\begin{equation*}
\E\{ B(X_{t_1}) ... B(X_{t_n})\} = \bigl[\B e^{-D\Lambda (t_2-t_1)} \B ~ ... ~ 
\B e^{-D\Lambda(t_n-t_{n-1})} \B \bigr]_{0,0} .
\end{equation*}
For instance, the second moment is 
\begin{equation}
\label{eq:phi2}
\E\{\phi^2/2\} = \< \sum\limits_m \B_{0,m}~ e^{-D\lambda_m(t_2-t_1)} ~ \B_{m,0} \>\2 ,
\end{equation}
where $\< ...\>\2$ denotes the time integral in
Eq.~(\ref{eq:time_integral}), for instance,
\begin{equation*}
\< e^{-D\lambda_m(t_2-t_1)} \>\2 = \frac{1}{t^2}\int\limits_0^t dt_1 ~ f(t_1) \int\limits_{t_1}^t dt_2 ~ f(t_2)~
e^{-D\lambda_m(t_2-t_1)} .
\end{equation*}
In this formal way, the original problem of finding the macroscopic
signal of diffusing spins is entirely reduced to the analysis of the
Laplace operator eigenmodes, and is thus solved as a physical problem.
In what follows, we show how this mathematical basis can be applied
for a theoretical analysis in many cases of particular interest.

\subsection{ NMR survey of restricted diffusion }
\label{sec:survey}

In this section, we present various results on restricted diffusion in
magnetic fields that we could obtain by using the Laplace operator
eigenfunctions \cite{Grebenkov07}.

\subsubsection{ Simple geometries }

Restricted diffusion in a slab (an interval), a cylinder (a disk), and
a sphere is a classical problem \cite{Crank,Carslaw}.  Since the
Laplace operator eigenfunctions in these domains are known explicitly,
many diffusion characteristics can be calculated analytically.  For
instance, the eigenfunctions and eigenvalues of the Laplace operator
in the unit disk are
\begin{equation*}
u_{nk}(\r) = \frac{\ve_n}{\sqrt{\pi}} ~ \frac{\beta_{nk}}{J_n(\alpha_{nk})} ~ J_n(\alpha_{nk}r) \cos n\varphi ,
\hskip 10mm  \lambda_{nk} = \alpha_{nk}^2,
\end{equation*}
where $\ve_n = \sqrt{2-\delta_{n,0}}$, the constants $\beta_{nk} =
\sqrt{\lambda_{nk}/(\lambda_{nk} - n^2 + h^2)}$ guarantee the
normalization (\ref{eq:normalization}), $J_n(z)$ are the Bessel
functions of the first kind, and $\alpha_{nk}$ are all the positive
roots of the equations
\begin{equation}
\label{eq:roots}
z J'_n(z) + h J_n(z) = 0 ,
\end{equation} 
coming from the Robin boundary condition (\ref{eq:eigen2}), with $h = \K/D$
($L=1$).  The factorization of the dependences on the radial and
angular coordinates $r$ and $\varphi$ is a direct consequence of the
rotational invariance.  It is worth stressing that here, the single
index $m$ is replaced by a double index $nk$ (with $n=0,1,2,3...$ and
$k=0,1,2,3...$) coming from enumeration of the positive roots
$\alpha_{nk}$.  We shall use the double index as a convenient notation
to enumerate the eigenvalues, eigenfunctions and the elements of
matrices and vectors.

The explicit form of the eigenfunctions is employed to investigate the
signal attenuation due to restricted diffusion in inhomogeneous
magnetic fields (see \cite{Grebenkov07} and references therein).  This
knowledge facilitates the construction of the matrix $\B$ for a given
profile $B(\r)$.  In general, this would require numerical integration
in Eq.~(\ref{eq:B}), but for some choices of the function $B(\r)$, the
integration can be realized analytically.  In \cite{Grebenkov07}, we
provided, for the first time, the explicit formulas for the matrix
$\B$ in two cases of practical interest: a linear magnetic field
gradient, $B(\r) = x$, corresponding to usual experimental setup for
NMR measurements, and a parabolic magnetic field, $B(\r) = |\r|^2$.
For example, we obtained for the unit disk in a linear gradient:
\begin{equation}
\label{eq:B_2d}
\B_{nk,n'k'} = \delta_{n,n'\pm 1} (1 + \delta_{n,0} + \delta_{n',0})^{1/2} \beta_{nk} \beta_{n'k'}
\frac{\lambda_{nk} + \lambda_{n'k'} - 2nn' + 2h(h-1)}{(\lambda_{nk} - \lambda_{n'k'})^2} .
\end{equation}
This matrix turned out to be fully expressed in terms of the
eigenvalues $\lambda_{nk}$.  This means that the originally
complicated problem of computing the macroscopic signal attenuation is
reduced to finding roots of the Bessel functions in
Eq.~(\ref{eq:roots}).  This result allows one for efficient numerical
computation of the signal, as well as for a thorough theoretical
analysis of the moments, as illustrated below.

\subsubsection{ Time-dependent diffusion coefficient }
\label{sec:classical}

Apart from being a basis for efficient numerical computations, the
matrix formalism is a powerful analytical tool.  As a representative
example, we consider a linear magnetic field gradient in a given
direction ${\bf e}$ and of intensity $g$:
\begin{equation*}
\beta = gL,  \hskip 10mm  B(\r) = ({\bf e}\cdot \r)/L ,
\end{equation*} 
$L$ being the characteristic size of the domain.
For relatively small gradients (small $q$), the macroscopic signal in
Eq.~(\ref{eq:E_char}) is essentially determined by the second moment
$\E\{\phi^2/2\}$:
\begin{equation}
\label{eq:GPA}
E \simeq 1 - q^2 \E\{\phi^2/2\} + O(q^4) \simeq \exp[-q^2\E\{\phi^2/2\}] + O(q^4) 
\end{equation}
(the first and other odd moments that would determine the phase of the
signal are omitted).  This formula known as ``Gaussian phase
approximation'' (GPA) was widely employed in many theoretical,
numerical and experimental studies of restricted diffusion (see
\cite{Grebenkov07} and references therein).  It is worth noting that
this formula becomes exact for free (unrestricted) diffusion, for
which the second moment was calculated by Stejskal and Tanner for
arbitrary temporal profile $f(t)$ of the applied gradient
\cite{Stejskal65}
\begin{equation}
\E\{\phi^2/2\}_0 = \frac{D}{L^2} \<(t_1 - t_2)\>\2 .
\end{equation}
Deviation of the second moment $\E\{\phi^2/2\}$ for restricted
diffusion from its value $\E\{\phi^2/2\}_0$ for free diffusion can be
used to characterize a confining domain.  Actually, their ratio is
directly related to apparent, effective or time-dependent diffusion
coefficient $D(t)$ measured in NMR experiments%
\footnote{
It is worth noting that this NMR definition of the time-dependent
diffusion coefficient is different from its classical form, the 
latter being obviously independent of the applied
magnetic field (see \cite{Grebenkov07} for further discussion). }
%
\begin{equation*}
D(t) = D~ \frac{\E\{\phi^2/2\}}{\E\{\phi^2/2\}_0} . 
\end{equation*}
According to Eq.~(\ref{eq:GPA}), this quantity is simply proportional
to the logarithm of the macroscopic signal and is thus easily
accessible in experiments.  A number of studies related the behavior
of the time-dependent diffusion coefficient and the geometry of a
diffusion-confining domain \cite{Sen04}.

If there is no surface relaxation, Eq.~(\ref{eq:phi2}) yields a
general representation of the time-dependent diffusion coefficient for
{\it any} confining domain and {\it arbitrary} temporal profile
\cite{Grebenkov07,Grebenkov08a}
\begin{equation}
\label{eq:ADC}
\frac{D(t)}{D} = \sum\limits_m L^2 \lambda_m \B_{0,m}^2 ~ w_f(D\lambda_m t) ,
\end{equation}
where the function $w_f(p)$ accounts for a given temporal profile
$f(t)$:
\begin{equation}
w_f(p) = \frac{\< e^{-p(t_2 - t_1)/t} \>\2}{\< -p(t_2 - t_1)/t\>\2} .
\end{equation}
It is easy to check the following normalization property for linear
gradients (taking $L=1$ for simplicity):
\begin{equation*}
\sum\limits_m \lambda_m \B_{0,m}^2 = 1
\end{equation*}
so that $\lambda_m \B_{0,m}^2$ can be considered as the weight of the
$m$th eigenmode to the time-dependent diffusion coefficient.  In turn,
the eigenvalue $\lambda_m$ determines a characteristic scale at which
the contribution appears in Eq.~(\ref{eq:ADC}) through the function
$w_f(p)$.  It is important to stress that Eq.~(\ref{eq:ADC}) allows
one to distinguish the roles of various ``ingredients'' of the
problem.  So, the spatial profile of the magnetic field enters only
into the weights $\lambda_m \B_{0,m}^2$, while the temporal profile is
fully taken into account by the function $w_f(p)$.  Finally, the
geometry determines both the time scales $(D\lambda_m)^{-1}$ and the
weights $\lambda_m \B_{0,m}^2$.  This connects directly the spectral
properties of the Laplace operator to measurable characteristics of
complex media.

The general representation (\ref{eq:ADC}) allowed us to retrieve and
extend many classical results.  In the long-time diffusion regime,
when the diffusion length $\sqrt{Dt}$ is much larger than $L$, the
function $w_f(p)$ behaves as
\begin{equation}
\label{eq:w_p_large}
w_f(p) \simeq p^{-2}  \frac{\int\nolimits_0^t dt ~ f^2(t)}{\< (t_1 - t_2)\>\2}  + O(p^{-3})  
\hskip 10mm  (p\gg 1) ,
\end{equation}
yielding
\begin{equation*}
E \simeq \exp\left[ - \frac{\gamma^2 g^2 L^4}{D} ~ \zeta_{-1} ~\int\limits_0^t dt ~ f^2(t) \right]
\end{equation*}
for any bounded domain, with the geometry-dependent dimensionless
constant
\begin{equation*}
\zeta_{-1} = \sum\limits_m \B_{0,m}^2 (L^2\lambda_{m})^{-1} .
\end{equation*}
This is an extension of the classical results by Robertson and Neuman
\cite{Robertson66,Neuman74}.  It  shows a high sensitivity of
diffusion-weighted measurements to the size $L$ of a confining domain
in the long-time regime.

In the opposite short-time diffusion regime ($p\ll 1$), only a small
fraction of particles near the boundary can ``feel'' their geometrical
confinement, while the remaining majority of particles diffuse as they
were in free (unbounded) space.  The fraction of ``restricted''
particles can be estimated as the ratio between the volume of the
diffusion layer of width $\sqrt{Dt}$ near the boundary and the total
volume: $S
\sqrt{Dt}/V$, $S$ being the total surface area.  The deviation of the
time-dependent diffusion coefficient from its nominal (free diffusion)
value $D$ is then quantified by this ratio.  A more rigorous analysis
based on Eq.~(\ref{eq:ADC}) gives \cite{Grebenkov07}
\begin{equation}
\frac{D(t)}{D} \simeq 1 - \sqrt{Dt} \underbrace{\zeta_{3/2} \frac{S}{V}}_{\rm geometry} 
\underbrace{\frac{\<(t_2 - t_1)^{3/2}\>\2}{\<(t_2 - t_1)\>\2~\sqrt{t}}}_{\rm temporal~ profile}  + O(t) ,
\end{equation}
where $\zeta_{3/2}$ is a geometry-dependent constant.  This is an
extension of the classical result by Mitra {\it et al.}
\cite{Mitra92,Mitra93,deSwiet94}.  This relation was suggested for
experimental measurement of the surface-to-volume ratio $S/V$ in
porous media.

\subsubsection{ Different diffusion regimes }
\label{sec:diagram}

In many cases of practical interest, the knowledge of the second
moment is enough for an accurate approximation of the macroscopic
signal.  When the gradient intensity increases, the fourth and
higher-order moments become progressively more and more significant
(see also Sect.~\ref{sec:cosine}).  In particular, Stoller and
co-workers obtained the asymptotic behavior of the macroscopic signal
at high gradient intensity for a slab geometry \cite{Stoller91}:
\begin{equation}
\label{eq:localization}
E \propto \exp\left[- \frac{a_1}{2} (D\gamma^2 g^2 t^3)^{1/3}\right] ,
\end{equation}
where $a_1\simeq 1.0188$ is the absolute value of the first zero of
the derivative of the Airy function.  In this so-called localization
regime, the signal is essentially formed by the nuclei diffusing (or
``localized'') near the boundary and thus acquiring less dephasing
than the nuclei in the bulk.  Four years later, this theoretical
prediction was experimentally confirmed by H\"urlimann {\it et al.}
\cite{Hurlimann95}.  Measuring the signal attenuation from water
molecules diffusing between two parallel plates, H\"urlimann and
co-workers observed a spectacular deviation from the Gaussian $g^2$
dependence of $\ln E$ at gradient intensities higher than $15$~mT/m.

\begin{figure}
\begin{center}
\includegraphics[width=85mm]{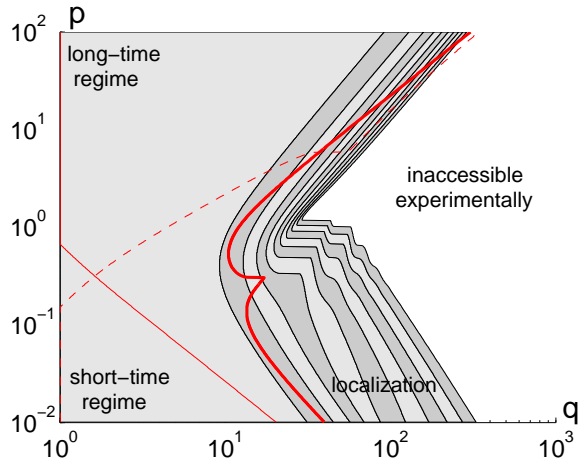}
\end{center}
\caption{ 
Different regimes of restricted diffusion in a slab under a linear
magnetic field gradient.  The signal is attenuated by a factor of $2$
at each line separating two adjacent gray-scale regions (appearing as
pale and dark stripes).  The first large pale region on the left is
composed of points $(q,p)$ for which the signal $E$ lies between $1/2$
and $1$.  The next dark stripe regroups points $(q,p)$ for which $1/4
\leq E\leq 1/2$, and so on.  The white area on the right corresponds
to pairs $(q,p)$ for which the signal is below $10^{-3}$.  Since such
a small signal is often comparable to noise, this area is referred to
as inaccessible experimentally.  On this $pq$ diagram, the bold line
delimits the region on the left in which the GPA predictions of the
signal attenuation are valid with an accuracy of at least $5\%$. }
\label{fig:diagram}
\end{figure}

In our review \cite{Grebenkov07}, we summarized different diffusion
regimes (Fig.~\ref{fig:diagram}).  Since diffusion and encoding in a
spin-echo experiment can be set {\it independently}, we plot the
signal as a function of the two dimensionless parameters $p =Dt/L^2$
and $q=\gamma g L t$.  From this plot, one can determine which kind of
restricted diffusion is to be expected, and which formula should be
applied to fit experimental data.  It is worth noting, however, that
this information is rather qualitative since the particular location
of different regions on the diagram depends on the diffusion-confining
geometry and applied magnetic field.

\subsubsection{ Rigorous results for relaxing boundaries }

For three simple domains (interval, disk and sphere), many relevant
quantities can be found analytically even in the presence of surface
relaxation.  Rigorous computations rely on explicit formulas for the
elements of the matrices $\B$ and $\Lambda$ that were shown to be
rational functions of the eigenvalues (e.g., Eq.~(\ref{eq:B_2d})).
Using the Laplace transform summation technique, we derived exact and
explicit representations for the zeroth and second moments in the
presence of surface relaxation \cite{Grebenkov07b}.  Within the
Gaussian phase approximation, these two moments determine the
reference and diffusion-weighted signals, respectively.

Without presenting the details, we mention that the computation was
essentially based on two observations:
\begin{itemize}
\item
The Laplace transform of the even-order moments can be reduced to a
combination of multiple sums of the form
\begin{equation*}
\sum\limits_{m_1,...,m_n} \frac{1}{(s_1 - \lambda_{m_1})^{\beta_1}}
\frac{1}{(\lambda_{m_1} - \lambda_{m_2})^{\beta_{12}}} 
\frac{1}{(s_2 - \lambda_{m_2})^{\beta_2}} 
\frac{1}{(\lambda_{m_2} - \lambda_{m_3})^{\beta_{23}}}  ~ ... ~
\frac{1}{(s_n - \lambda_{m_n})^{\beta_n}}
\end{equation*}
where $\{s_k\}$ is a set of real parameters, and $\{\beta_k\}$ is a
set of positive integer numbers.

\item
Algebraic transformations and differentiations can reduce any sum of
this form to the function
\begin{equation}
\sum\limits_m \frac{1}{s - \lambda_m}
\end{equation}
which can be calculated explicitly for the eigenvalues $\lambda_m$ of
the Laplace operator in the interval, disk, or sphere.  

\end{itemize}
As a consequence, the Laplace transform of the even-order moments have
explicit, though cumbersome, analytical forms.  In particular, these
explicit forms allow one to study the asymptotic behavior of the
moments in different diffusion regimes.  In the short-time regime, the
series expansion in half-integer powers of the diffusion coefficient
was generalized to arbitrary temporal profile of a linear magnetic
field gradient.  In the long-time regime, it was shown how the
presence of surface relaxation modified classical Robertson's results
(for details, see \cite{Grebenkov07b}).

\subsubsection{ Cosine magnetic field in a slab }
\label{sec:cosine}

Although the above analysis could in principle be carried out for any
even-order moment, its practical implementation was already tedious even
for the second moment.  

But a better knowledge of the high-order moments $\E\{\phi^n\}$ would
help to understand the limits of applicability of the Gaussian phase
approximation and the transition towards non-Gaussian regimes.  We
addressed this problem for restricted diffusion between parallel
planes in a cosine magnetic field \cite{Grebenkov07d}.  The specific
choice of the spatial profile to be proportional to an eigenfunction
of the Laplace operator in this confining geometry considerably
simplified the underlying mathematics.  In fact, the eigenfunctions of
the Laplace operator on the unit interval ($L=1$) with reflecting
endpoints are simply $u_m(x) = \epsilon_m \cos (\pi m x)$, so that the
substitution of the spatial profile $B(x) = \cos (\pi x)$ into
Eq.~(\ref{eq:B}) yields a particularly simple form
\begin{equation*}
\B_{m,m'} = \frac{1}{\epsilon_m \epsilon_{m'}}\bigl(\delta_{m',m-1} + \delta_{m',m+1}\bigr) ,
\end{equation*}
where $\epsilon_m = \sqrt{2 - \delta_{m,0}}$ are the normalization
constants.  This subdiagonal structure of the matrix $\B$ suggests the
use of an analogy with reflected random walks to calculate the
multiple correlation functions $\E\{B(X_{t_1})...B(X_{t_n})\}$.  For
instance, the second moment with the cosine magnetic field is simply
\begin{equation}
\label{eq:E2_cosine}
\E\{\phi^2/2\} = \frac12 \< \exp[-D\pi^2(t_2 - t_1)]\>\2 .
\end{equation}
Exact and explicit relations for several higher-order moments were
also reported \cite{Grebenkov07d}.

\begin{figure}
\begin{center}
\includegraphics[width=100mm]{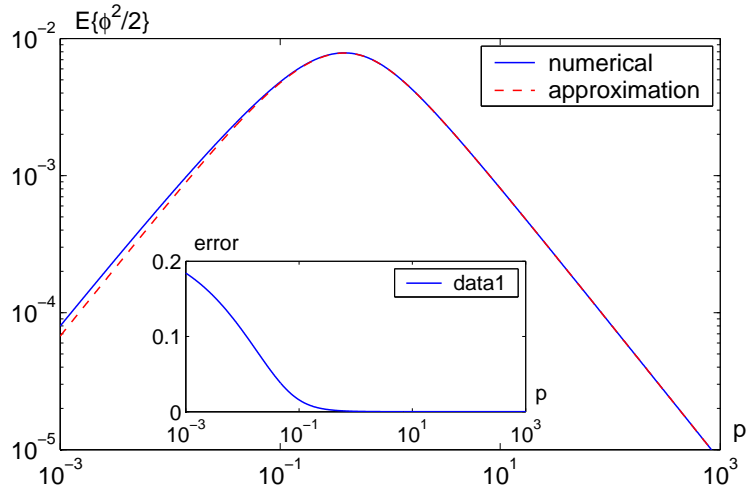}
\end{center}
\caption{
Numerically computed second moment $\E\{\phi^2/2\}$ for a linear
gradient with bipolar temporal profile as a function of $p = Dt/L^2$
and its explicit approximation derived from the exact solution 
(\ref{eq:E2_cosine}) for a cosine spatial profile.
The inset shows relative deviation between them.  }
\label{fig:E2_cosine}
\end{figure}

These results helped us to study the general structure and the
properties of the high-order moments, as well as deviations from the
GPA at magnetic fields of moderate or high intensity.  Although the
cosine magnetic field is indeed very specific (e.g., its experimental
realization presents a challenge in itself), the deduced properties of
the high-order moments provide a better understanding of the GPA and
its limitations in general.  Moreover, numerical simulations showed
that the explicit relations obtained for the cosine magnetic field can
be used as good approximations for linear magnetic field gradients (as
illustrated in Fig.~\ref{fig:E2_cosine} for the second moment).  Note
also that Zielinski and Sen considered the cosine spatial profile to
model and study susceptibility-induced local magnetic fields
\cite{Zielinski00}.

\subsubsection{ Diffusion in circular and spherical layers }
\label{sec:layers}

There still exists a substantial ``gap'' in understanding restricted
diffusion in simple geometries, on the one hand, and in natural porous
media such as sedimentary rocks, cement, or biological tissues, on the
other hand.  From a theoretical point of view, the complexity of
porous structures is related to multiple length scales, ranging from
the size of tiny pores (in the order of several microns in some rocks)
to the overall size of the sample (e.g., few centimeters).  Moreover,
porous media often present a hierarchical structure of pores so that
intermediate characteristic length scales emerge.  In this case, even
the introduction of representative dimensionless parameters such as
$p$, $q$, and $h$ becomes ambiguous as being depended on a geometrical
length $L$ used.  It is thus important to understand how classical
theories for single-scale shapes (like a slab) can be extended to
incorporate multiple scales.  Since an analytical calculation in
natural porous media is practically infeasible, some simplified
confining domains with two or multiple length scales become
particularly useful.

\begin{figure}
\begin{center}
\includegraphics[width=100mm]{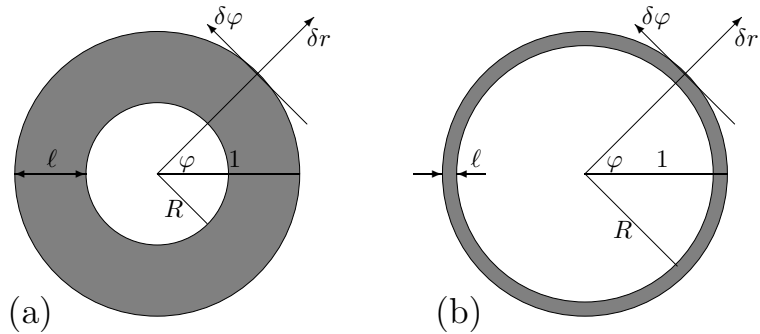}
\end{center}
\caption{
Circular layers of thickness $\ell = 1-R$ with inner radius $R$ and
outer radius $L=1$: $R=0.5$ (a) and $R=0.9$ (b).  Note that variations
in radial and angular coordinates are orthogonal to each other.  }
\label{fig:layers}
\end{figure}

In this light, circular and spherical layers (or shells) appear as
appropriate models (Fig.~\ref{fig:layers}).  Like for a disk and a
sphere, the Laplace operator eigenbasis in these domains is known
explicitly.  Using the properties of the underlying Bessel equation,
we derived explicit analytical formulas for the matrix $\B$
representing a linear magnetic-field gradient (or parabolic magnetic
field as well) in the Laplace operator eigenbasis \cite{Grebenkov08a}.
From the numerical point of view, the problem is completely reduced to
finding roots of the equations with Bessel functions.  Once these
roots are found, the governing matrices $\B$ and $\Lambda$ can easily
be constructed and used to compute the macroscopic signal for any
temporal profile of the magnetic field.

Bearing in mind multiscale structures, we particularly focus on thin
layers (Fig.~\ref{fig:layers}b), for which perturbative calculations
turned out to be surprisingly accurate (here we assume that there is
no surface relaxation).  For instance, the perturbation series of the
eigenvalues in powers of the small dimensionless thickness $\ell$ are
(in 2D):
\begin{equation*}
\begin{split}
\lambda_{n0} & = n^2 \biggl(1 + \ell + \frac56 \ell^2 + \frac23 \ell^3 - \frac{n^2-16}{30} \ell^4 + O(\ell^5)\biggr) , \\
\lambda_{nk} & = \frac{\pi^2 k^2}{\ell^2} \biggl(1 + \frac{n^2+3/4}{\pi^2 k^2} \ell^2 + O(\ell^3)\biggr) \hskip 5mm (k>0) ,\\
\end{split}
\end{equation*}
so that the analysis is indeed much simpler for thin layers than for
the unit disk (we took $L = 1$ so that the eigenvalues are
dimensionless).  The smallness of the thickness $\ell$ with respect to
the perimeter $2\pi$ results in a large gap between
$\lambda_{n0}\approx n^2$ and $\lambda_{nk}\approx
\pi^2 k^2/\ell^2$ with $k>0$.  The smaller eigenvalues $\{\lambda_{n0}\}$
describe large displacements in angular coordinates along the
boundaries (i.e., along the circumference of the circular layer).  In
turn, the larger eigenvalues $\{\lambda_{nk}\}$ (with $k > 0$)
describe small displacements in radial coordinate between the inner
and outer boundaries.  This situation is somehow analogous to
restricted diffusion in a thin rectangle with sides $\pi$ and $\ell$,
for which $\lambda_{nk}^{\rm rect} = n^2 +
\pi^2 k^2/\ell^2$.

\begin{figure}
\begin{center}
\includegraphics[width=75mm]{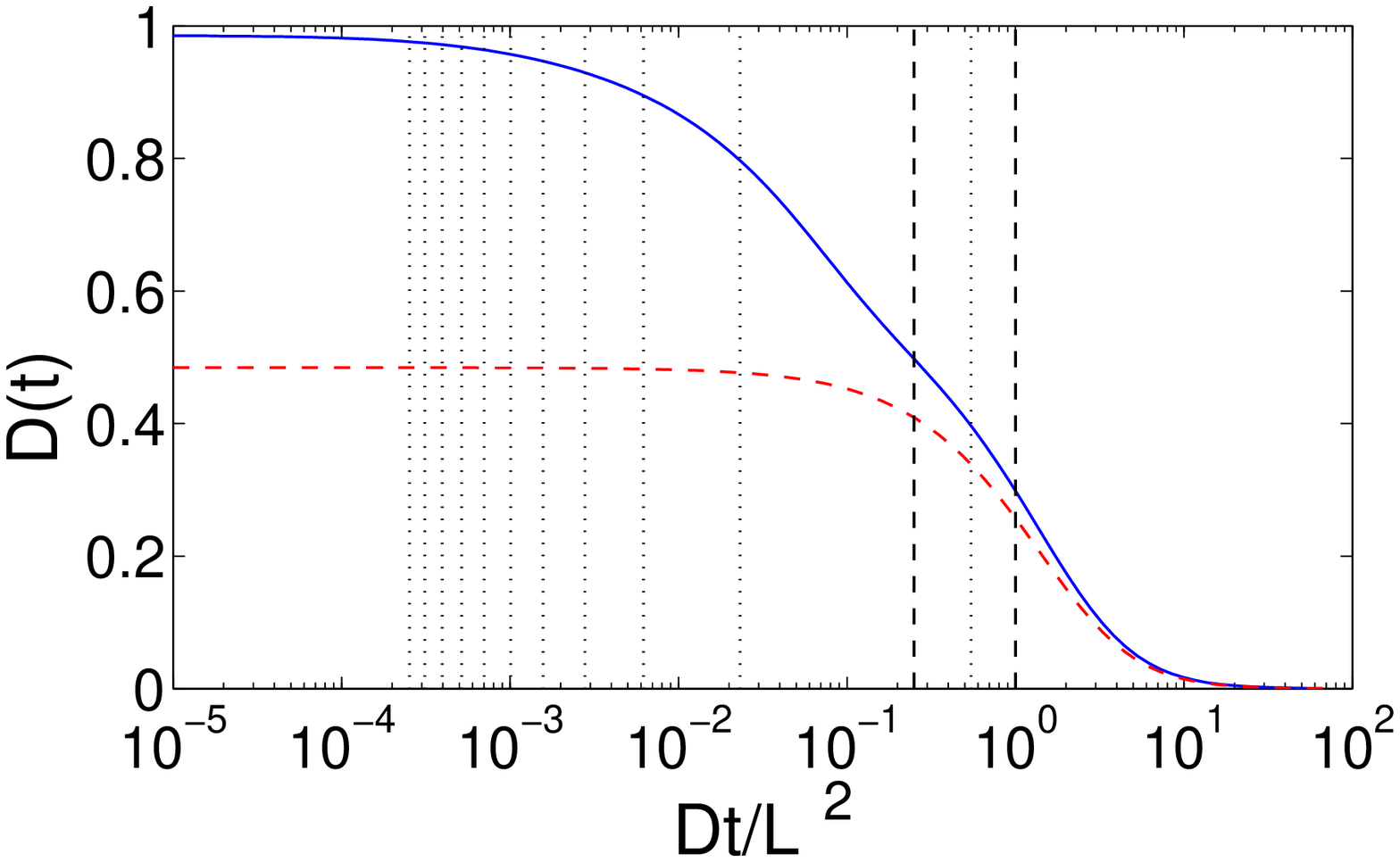}
\includegraphics[width=75mm]{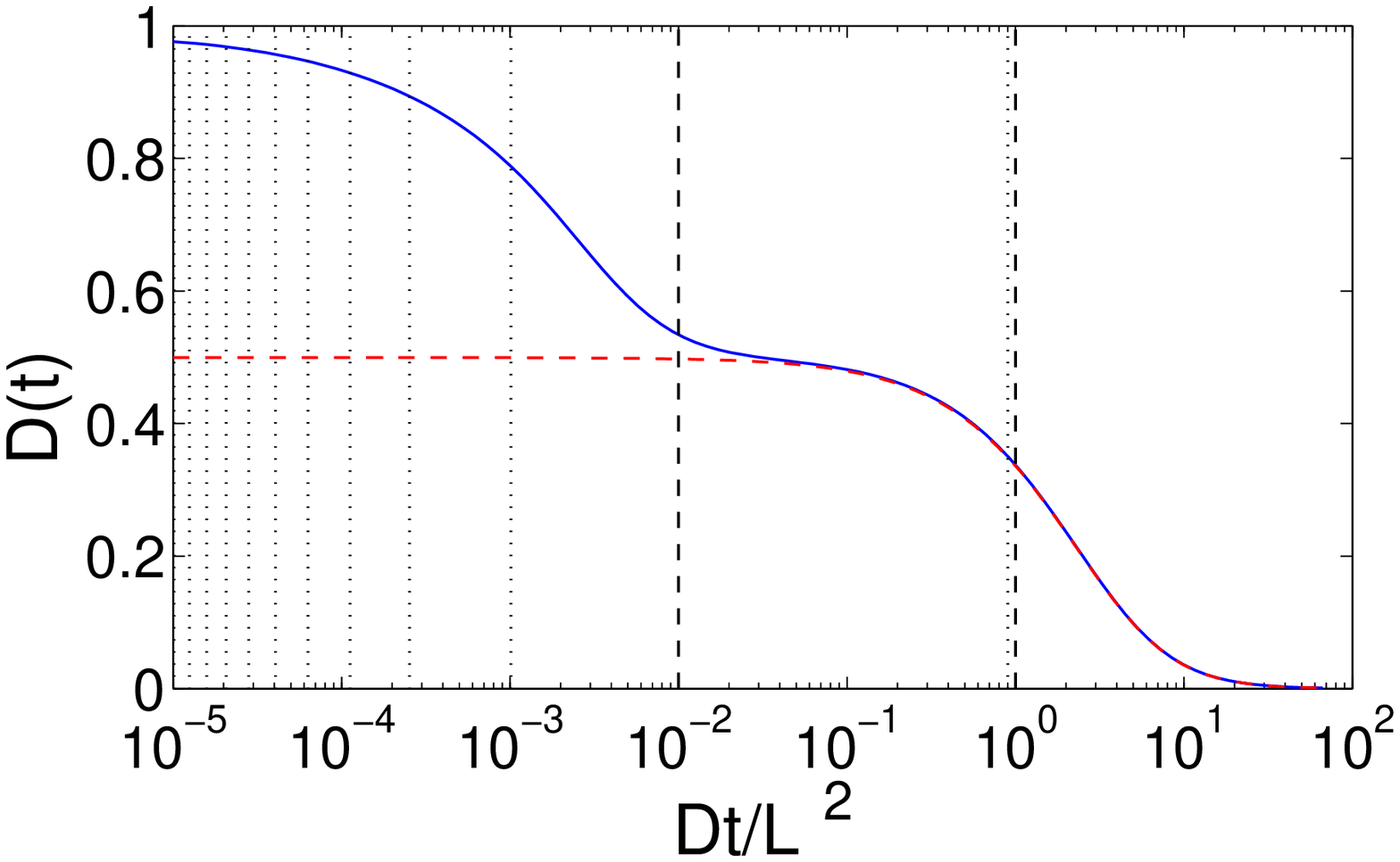}
\end{center}
\caption{
Normalized time-dependent diffusion coefficient $D(t)/D$ as a function
of the dimensionless parameter $p = Dt/L^2$ ($L = 1$) for circular
layers with inner radii $R = 0.5$ (left) and $R = 0.9$ (right).  The
solid line shows the exact computation via Eq.~(\ref{eq:ADC_layer}),
while the dashed line represents an approximate solution containing
only the first contributing eigenvalue $\lambda_{10}$ and thus failing
for small $t$.  Two vertical dashed lines (at $\ell^2$ and $1$)
roughly split the $p$ axis in three regions: short-time regime (on the
left, $p\ll \ell^2$), intermediate region, and long-time regime (on
the right, $p\gg 1$).  The vertical dotted lines show the scales
$\lambda_{1k}^{-1}$ ($k$ ranging from $0$ to $10$, with $k = 0$ on the
right and $k = 10$ on the left).  One can observe a plateau due to a
large separation between the two contributing eigenvalues
$\lambda_{10}$ and $\lambda_{11}$ (the thinner the layer, the wider
the plateau).  Note that 2D diffusion in a thick circular layer (a 2D
shape) is reduced to essentially one-dimensional motion in a thin
circular layer, yielding the reduction factor $1/2$ (the level of the
plateau). }
\label{fig:ADC_layers}
\end{figure}

Apart from the above perturbative analysis and derivation of the
explicit formulas for the matrix $\B$, the main result of
\cite{Grebenkov08a} concerned the behavior of the time-dependent
diffusion coefficient in these two-scale domains.  Although the
summation in Eq.~(\ref{eq:ADC}) was carried out over all eigenmodes
(here enumerated by double index $nk$), the rotation invariance of the
circular and spherical layers eliminates the contributions of all
eigenmodes except for $n=1$:
\begin{equation}
\label{eq:ADC_layer}
\frac{D(t)}{D} = \lambda_{10} \B_{10}^2 w_f(D\lambda_{10}t) + 
\sum\limits_{k=1}^\infty \lambda_{1k} \B_{1k}^2 w_f(D\lambda_{1k}t) .
\end{equation}
In this sum, the first term with small eigenvalue $\lambda_{10}\approx
1$ (in 2D) is explicitly separated from the remaining terms with large
eigenvalues $\lambda_{1k} \approx \pi^2 k^2/\ell^2$.  As we mentioned
in Sect.~\ref{sec:classical}, the function $w_f(p)$ slowly approaches
$1$ as $p$ goes to $0$, and has a power law decay (\ref{eq:w_p_large})
for $p$ going to infinity.  A large gap between $\lambda_{10}$ and
$\lambda_{11}$ creates a region of values $p$ such that
$\lambda_{11}^{-1} \ll p \ll \lambda_{10}^{-1}$, for which the first
term in Eq.~(\ref{eq:ADC_layer}) is nearly constant, while the other
terms can be neglected.  This suggests an emergence of an intermediate
diffusion regime with a nearly constant $D(t)$, which is analogous to
the tortuosity regime in porous media \cite{deSwiet96,Sen04}.  This is
a new, two-scale feature which could not be observed for one-scale
domains such as the unit disk or sphere (Fig.~\ref{fig:ADC_layers}).
A complete analytical description developed in \cite{Grebenkov08a}
allows one to study first the transition from the short-time diffusion
to the intermediate (or tortuosity) regime at $p$ around $\ell^2$, and
then the transition to the ultimate long-time regime when $p$ exceeds
$1$.

In general, the emergence of an intermediate region with a constant
$D(t)$ would be a sign of a multi-scale geometry.  Inversely, a
significant separation in length scales in statistically homogeneous
and isotropic porous media is expected to result in distinct constant
regions in the behavior of $D(t)$.  Since these regions can in
principle be observed by varying the diffusion time $t$, this
observation suggests an experimental way for detecting multiple
scales.  However, a reliable interpretation of such measurements still
requires a substantial study of the Laplace operator eigenbasis in
porous structures.

\subsection{ Residence times of reflected Brownian motion }
\label{sec:residence} 

\begin{figure}
\begin{center}
\includegraphics[width=70mm]{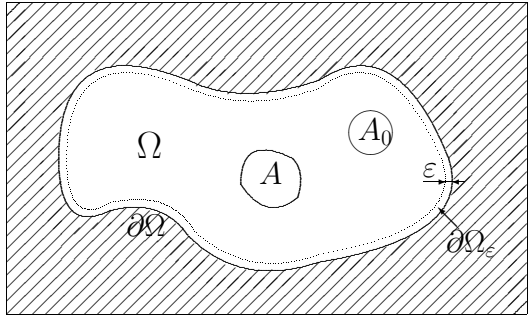}
\end{center}
\caption{
Diffusive motion of the particles restricted inside a bounded
confining domain $\Omega$ with reflecting boundary $\pa$.  The
residence time $\varphi$ indicates how long a diffusing particle,
started somewhere in the domain (e.g., in the subset $A_0$), spends in
a subset $A$ until time $t$.  The local time shows how long the
particle spends in a close neighborhood ($\ve$-vicinity or
$\ve$-sausage $\pa_\ve$) of the boundary $\pa$.  }
\label{fig:domain_resid}
\end{figure}

In the previous subsection, we focused on NMR-oriented applications,
when the choice of the ``observable'' $B(\r)$ of Brownian motion to be
a linear function of $\r$ was dictated by a widespread use of a linear
magnetic field gradient in modern NMR scanners.  The range of problems
that can be tackled by matrix formalism is much broader.  We consider
here another important choice for the ``observable'', when $B(\r)$ is
equal to the indicator function $\I_A(\r)$ of a subset $A$ of the
confining domain: $\I_A(\r) = 1$ for $\r\in A$, and $0$ otherwise
(Fig.~\ref{fig:domain_resid}).  In this case, the random variable
$\varphi$ from Eq.~(\ref{eq:phi}) can be thought of as a ``counter''
which is turned on whenever the diffusing particles resides in $A$
\cite{Grebenkov07a}.  This is so-called residence or occupation time
which is relevant for various diffusion-influenced reactions, for
which the net outcome and the whole functioning of the system strongly
depend on how long the diffusing particles remain in reactive zones
(e.g., in a chemical reactor or biological cell) \cite{Agmon84}.  For
instance, one can estimate the ``trapping'' time that particles spend
in deep ``fjord-like'' pores of a catalyst.  One can also mention
surface relaxation processes in NMR \cite{Brownstein79} or kinetics of
binding of ligands to a cell partially covered by receptors in
microbiology \cite{Zwanzig91}.

The characteristic function $\E\{e^{iq\varphi}\}$ of the residence
time $\varphi$ is determined by Eq.~(\ref{eq:exp-surv}), in which the
matrix $\B$ is defined according to Eq.~(\ref{eq:B}) with $\B(\r) =
\I_A(\r)$:
\begin{equation}
\label{eq:B_resid}
\B_{m,m'} = \int\limits_\Omega d\r ~ u_m^*(\r) ~ \I_A(\r) ~ u_{m'}(\r) = 
\int\limits_A d\r ~ u_m^*(\r) ~ u_{m'}(\r) .
\end{equation}
The sampling function $\tilde{\rho}(\r)$, determining the vector
$\tilde{U}$, provides a way to weight or to condition the arrival
points.  In the simplest case with no conditioning, one uses
$\tilde{\rho}(\r) = 1$.  If one aims to investigate the survival
probability of only those particles that arrive in some subregion
$\Omega_0$ of $\Omega$, one takes $\tilde{\rho}(\r) \propto
\I_{\Omega_0}(\r)$ (Fig. \ref{fig:domain_resid}).

With the matrix formalism, one can easily assess much finer and more
detailed statistics of residence times.  For instance, to compute the
residence time of the diffusing particle between two times $0 < t_1 <
t_2 < t$, a ``counter'' should be forced to turn on only during this
period.  Since the evolution of the system during two other periods
$[0,t_1]$ and $[t_2,t]$ is unperturbed (no ``counting'' or
``interaction''), Eq.~(\ref{eq:exp-surv}) can be modified as
\begin{equation*}
\E\{e^{iq\varphi}\} = \bigl(U ~ e^{-D\Lambda t_1}~ e^{-(D\Lambda -iq\B)(t_2 - t_1)} ~ e^{-D\Lambda(t-t_2)} ~ \tilde{U} \bigr) .
\end{equation*}

Three matrix exponentials represent three successive time periods of
the evolution.  This ``matrix product rule'' can be applied in
general, when one studies the residence time for a sequence of time
intervals.  Moreover, one can change the region of interest (set $A$)
between different time periods in order to describe exchange processes
between subsets (e.g., two pores of a medium).

\begin{figure}
\begin{center}
\includegraphics[width=100mm]{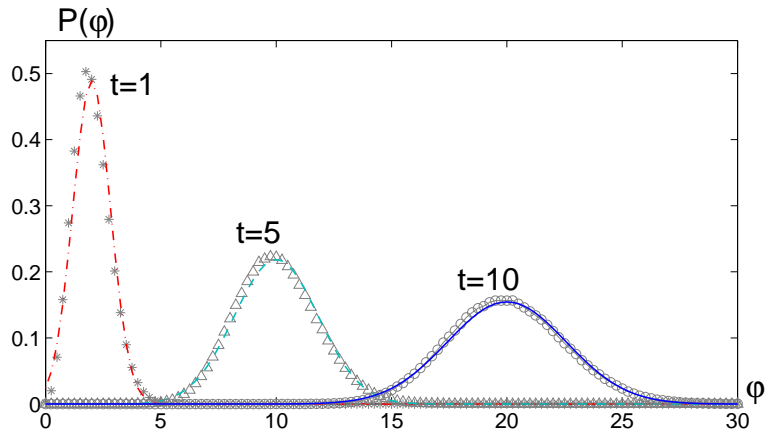}
\end{center}
\caption{
The probability distribution of the local time of reflected Brownian
motion on the unit interval at times $t = 1, 5, 10$ (stars, triangles,
and circles, respectively) obtained by Monte Carlo simulations, and
its Gaussian approximation with mean $b_{1,1} = 2$ and variance
$b_{2,1} = 2/3$ (dashed-dotted, dashed, and solid lines,
respectively).  }
\label{fig:residence_Gauss}
\end{figure}

We employed the matrix representation of Sect.~\ref{sec:moments} to
investigate the long-time behavior of the moments of the residence
time and to show that $\E\{\varphi^n\}$ is a polynomial of degree $n$
as $t$ goes to infinity \cite{Grebenkov07a}.  The coefficients of this
polynomial are expressed through the two governing matrices $\B$ and
$\Lambda$, and vectors $U$ and $\tilde{U}$, using a diagrammic
representation.  In addition, we considered cumulant moments $\prec
\varphi^n \succ$ defined as coefficients of the series
\begin{equation*}
\ln \E\{ e^{iq\varphi}\} = \sum\limits_{n=1}^\infty \frac{(iq)^n}{n!} \prec \varphi^n \succ .
\end{equation*}
We argued that all the cumulant moments behave as linear functions of $t$
at long times:
\begin{equation*}
\prec \varphi^n\succ \simeq b_{n,1} t + b_{n,0} ,
\end{equation*}
where the coefficients $b_{n,k}$ were explicitly given by using a
diagrammic representation.  This statement was explicitly demonstrated
for the cumulant moments up to the order $4$ and checked numerically
for several higher orders.  Since the variance $\prec \varphi^2 \succ$
linearly increase in time, it is natural to renormalize $\varphi$ by
$\sqrt{t}$.  The first cumulant moment of the new random variable
$\varphi/\sqrt{t}$ is still increasing in time, while its variance
approaches a constant.  In contrast, higher-order cumulant moments
with $n>2$ go to $0$.  In the long-time limit, the probability
distribution of the normalized random variable $\varphi/\sqrt{t}$
becomes closer and closer to a Gaussian distribution with mean
$b_{1,1} \sqrt{t}$ and variance $b_{2,1}$, where
\begin{equation*}
b_{1,1} = \B_{0,0},  \hskip 10mm  b_{2,1} = 2\sum\limits_m \B_{0,m}^2 \lambda_m^{-1} .
\end{equation*}
As a consequence, the normalized residence time $\varphi/\sqrt{t}$ is
getting closer and closer to a Gaussian variable as time $t$ grows.
These theoretical results are successfully confronted with Monte Carlo
simulations (Fig.~\ref{fig:residence_Gauss}).

\subsubsection{ Local time on the boundary }

The definition of reflected Brownian motion in
Sect.~\ref{sec:reflected} involves the local time $\ell_t$ which
characterizes the residence time on the boundary
\begin{equation*}
\ell_t = \lim\limits_{\ve \to 0} \frac{1}{\ve} \int\limits_0^t ds~ \I_{\pa_\ve}(X_s) ,
\end{equation*}
where $\pa_\ve$ is the $\ve$-vicinity of the boundary $\pa$: $\pa_\ve
= \{ \r \in \Omega ~:~ |\r - \pa| \leq \ve\}$
(Fig.~\ref{fig:domain_resid}).  The local time $\ell_t$ is also
related to the statistics of finite-distance reflections from the
boundary, which is crucial for intermittent Brownian dynamics
\cite{Bychuk95,Stapf95,Levitz05b,Benichou06}.  The characteristic
function of the local time is given by Eq.~(\ref{eq:exp-surv}) with
the matrix $\B$ determined by substituting $B(\r) =\I_{\pa_\ve}/\ve$
in Eq.~(\ref{eq:B_resid}).  When normalized by $\ve$, the integral
over the $\ve$-vicinity $\pa_\ve$ converges to the surface integral
over the (smooth) boundary $\pa$, yielding
\begin{equation}
\label{eq:Bs}
\tilde{\B}^s_{m,m'} = \int\limits_{\pa} d\r ~ u_m^*(\r) ~ u_{m'}(\r) .
\end{equation}
Here, the passage to the limit $\ve\to 0$, which is in general
delicate and time-consuming for other numerical techniques is
implemented intrinsically.  This is one of the crucial advantages of
the matrix formalism for computing local times.

\subsubsection{ Surface relaxation mechanism }

The matrix formula (\ref{eq:exp-surv}) for the macroscopic signal
exhibits the explicit dependence on the physical parameters $D$ and
$\gamma \beta$, characterizing respectively diffusion and encoding.
In turn, the dependence on the surface relaxivity $\K$ remains
implicit, as being incorporated via the Robin boundary condition
(\ref{eq:eigen2}) for the eigenfunctions.  As a consequence, the
eigenfunctions, the eigenvalues, and the governing matrices $\Lambda$
and $\B$ have to be recalculated for each new value of the surface
relaxivity $\K$.  Since their computation is in general the most
time-consuming step, such an implicit dependence on $\K$ can be
considered as a drawback of matrix formalisms.

This drawback can be overcome by introducing relaxation mechanisms via
an inhomogeneous distribution of relaxation rates $\tilde{B}(\r)$ in
Eq.~(\ref{eq:diffusion2}) with {\it Neumann boundary condition}.
Surface relaxation can be implemented via a distribution
$\tilde{B}(\r)$ localized near the boundary, e.g., $\tilde{B}_\ve(\r)
= \ve^{-1} \I_{\pa_\ve}(\r)$.  In the limit $\ve$ going to $0$, the
volume integral in Eq.~(\ref{eq:B}), determining the governing matrix
$\tilde{B}$, is reduced to the boundary integral (\ref{eq:Bs}).

Since the mechanisms of gradient encoding and of bulk or surface
relaxations are independent, their effects are simply superimposed as
a linear combination of the corresponding terms in
Eq.~(\ref{eq:diffusion2}).  Consequently, the above expressions for
the signal can be easily modified to include different attenuation
mechanisms.  For instance, Eq.~(\ref{eq:exp-surv}) for the FID in the
presence of surface relaxation becomes
\begin{equation*}
E = \bigl(U ~ e^{-(D\Lambda + i\gamma\beta \B + \K\tilde{\B}^s)t} \tilde{U}\bigr) .
\end{equation*} 
The advantage of this relation is the explicit dependence on all three
physical parameters $D$, $\gamma \beta$, and $\K$.  The structure of
each term has a clear physical interpretation: $D\Lambda$ describes
restricted diffusion, $i\gamma\beta\B$ represents the dephasing, and
$\K\tilde{\B}^s$ accounts for surface relaxation.  One can similarly
extend other matrix formulas for the CPMG sequence or any temporal
profile.  Nonuniform bulk or surface relaxations can also be
incorporated.

It is crucial to stress that here the eigenvalues and eigenfunctions
are defined for Neumann boundary condition, whatever the value of
surface relaxivity $W$ is.  As a consequence, these eigenfunctions, as
well as the governing matrices $\Lambda$, $\B$, and $\tilde{\B}^s$,
depend only on the diffusion-confining geometry and have to be
constructed only once for a given confining domain.  The introduction
of the matrix $\tilde{\B}^s$ is therefore an alternative way to
incorporate uniform surface relaxation.  Moreover, it is a general
frame for dealing with a nonuniform distribution $\tilde{B}(\r)$ of
the relaxation rates, either in the bulk, or on the boundary.  This
concept is easily extendable for a superposition of various
attenuation mechanisms.  For instance, one can study the combined
effect of the surface and bulk relaxations, gradient encoding,
presence of dipolar magnetic field, etc.


\newpage
\section*{ Conclusion }
\addcontentsline{toc}{section}{{Conclusion}}

In this work, we presented theoretical and numerical results that we
achieved in the course of the last five years.  In order to study
restricted diffusion in complex geometries, we combined probabilistic
tools with spectral analysis.

An implementation of Monte Carlo techniques that we specifically
adapted to several model geometries allowed us to shed a new light
onto diffusion in complex media.  With these tools, we tackled
such problems as multifractal properties of the harmonic measure 
in 2D and 3D (Sect.~\ref{sec:harmonic}), scaling properties of the 
spread harmonic measures and the role of the exploration length
(Sect.~\ref{sec:spread}), first passage statistics
(Sect.~\ref{sec:first_passage}), passivation processes
(Sect.~\ref{sec:passivation}), diffusion-weighted imaging of the 
lungs (Sect.~\ref{sec:Kitaoka}).  These problems come from 
different fields (harmonic analysis, heterogeneous catalysis, heat
transfer, nuclear magnetic resonance, physiology, medicine, etc.),
in which geometrical complexity plays a central role.

Monte Carlo techniques successfully answer the question {\it how}
particles diffuse in a given medium.  However, they may fail to
explain {\it why} some features of diffusion are common while others 
are too specific or geometry-dependent.  In other words, what differentiates
various confining media in respect of diffusion?  This question is 
tightly related to inverse and optimization problems which consist 
in determining or designing geometry according to some diffusion
characteristics.  A spectral approach provides a unifying mathematical
``language'' for tackling these problems.  In fact, many diffusion 
characteristics can be expressed in terms of the Laplace operator 
eigenfunctions whose properties are therefore a source of potentially
important information.  For instance, the interplay between these 
eigenfunctions and a linear magnetic field gradient yielded an 
intermediate diffusion regime for thin circular and spherical layers 
(Sect.~\ref{sec:layers}).  More generally, the
spectral approach was successfully applied to retrieve, extend, and
critically discuss various results on restricted diffusion in NMR
(Sect.~\ref{sec:survey}).  Residence time and other functionals of
reflected Brownian motion could as well be analyzed
(Sect.~\ref{sec:residence}).  The investigation of restricted
diffusion in complex geometries based on the study of the properties
of the underlying Laplace operator eigenbasis is our principal 
guideline.

Many questions about diffusion in complex geometries remain open.  On
the one hand, a further study of restricted diffusion in model
geometries (such as packs of spheres or cylinders) will reveal and
help to better understand various features of this process.  On the
other hand, it is possible to apply the developed arsenal of
theoretical and numerical methods to complex structures in nature and
industry: a three-dimensional porous morphology of a cement paste
reconstructed by X-ray microtomography; skin interstitial space; the
human placenta; the lungs, to name a few.  Although a complete
comprehension of diffusion in natural geometries is irrealistic
because of their complexity, this very complexity may help in
averaging out specific features of the confining media and of
transport processes.  A practical goal of this study is to reveal
relevant characteristics of the medium that would essentially
determine and control diffusive transport.


\newpage
\section*{ Acknowledgments }
\addcontentsline{toc}{section}{{Acknowledgments}}

A cross-disciplinary character of the presented works could not be
maintained without collaborations.  The presented results have been
obtained together with J. S. Andrade, M. Filoche, G. Guillot,
P. Levitz, B. Sapoval, and M. Zinsmeister.  I'm grateful to
B. Duplantier, T. Iakovleva, P. Levitz and M. Plapp for careful
proofreading of the manuscript and numerous advices.


\newpage
\addcontentsline{toc}{section}{{Bibliography}}

\end{document}